\documentclass[preprint,review,3p]{elsarticle}
\usepackage{graphicx}
\usepackage{comment}
\usepackage{algorithm,algorithmic}
\usepackage{ctable}
\usepackage{caption}
\usepackage[labelformat = empty,position=top]{subcaption}
\usepackage[export]{adjustbox}
\usepackage{xcolor}
\usepackage{amssymb}
\usepackage{natbib}
\usepackage{amsmath}
\usepackage{bbm}
\usepackage{wrapfig}
\usepackage{cases}
\usepackage{mathtools}
\usepackage{siunitx}
\usepackage[flushleft]{threeparttable}
\newcommand{\ud}{\mathrm{d}}
\newcommand{\RN}[1]{%
  \textup{\uppercase\expandafter{\romannumeral#1}}%
}

\journal{Journal of the Mechanics and Physics of Solids}

\begin{document}

\begin{frontmatter}

\title{Quantifying the effect of hydrogen on dislocation dynamics: A three-dimensional discrete dislocation dynamics framework}
%% Group authors per affiliation:
 \author{Yejun Gu\corref{cor1}}
 \ead{yejungu@jhu.edu}
 \author{Jaafar A. El-Awady\corref{cor1}}
 \ead{jelawady@jhu.edu}

 \address{Department of Mechanical Engineering, Whiting School of Engineering, The Johns Hopkins University, Baltimore, MD 21218, USA}\cortext[cor1]{Corresponding author}

\begin{abstract}
We present a new framework to quantify the effect of hydrogen on dislocations using large scale three-dimensional (3D) discrete dislocation dynamics (DDD) simulations. In this model, the first order elastic interaction energy associated with the hydrogen-induced volume change is accounted for. The three-dimensional stress tensor induced by hydrogen concentration, which is in equilibrium with respect to the dislocation stress field, is derived using the Eshelby inclusion model, while the hydrogen bulk diffusion is treated as a continuum process. This newly developed framework is utilized to quantify the effect of different hydrogen concentrations on the dynamics of a glide dislocation in the absence of an applied stress field as well as on the spacing between dislocations in an array of parallel edge dislocations. A shielding effect is observed for materials having a large hydrogen diffusion coefficient, with the shield effect leading to the homogenization of the shrinkage process leading to the glide loop maintaining its circular shape, as well as resulting in a decrease in dislocation separation distances in the array of parallel edge dislocations. On the other hand, for materials having a small hydrogen diffusion coefficient, the high hydrogen concentrations around the edge characters of the dislocations act to pin them. Higher stresses are required to be able to unpin the dislocations from the hydrogen clouds surrounding them. Finally, this new framework can open the door for further large scale studies on the effect of hydrogen on the different aspects of dislocation-mediated plasticity in metals. With minor modifications of the current formulations, the framework can also be extended to account for general inclusion-induced stress field in discrete dislocation dynamics simulations.
\end{abstract}

\end{frontmatter}

%###############################################################
\section{Introduction}
Hydrogen (H) embrittlement is one of the most common types of environmentally induced damage that affects the reliable performance of structural materials in many applications~\cite{robertson2001,pundt2006,barnoush2008,barnoush2010,song2013, huang2017}. One of the earliest studies on this topic can be dated back to 1874, when Johnson first reported the reduction in ductility and fracture stress in iron and steels caused by the presence of hydrogen~\cite{johnson1874}. Since then, numerous investigations have been conducted to understand the H embitterment (HE) of metals. In metals, pre-existing dislocations are ubiquitous, and H-dislocation interactions will impact almost all stages of plastic deformation. Generally, three main mechanisms have been proposed to explain HE in metals, namely, stress-induced hydride formation~\cite{takano1974,gahr1977}, hydrogen-enhanced decohesion (HEDE)~\cite{zapffe1941,oriani1972,oriani1979} and hydrogen-enhanced localized plasticity (HELP)~\cite{beachem1972,birnbaum1994}.

For decades, it was widely accepted that intergranular fracture is mainly driven by HEDE at the grain boundary (GB) in non hydride-forming systems in high-pressure H environments. However, recent experimental observations have indicated the importance of dislocation plasticity and HELP mechanism on the observed failure in such environments. Martin et al. have reported the observation of very dense dislocation microstructures forming underneath the brittle-like fractured surfaces in Ni~\cite{martin2012} and Fe~\cite{martin2011,martin2013,wang2014} deformed in high-pressure H environments, even though the samples failed at relatively low strains. Such dense dislocation micrstructures are only plausible at a much higher strain levels in the absence of H~\cite{keller2010}. These studies have instigated the need to quantify the effect of H on the collective behavior of dislocations. 

Atomistic simulations have been commonly used over the past two decades to examine the effect of H on the different aspects of deformation in metals. In particular, molecular dynamics (MD) simulations have been performed to examine the effect of H on dislocation emission~\cite{zhou1998,wen2009}, dislocation nucleation~\cite{wen2009} and dislocation mobility~\cite{song2014,nedelcu2002}. In addition, molecular statics (MS) simulations~\cite{von2011,song2011,tang2012} and ab initio calculations \cite{lu2001,itakura2013} were also conducted to evaluate the interactions of H with pre-existing dislocations. Nevertheless, all those simulations are only able to quantify the effect of H on either dislocation nucleation from surface, which is a debatable mechanism in bulk single crystals or coarse-grained polycrystals, or investigate the interactions of H with individual dislocations due to the limited time and length scales in these simulations. The typical volume in molecular simulations being less than $\SI{e6}{\cubic\nano\metre}$ is too small to obtain a definitive understanding on the mechanisms associated with dislocation network evolutions in the presence of H. On the other hand, existing continuum models accounting for the effect of H~\cite{sofronis2001,bammann2005,novak2009} smear out dislocation networks and relay on empirical fitting with some ad-hoc assumptions in the constitutive equations typically used, therefore cannot accurately probe dislocation microstructure evolutions in H environment.

As such, it is necessary to develop a simulation framework at the mesosacle to be able to study the influence of H on the collective behavior of dislocations. Large scale three-dimensional discrete dislocation dynamics simulations provide an avenue for such studies. In DDD, the atomistic details of the material are averaged out by focusing on the defects at the mesoscale~\cite{kubin1993,zbib1998,ghoniem2000,arsenlis2007}. Unlike other mesoscale models of plasticity which consider the dislocation density in terms of a homogenized field, 3D DDD simulations models the collective motion of many dislocations explicitly so that individual dislocation-dislocation interactions can be properly evaluated. These discrete models have typical length and time scales on the order of microns and seconds, reaching microscopic experimental scales that dislocations are observed in real time~\cite{bulatov2006}. Over the past two decades, DDD simulations have been employed successfully to study many aspects of dislocation-mediated plasticity in metals (e.g.~\cite{bulatov_2006,el2009,fan2015,el2016}), and by incorporating the effect of H into this framework, many aspects of the HE problem can be effectively studied.
%There have been numerous applications of DDD methods in accounting for interesting phenomena, such as strain hardening of crystals~\cite{bulatov_2006, arsenlis2007, devincre2008,greer2008,el2009,el2009a}, crack tip plasticity and dislocation-crack interaction~\cite{van2002,mastorakos2006} and dislocation climb related phenomena~\cite{Raabe1998, ghoniem2000, Xiang2003, Xiang2006, arsenlis2007, Mordehai2008a,Mordehai2011, Keralavarma2012,Raabe2013}.

Here, we present a new 3D DDD framework that accounts for both H-dislocation interactions as well as H bulk diffusion. The paper is organized as follows. In Section~\ref{sec:hdis}, a mathematical formulation of the H distribution in equilibrium with respect to the dislocation stress field is derived. In Section~\ref{sec:stress}, a new analytical formulation for the H-induced stress field in three-dimensional space is developed, in which H concentration is treated as a continuum variable. This is a generalization of an earlier two dimensional space formulation \cite{cai2014}. In Section~\ref{sec:impddd}, we introduce the numerical details of incorporating H bulk diffusion and H-dislocation interaction formulation and hydrogen diffusion process appropriate for use in the 3D DDD code ParaDiS. Several numerical applications of this newly developed H-induced DDD framework are discussed in Section~\ref{sec:example}. Finally, the conclusions of this work are summarized in Section~\ref{sec:conc}.

%###############################################################
\section{Hydrogen distribution in an infinite solid}\label{sec:hdis}

The reference H volume concentration in a solid in the absence of any internal stresses other than those induced by H atoms can be defined as follows,

\begin{align}
c_0 = \frac{N_{{\rm H},0}}{V},
\end{align}

\noindent where $N_{{\rm H},0}$ is the number of atomic sites occupied by H atoms, and $V$ is the total volume. The reference mole fraction is then defined as

\begin{align}
\chi_0 = \frac{N_{{\rm H},0}}{N_{\rm max}},
\end{align}

\noindent where $N_{\rm max}$ is the total number of possible occupation sites for H atoms in the solid. By assuming that two H atoms cannot occupy the same solute site, the maximum H volume concentration is $c_{\rm max}=N_{\rm max}/V$.

For an infinite solid subject to a uniform, non-zero external loading, $P^{\rm A}=-\sigma^{\rm A}_{kk}/3$, the chemical potential of H atoms in an otherwise perfect bulk can be expressed as

\begin{align}
\mu = \mu_0 + P^{\rm A}\Delta V + k_{\rm B}T\ln\frac{\chi}{1 - \chi},
\end{align}

\noindent where $\mu_0$ is the reference chemical potential for a H-atom inclusion in an infinite medium in a zero pre-existing stress state, and $\Delta V$ is the free expansion of the matrix due to the H-inclusion. The reference H mole fraction can be computed for the case of $\mu = 0$ and is

\begin{align}
\chi_0 = \left[1 + \exp\left(\frac{\mu_0 + P^{\rm A}\Delta V}{k_BT}\right)\right]^{-1}.
\label{eq:chi0}
\end{align}

It should be noted that a H-atom inclusion in the matrix does not affect the work required to insert another H atom in the solid, since the stress field of a H-atom is purely deviatoric outside the H-atom inclusion.

In the presence of dislocations, the H-dislocation interaction energy, $W_{\rm int}$, must be accounted for in the chemical potential expression, which would now be expressed as

\begin{align}
\mu = \mu_0 + P^{\rm A}\Delta V + k_{\rm B}T\ln\frac{\chi}{1 - \chi} + W_{\rm int}.
\end{align}

\noindent Thus, the H-flux, $\mathbf{J}$, in the solid can be directly computed from Fick's first law of diffusion:

\begin{align}
\mathbf{J} = -\frac{D_Hc_{\max}\chi}{k_{\rm B}T}\nabla\mu,
\end{align}

\noindent where $D_H$ is the H bulk diffusion coefficient, and $\nabla = \left(\frac{\partial}{\partial x_1}, \frac{\partial}{\partial x_2}, \frac{\partial}{\partial x_3}\right)$ is the gradient with respect to the coordinate system $\mathbf{x} = \left(x_1, x_2, x_3\right)$. In addition, the rate of change of the H mole fraction can be computed from Fick's second law of diffusion: %% Ref: DIFFUSION APPROXIMATION AND HOMOGENIZATION OF THE SEMICONDUCTOR BOLTZMANN EQUATION, SIAM MMM,2005

\begin{align}
\frac{\partial \chi}{\partial t} = -\nabla\cdot \mathbf{J}/c_{\max} = D_H\nabla\cdot\left(\frac{\nabla\chi}{1 - \chi}\right) + \frac{D_H}{k_{\rm B}T}\nabla\cdot(\chi\nabla W_{\rm int}).
\label{eq:Hdiffusion}
\end{align}

\noindent For a H distribution in equilibrium with respect to $W_{\rm int}$, Eq. \eqref{eq:Hdiffusion} reduces to

\begin{align}
\nabla\cdot\left(\frac{\nabla\chi}{1-\chi}\right)+\frac{1}{k_{\rm B}T}\nabla\cdot(\chi\nabla W_{\rm int})=0.
\label{eq:Hpos}
\end{align}

Finally, the solution to Eq. \eqref{eq:Hpos} yields the H equilibrium concentration in an infinite solid, which obeys the Fermi-Dirac distribution \cite{cai2014} and is given by

\begin{align}
\chi(\mathbf{x}) = \left[1 + \frac{1 - \chi_0}{\chi_0}\exp\left(\frac{W_{\rm int}}{k_BT}\right)\right]^{-1}.
\label{eq:sol-Hpos}
\end{align}

It should be noted that here only the H bulk diffusion (hereinafter referred to as ``H diffusion" for the sake of simplicity) is accounted for, while the effect of H pipe diffusion will be addressed elsewhere.
%###############################################################
\section{Hydrogen-dislocation interactions formulations}\label{sec:stress}

The interaction energy, $W_{\rm int}$, associated with a H-atom inclusion can be decomposed into several terms \cite{sofronis1995b}, namely, the first order elastic interaction energy between H atoms and dislocations, $W^{(1)}_{\rm int}$ \cite{eshelby1957}, the second order elastic interaction energy that arises from the moduli change due to the introduction of H atoms, $W^{(2)}_{\rm int}$ \cite{eshelby1955}, as well as higher order energy terms (e.g. H-H interaction \cite{von2011} etc.).

Here, as a first attempt, we will focus on the first order interaction energy to understand the effect of H-atom inclusions on the dislocation dynamics and dislocation-mediated plasticity. This is consistent with the assumptions in the Eshelby inclusion model, in which the elastic constants of the material are not affected by the inclusion \cite{eshelby1957,eshelby1959}. Accordingly, in the following a H atom is modeled as an Eshelby inclusion \cite{eshelby1957,eshelby1959} with the following assumptions:

\begin{enumerate}
\item The matrix is a homogeneous and isotropically elastic medium with a shear modulus $G$, Poisson's ratio $\nu$ and bulk modulus $K$.
\item The H atom has a spherical shape and produces purely dilatational eigenstrain. That is to say, a spherical occupation-site with a radius $r_0$ is expanded to be a sphere with a radius $(1 + \varepsilon)r_0$ upon the introduction of a H atom, where $\varepsilon$ is a small positive number related to the volume of a H atom. The associated unconstrained volume change (i.e. the free expansion of the H inclusion out of the matrix) is
    \begin{align}
    \Delta V& = 4\pi\left[(1 + \varepsilon)^3 - 1\right]r_0^3/3 \approx 4\pi r_0^3\varepsilon.
    \end{align}
\item The H inclusion is elastically isotropic and its elastic constants are identical to those of the surrounding matrix.
\item The H atoms are well-separated such that no two atoms can overlap.
\end{enumerate}

This simple Eshelby inclusion model will help in the derivation of an analytical formulation to quantify the effect of H atoms on the mechanical properties of materials. According to Eshelby's solution for a single misfitting inclusion in an infinite medium \cite{eshelby1957,eshelby1959,eshelby1961}, the first order interaction energy of a H-atom inclusion in the presence of a dislocation stress field $\sigma_{ij}^d$ can be given by

\begin{align}
W_{\rm int}^{(1)} = P\Delta V,
\label{eq:int_ene1}
\end{align}

\noindent where $P = -\frac{1}{3}\sigma_{kk}^d$ is the hydrostatic pressure associated with the dislocation stress field. By substituting Eq. \eqref{eq:int_ene1} into Eq. \eqref{eq:Hpos}  the H equilibrium concentration in the bulk in the presence of a dislocation stress field, can be evaluated as

\begin{align}
\chi(\mathbf{x}) = \left[1 + \frac{1-\chi_0}{\chi_0}\exp\left(-\frac{\sigma_{kk}^{d}(\mathbf{x})\Delta V}{3k_{\rm B}T}\right)\right]^{-1}.
\label{eq:chi}
\end{align}

\noindent Linearizing Eq. \eqref{eq:chi}, the H concentration in the solid can be approximated by

\begin{align}
c(\mathbf{x}) - c_0 \approx &c_{max}\chi_0(1 - \chi_0)\frac{\sigma_{kk}^d(\mathbf{x})\Delta V}{3k_{\rm B}T},
\label{eq:c-linear}
\end{align}

\noindent where $c_0=c_{\max}\chi_0$ is the reference H concentration. While this linearized approximation is valid for the case of a small hydrostatic stress, it would fail at the dislocation core. Nevertheless, the usefulness of this method can be realized here given the smearing out of the dislocation core in discrete dislocation dynamics simulations. This is analogous to the linearlized approximation used in the dislocation climb problem \cite{hirth1982theory,gu2015}.

Now, recall that the stress tensor of an arbitrary dislocation, $\Gamma$, can be expressed as \cite{hirth1982theory}

\begin{align}
\boldsymbol{\sigma}^d=&\frac{G}{4\pi}\int_{\Gamma}\left(\mathbf{b}\times\nabla'\right)\frac{1}{R}\otimes\ud\mathbf{l}'+\frac{G}{4\pi}\int_{\Gamma}\ud\mathbf{l}'\otimes\left(\mathbf{b}\times\nabla'\right)\frac{1}{R}\nonumber\\
&-\frac{G}{4\pi(1-\nu)}\int_{\Gamma}\nabla'\cdot\left(\mathbf{b}\times\ud \mathbf{l}'\right)\left(\nabla\otimes\nabla-\mathbf{I}\nabla^2\right)R,
\label{eq:stress-d}
\end{align}

\noindent where $R = \sqrt{(x_1-x_1')^2+(x_2-x_2')^2+(x_3-x_3')^2}$, and $\nabla' = \left(\frac{\partial}{\partial x_1'}, \frac{\partial}{\partial x_2'}, \frac{\partial}{\partial x_3'}\right)$, is the gradient with respect to $\mathbf{x}'$. Thus, the dislocation-induced hydrostatic stress is 

\begin{align}
\frac{\sigma^d_{kk}}{3}=\frac{G(1+\nu)}{12\pi(1-\nu)}\int_\Gamma b_m\varepsilon_{lmk}\frac{\partial}{\partial x_l'}\nabla'^2R\ud x_k'.
\label{eq:hydros-stress-d}
\end{align}

On the other hand, the H-induced stress field can be expressed as follows (see \ref{sec:app} for detailed derivation as well as Refs. \cite{pap1932,neuber1934,wolfer1985}),

\begin{align}
\sigma_{ij}^{\rm H} = &-\frac{G}{2(1-\nu)}B_{0,ij} - \frac{2G(1+\nu)}{3(1-\nu)}\delta_{ij}(c - c_0)\Delta V,
\label{eq:sigH}
\end{align}

\noindent where $\delta_{ij}$ is the Kronecker delta, and the Papkovich-Neuber scalar potential, $B_0$, is the solution of the Poisson's equation (see Eq. \eqref{eq:app6} for detailed derivation of this Poisson's equation):

\begin{align}
\nabla^2 B_0 = -\frac{4}{3}(1+\nu)(c - c_0)\Delta V.
\label{eq:pn-potential}
\end{align}

\noindent Substituting Eq. \eqref{eq:c-linear} with Eq. \eqref{eq:hydros-stress-d} into Eq. \eqref{eq:pn-potential}, one obtains the following Poisson's equation:

\begin{align}
\nabla^2 B_0=&-\frac{G(1+\nu)^2\Delta V^2}{9\pi(1-\nu)k_{\rm B}T}c_{max}\chi_0(1-\chi_0)\int_\Gamma b_m\varepsilon_{lmk}\frac{\partial}{\partial x_l'}\nabla'^2R\ud x_k',
\label{eq:poisson-pn}
\end{align}

\noindent By utilizing the identity $\nabla^2 R = \nabla'^2R$, the solution to Eq. \eqref{eq:poisson-pn} is

%\begin{align}
%\int_\Gamma b_m\varepsilon_{lmk}\frac{\partial}{\partial x_l'}\nabla'^2R\ud x_k'=\nabla^2\int_\Gamma b_m\varepsilon_{lmk}\frac{\partial}{\partial x_l'}R\ud x_k'.
%\label{eq:identity1}
%\end{align}

\begin{align}
B_0=-\frac{G(1+\nu)^2\Delta V^2}{9\pi(1-\nu)k_{\rm B}T}c_{max}\chi_0(1-\chi_0)\int_{\Gamma}b_m\varepsilon_{lmk}\frac{\partial R}{\partial x_l'}\ud x_k'.
\label{eq:sol2poisson-pn}
\end{align}

\noindent From Eqs \eqref{eq:sigH} and \eqref{eq:sol2poisson-pn}, the H-induced stress tensor can be expressed as follows,

\begin{align}
\sigma_{ij}^{H} = &\frac{G^2(1+\nu)^2\Delta V^2}{18\pi(1-\nu)^2k_{\rm B}T}c_{max}\chi_0(1-\chi_0)\int_{\Gamma}b_m\varepsilon_{lmk}\left(\frac{\partial^3 R}{\partial x_l'\partial x_i'\partial x_j'}-\delta_{ij}\frac{\partial}{\partial x_l'}\nabla'^2R\right)\ud x_k'
\label{eq:sigma_H}
\end{align}

It is clear by comparing Eqs \eqref{eq:sigma_H} and \eqref{eq:stress-d} that the functional forms of the line integrals in both equations for the dislocation-induced stress tensor and the H-induced stress tensor are identical, which indicates that the computational costs to numerically compute both stresses are of the same order.

%$\sigma_{13}^{\rm H}$ and $\sigma_{23}^{\rm H}$ are both $0$ due to the symmetry with respect to $xy-$plane, e.g.
%\begin{align}
%\sigma^{\rm H}_{13}(\mathbf{x})&=0.
%\end{align}
%Thus the introduction of hydrogen concentration which is uniform along $z-$ axis has no contribution to screw dislocation glide.

%This $3D$ hydrogen-induced stress formulation is reduced to Eq.~(B.7-10) in Ref.~\cite{cai2014} for an infinitely long straight edge dislocation along $z-$axis with the Burgers vector $\mathbf{b}=(b_1,0,0)$ in an infinite medium, which are listed below,
%\begin{align}
%&\sigma_{12}^H=\frac{G A_1}{1-\nu}\frac{x_1(x_1^2-x_2^2)}{(x_1^2+x_2^2)^2},\\
%&\sigma_{11}^H=\frac{G A_1}{1-\nu}\frac{x_2(x_2^2-x_1^2)}{(x_1^2+x_2^2)^2}-\frac{4A_2x_2}{x_1^2+x_2^2},\\
%&\sigma_{22}^H=\frac{G A_1}{1-\nu}\frac{x_2(3x_1^2+x_2^2)}{(x_1^2+x_2^2)^2}-\frac{4A_2x_2}{x_1^2+x_2^2},\\
%&\sigma_{11}^H-\sigma_{22}^H=-\frac{G A_1}{1-\nu}\frac{4x_1^2x_2}{(x_1^2+x_2^2)^2}.
%\end{align}

The overall stress at any field point, $\mathbf{x}$, can thus be computed as follows,

\begin{align}
\sigma_{ij}=&\sigma_{ij}^d+\sigma_{ij}^H\nonumber\\
=&-\frac{G}{8\pi}\int_{\Gamma}b_m\varepsilon_{lmi}\frac{\partial}{\partial x_l'}\nabla'^2R\ud x_{j}'-\frac{G}{8\pi}\int_{\Gamma}b_m\varepsilon_{lmj}\frac{\partial}{\partial x_l'}\nabla'^2R\ud x_{i}'\nonumber\\
&+\left[\frac{G^2(1+\nu)^2\Delta V^2}{18\pi(1-\nu)^2k_{\rm B}T}c_{max}\chi_0(1-\chi_0)-\frac{G}{4\pi(1-\nu)}\right]\int_{\Gamma}b_m\varepsilon_{lmk}\Bigg(\frac{\partial^3R}{\partial x_l'\partial x_i'\partial x_j'}-\delta_{ij}\frac{\partial}{\partial x_l'}\nabla'^2R\Bigg)\ud x_k'.
\label{eq:stress-overall}
\end{align}

%#############################################################################################
\section{Accounting for hydrogen in the framework of three-dimensional discrete dislocation dynamics simulations}\label{sec:impddd}

All implementations performed here are conducted using an in-house, modified version of the 3D DDD open source code ParaDiS \cite{arsenlis2007} that guarantees all dislocation reactions are planar, and incorporates a set of atomistically-informed and physics-based cross-slip rules, the details of which are described elsewhere \cite{HUSSEIN2015}. In the following, only the relevant points that pertain to the incorporation of H into this 3D DDD framework are discussed.

\subsection{Dislocation nodal forces}

In this 3D DDD framework, each dislocation is discretized into connected straight segments. The endpoints of each segment are referred to as dislocation nodes. The time evolution of each dislocation segment is computed by utilizing a mobility law, which relates the dislocation nodal velocities to the nodal forces. As shown by Eq. \eqref{eq:sigma_H}, H atoms induce an additional stress field, $\sigma^H$, in the simulation cell, that should be evaluated when computing the dislocation nodal forces. 

\begin{figure}[!htb]
\centering
{\includegraphics[width=12cm]{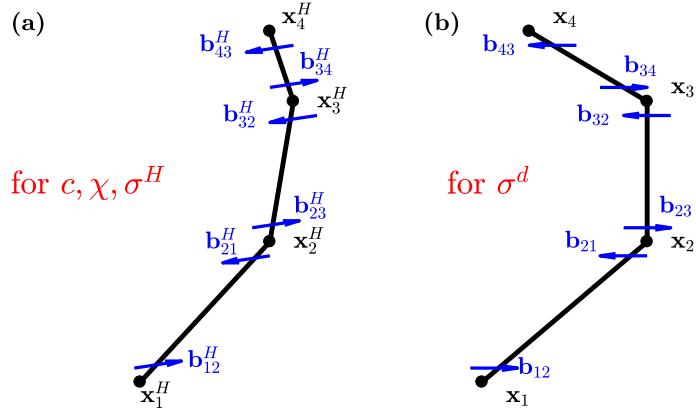}
\caption{Schematic diagram of the: (a) H-related dislocation nodes; and (b) actual dislocation nodes.}
\label{fig:node-notation}}
\end{figure}

The evaluation of the H-induced stress field requires full knowledge of the H distribution in the simulation cell at every simulation time step. Accordingly, a new set of ``virtual'' dislocation nodes that pertain only to the calculations of the H distribution  are introduced here. These nodes will be termed hereafter ``\textit{H-related dislocation nodes}'' in order to distinguish them from the conventional (i.e. actual) dislocation nodes. It should be noted that the H-related dislocation nodes do not necessarily overlap with the conventional dislocation nodes, as will be discussed in Sec. \ref{subsec:hdiff}. The conventional dislocation nodes and the H-related nodes are shown schematically in Fig. \ref{fig:node-notation}.

%--------------------------
%--------------------------

The stress tensor at any point $\mathbf{x}$ in the simulation cell induced by a dislocation segment having Burgers vector $\mathbf{b}_{12}$ and bounded by the two conventional dislocation nodes $\mathbf{x}_1$ and $\mathbf{x}_2$ (see Fig. \ref{fig:node-notation}) can be expressed as \cite{cai2006,arsenlis2007}

\begin{align}
\boldsymbol{\sigma}^{d,\{12\}}(\mathbf{x})=&-\frac{G}{8\pi}\int_{\mathbf{x}_1}^{\mathbf{x}_2}\left(\frac{2}{R_a^3}+\frac{3a^2}{R_a^5}\right)\left[(\mathbf{R}\times\mathbf{b}_{12})\otimes \ud\mathbf{x}'	+\ud\mathbf{x}'\otimes(\mathbf{R}\times\mathbf{b}_{12})\right]\nonumber\\
&+\frac{G}{4\pi(1-\nu)}\int_{\mathbf{x}_1}^{\mathbf{x}_2}\left(\frac{1}{R_a^3}+\frac{3a^2}{R_a^5}\right)\left[\left(\mathbf{R}\times\mathbf{b}_{12}\right)\cdot\ud\mathbf{x}'\right]\mathbf{I}\nonumber\\
&-\frac{G}{4\pi(1-\nu)}\int_{\mathbf{x}_1}^{\mathbf{x}_2}\frac{1}{R_a^3}\big(\left(\mathbf{b}_{12}\times\ud\mathbf{x}'\right)\otimes\mathbf{R}+\mathbf{R}\otimes\left(\mathbf{b}_{12}\times\ud\mathbf{x}'\right)\big)\nonumber\\
&+\frac{G}{4\pi(1-\nu)}\int_{\mathbf{x}_1}^{\mathbf{x}_2}\frac{3}{R_a^5}\left[\left(\mathbf{R}\times\mathbf{b}_{12}\right)\cdot\ud\mathbf{x}'\mathbf{R}\otimes\mathbf{R}\right],
\end{align}

\noindent where $a$ is the core width, $\mathbf{I}$ is the second order identity tensor, $\mathbf{R}=\mathbf{x}-\mathbf{x}'$, and $R_a=\sqrt{\mathbf{R}\cdot\mathbf{R}+a^2}$.

%Note that the linearized equilibrium hydrogen concentration in Eq.~\eqref{eq:c-linear} has the linear property with dislocation segments. For instance, the hydrogen concentration due to both the H-related dislocation segment from H-related dislocation node $\mathbf{x}_1^H$ to H-related dislocation node $\mathbf{x}_2^H$ (denoted by ``H-Seg 1") and the H-related dislocation segment from $\mathbf{x}_3^H$ to $\mathbf{x}_4^H$ (denoted by ``H-Seg 2") is equal to the summation of the hydrogen concentration due to H-Seg 1 and the concentration in response to H-Seg 2, i.e. $c(\text{due to H-Seg 1 and H-Seg 2})=c(\text{due to H-Seg 1})+c(\text{due to H-Seg 2})$. Hence the H-induced stress at a point $\mathbf{x}$ can be decomposed into a summation of the parts corresponding to each H-related segment.

One the other hand, the H-induced stress tensor corresponding to a virtual dislocation segment with a Burgers vector $\mathbf{b}_{12}^H$ and bounded by the H-related dislocation nodes $\mathbf{x}_1^{H}$ and $\mathbf{x}_2^{H}$ (see Fig. \ref{fig:node-notation}) can be expressed as

\begin{align}
\boldsymbol{\sigma}^{H,\{12\}}(\mathbf{x})=&-\frac{G^2(1+\nu)^2\Delta V^2}{18\pi(1-\nu)^2k_{\rm B}T}c_{max}\chi_0(1-\chi_0)\int_{\mathbf{x}_1^H}^{\mathbf{x}_2^H}\left(\frac{1}{R_a^3}+\frac{3a^2}{R_a^5}\right)\left[\left(\mathbf{R}\times\mathbf{b}_{12}^H\right)\cdot\ud\mathbf{x}'\right]\mathbf{I}\nonumber\\
&+\frac{G^2(1+\nu)^2\Delta V^2}{18\pi(1-\nu)^2k_{\rm B}T}c_{max}\chi_0(1-\chi_0)\int_{\mathbf{x}_1^H}^{\mathbf{x}_2^H}\frac{1}{R_a^3}\big(\left(\mathbf{b}_{12}^H\times\ud\mathbf{x}'\right)\otimes\mathbf{R}+\mathbf{R}\otimes\left(\mathbf{b}_{12}^H\times\ud\mathbf{x}'\right)\big)\nonumber\\
&-\frac{G^2(1+\nu)^2\Delta V^2}{18\pi(1-\nu)^2k_{\rm B}T}c_{max}\chi_0(1-\chi_0)\int_{\mathbf{x}_1^H}^{\mathbf{x}_2^H}\frac{3}{R_a^5}\left[\left(\mathbf{R}\times\mathbf{b}_{12}^H\right)\cdot\ud\mathbf{x}'\mathbf{R}\otimes\mathbf{R}\right].
\end{align}

It should be noted that for H-related dislocation segment, $\ud\mathbf{x}'$, that has a screw character (i.e. $\ud \mathbf{x}'\parallel \mathbf{b}_{12}^H$), both $(\mathbf{R}\times\mathbf{b}_{12}^H)\cdot\ud\mathbf{x}'=0$ and $\mathbf{b}_{12}^H\times\ud\mathbf{x}'=0$, which indicates that the H-related dislocation segment having a screw character do not contribute to the H-induced stress.

At the dislocation node $\mathbf{x}_4$ on the dislocation segment bounded by the conventional dislocation nodes $\mathbf{x}_3$ and $\mathbf{x}_4$ with a Burgers vector $\mathbf{b}_{43}$, the nodal force $\mathbf{f}^{d,\{12\}}(\mathbf{x}_4)$ resulting from the stress field $\boldsymbol{\sigma}^{d,\{12\}}$ is

\begin{align}
\mathbf{f}^{d,\{12\}}(\mathbf{x}_4)=&\int_0^L\frac{l}{L}\left[\boldsymbol{\sigma}^{d,\{12\}}\left(\Big(1-\frac{l}{L}\Big)\mathbf{x}_3+\frac{l}{L}\mathbf{x}_4\right)\cdot\mathbf{b}\right]\times\mathbf{t}~\ud l,\label{eq:nodalstr}
\end{align}
while the nodal force $\mathbf{f}^{H,12}$ arising from the stress field $\boldsymbol{\sigma}^{H,12}$ is
\begin{align}
\mathbf{f}^{H,\{12\}}(\mathbf{x}_4)=&\int_0^L\frac{l}{L}\left[\boldsymbol{\sigma}^{H,\{12\}}\left(\Big(1-\frac{l}{L}\Big)\mathbf{x}_3+\frac{l}{L}\mathbf{x}_4\right)\cdot\mathbf{b}\right]\times\mathbf{t}~\ud l,\label{eq:hnodalstr},
\end{align}

\noindent where $L=|\mathbf{x}_4-\mathbf{x}_3|$, and $\mathbf{t}=(\mathbf{x}_4-\mathbf{x}_3)/L$. Thus, the total nodal force at any dislocation node $\mathbf{x}_k$ would be

\begin{align}
\mathbf{f}(\mathbf{x}_k)=\sum_{\mathbf{x}_l}\sum_{(\mathbf{x}_i,\mathbf{x}_j)}\mathbf{f}^{d,\{ij\}}+\sum_{\mathbf{x}_l}\sum_{(\mathbf{x}_i^H,\mathbf{x}_j^H)}\mathbf{f}^{H,\{ij\}}+\mathbf{f}_{\rm ext}+\mathbf{f}_{\rm core},\label{eq:nodalf}
\end{align}
the summation symbol $\sum_{\mathbf{x}_l}$ is over all dislocation nodes $\mathbf{x}_l$ connected to node $\mathbf{x}_k$, $\sum_{(\mathbf{x}_i,\mathbf{x}_j)}$ is over all dislocation segments bounded by dislocation nodes $\mathbf{x}_i$ and $\mathbf{x}_j$ in the simulation cell, $\sum_{(\mathbf{x}_i^H,\mathbf{x}_j^H)}$ are over all H-related dislocation segments bounded by H-related dislocation nodes $\mathbf{x}^{H}_i$ and $\mathbf{x}^{H}_j$, $\mathbf{f}_{\rm ext}=(\boldsymbol{\sigma}_{\rm ext}\cdot\mathbf{b})\times\boldsymbol{\xi}$, and $\mathbf{f}_{\rm core}$ is the force term that accounts for the dislocation core energy.

%******************************************************************************************************
\subsection{Hydrogen diffusion}\label{subsec:hdiff}

In Sec. \ref{sec:hdis} and Sec. \ref{sec:stress}, the formulation for computing the H distribution in equilibrium with the dislocations 3D stress field was derived. Thus, to compute the H distribution in the framework of 3D DDD, three cases will be considered depending on the H diffusion coefficient of the solid being modeled. For reference, H diffusion parameters for different metals are summarized in Table \ref{table:diff}. If the time scale for H diffusion is on the same order as that for dislocation motion, the H distribution will almost always be in equilibrium with the evolving dislocation network at every time step of the simulations. On the other hand, if the time scale for H diffusion is smaller than that for dislocation motion, then a special computational considerations must be imposed to compute the H distribution every time step. 
%------------------------------------------
\begin{table}[!htb]
\caption{Hydrogen diffusion parameters~\cite{volkl1975}}
\label{table:diff}
\centering
\begin{threeparttable}
\begin{tabular}{l l l l l}
\hline
Material & Pre-exponetial  & Activation energy  & $D_H$ at $\SI{300}{\K}$ & $v_c\sqrt{\Delta_t/D_H}^{\dagger}$\\
& $D_{H0}~(\si{\square\metre\per\second})$ &$Q~(\SI{}{\J})$ & ($\si{\square\metre\per\second}$) & \\\hline
Pd & $2.90\times10^{-7}$ & $3.685\times10^{-20}$ & $3.90\times10^{-11}$ & 1.64\\
Ni~($T<\SI{627}{\K}$) & $4.76\times10^{-7}$ & $6.569\times10^{-20}$ & $5.97\times10^{-14}$ &40.9273\\
$\alpha-$Fe$^{\mathsection}$ & $4.00\times10^{-8}$ & $7.530\times10^{-21}$ & $6.47\times10^{-9}$ &0.1243\\
& $7.50\times10^{-8}$ & $1.410\times10^{-20}$ & $2.48\times10^{-9}$ & 0.2088\\
Nb~($T>\SI{273.15}{\K}$) & $5.0\times10^{-8}$ & $1.698\times10^{-20}$ & $8.22\times10^{-10}$&0.3488\\
Ta & $4.4\times10^{-8}$ & $2.243\times10^{-20}$ & $1.94\times10^{-10}$&0.7180\\
Va & $2.9\times10^{-8}$ & $6.889\times10^{-21}$ & $5.48\times10^{-9}$&0.1351\\
\hline
\end{tabular}
 \begin{tablenotes}
      \small
      \item The H diffusion coefficient is defined as $D_H=D_{H0}\exp(-\frac{Q}{k_BT})$.
      
      ${\dagger}$: $v_c\sim 1~m/s$ and $\Delta_t\sim 10^{-10}~s$.

      ${\mathsection}$: Two sets of parameters are reported.
         \end{tablenotes}
\end{threeparttable}
\end{table}
%------------------------------------------

For a given simulation time step, $\Delta_t$, the average distance traversed by dislocations in the simulation cell is $v_c\Delta_{t}$, while that traversed by H atoms is $\sqrt{D_H\Delta_{t}}$. Accordingly, when the ratio between both distances, $v_c\sqrt{\Delta_t/D_H}$, is much less than $1.0$ (i.e. in solids having a large H diffusion coefficient such as $\alpha$-Fe), it can be assumed that the H concentration is always in equilibrium with the evolving dislocation microstructure at every time step of the simulation. This condition will be referred to as ``Case \RN{1}'' hereafter. In this case, H concentration is solved every time step with the current dislocation stress field, and thus, the H-related dislocation nodes overlap with the actual dislocation nodes.

When $v_c\sqrt{\Delta_t/D_{\rm H}}\gg 1$ (i.e. in metals having a small $D_H$, such as Ni at room temperature), the H concentration evolves very slowly, if at all, in one DDD simulation time step and the new concentration profile does not differ much from that at the previous time step. Thus, it is safe to assume that the H concentration is fixed. This condition will be referred to as ``Case \RN{2}'' hereafter. When conducting quasi-static DDD simulations it can be assumed that the H concentration will reach equilibrium with the current dislocation microstructure when incremental plastic strain or alternatively the maximum dislocation node velocity is less than a small critical value. The exact values of the incremental plastic strain and/or dislocation nodal velocity must be chosen such that the overall error introduced due to the assumption that the H field remains static between every update is minimized. Such a criterion is commonly used in two-dimensional dislocation dynamics simulations of phase separation \cite{Quek2013} and dislocation climb \cite{Keralavarma2012}, where the time scale of dislocation glide and diffusion are different. It should be noted that in the illustrative simulations to be discussed in Sec.~\ref{sec:example} such a criterion is not used since the total simulation time is small and no relevant H-diffusion will occur. It should be noted that in this case the H-related dislocation nodes are not always the same as the actual dislocation nodes. This is due to the lag in the H-induced DDD algorithm since the H distribution does not evolve at all until a certain criterion is met, while the dislocation motion takes place at every simulation time step. As the criterion is triggered to update H-related dislocation nodes, all of the geometry properties (positions, connectivities, Burgers vectors, etc.) of H-related dislocation nodes are updated by the actual dislocation nodes at the current step.  

Finally, when $v_c\sqrt{\Delta_t/D_{\rm H}}\sim 1$(e.g. in Pd at room temperature), the time scale for H diffusion is comparable to that for dislocation motion. This will be referred to as ``Case \RN{3}'' hereafter. In this case, to accurately describe the H diffusion process, the diffusion equation, Eq. \eqref{eq:Hdiffusion}, must be modified to add on a velocity related term to account for the solute drag effect \cite{sills2016} and be integrated every time step in the DDD framework coupled with the finite element method (FEM) \cite{crone2014} or boundary element method (BEM) \cite{el2008}. However, in practice this will be computationally expensive due to the high resolution requirement at dislocation cores. This case will not be simulated in the following numerical examples and will be addressed elsewhere.

%#####################################################################
\section{Numerical examples}\label{sec:example}

The above implementation of H in the 3D DDD framework is utilized to study the effect of H on two dislocation dynamics problems of general interest. The material properties used in the current simulations are summarized in Table \ref{table:parameter}, which are those of Ni. However, to give a quantitative understanding of the effect of the magnitude of the H diffusion coefficient on the dislocation dynamics, the simulations were performed for both a large H diffusion coefficient (mimicking Ni at room temperature) and a small H diffusion coefficient (mimicking $\alpha$-Fe). 

%------------------------------------------
\begin{table}[!htb]
\centering
\begin{threeparttable}
\caption{Material properties adopted in all simulations}
\label{table:parameter}
\begin{tabular}{l l l}
\hline
Property$^\dagger$ & Symbol & Value\\
\hline
Temperature & T & $\SI{300}{\K}$\\
Lattice parameter & $a_0$ & $\SI{3.524}{\angstrom}$ \\
Burgers vector magnitude & $b$ & $\SI{2.49}{\angstrom}$\\
%$k_B$ & $1.381\times10^{-23}~$\SI{}{\J}$\cdot$\SI{}{\K}$^{-1}$\\
Shear Modulus & $G$ & $\SI{74.73}{\GPa}$ \cite{wen2011} \\
Poisson ratio & $\nu$ & $0.318$ \cite{wen2011} \\
Change in volume due to a H-atom inclusion & $\Delta V$ & $\SI{1.4}{\cubic\angstrom}$ \cite{thomas1983} \\
Atomic volume & $\Omega$	  & $\SI{10.77}{\cubic\angstrom}$ \cite{thomas1983} \\          % This should be |Omega only since Delta V is given in the previous line
%$M_{\rm g}$ & $4.0\times 10^4~$\SI{}{\m}$\cdot$\SI{}{\N}$^{-1}\cdot$\SI{}{\s}$^{-1}$  & \\
\hline
\end{tabular}
 \begin{tablenotes}
      \small
    \item $\dagger$: The mechanical properties associated with H concentration changes are assumed to be negligible in nickel \cite{haftbaradaran2011}.
      \end{tablenotes}
\end{threeparttable}
\end{table}
\subsection{Effect of hydrogen on the dynamics of a circular dislocation glide loop}

\begin{figure}[!htb]
\centering
\includegraphics[width=8cm]{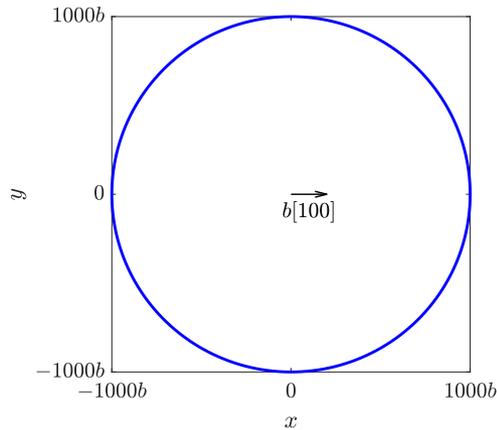}
\caption{A circular glide loop having an initial radius $R_0 = 1000b$ lying on a $(001)$ plane with a Burgers vector $b[100]$.}
\label{fig:gloop_sch}
\end{figure}

\begin{figure}[!htb]
        \centering
        %\subfigure[$\chi_0=0.1$]
        \begin{subfigure}[t]{0.03\textwidth}
        \textbf{(a)}
        \end{subfigure}
        \begin{subfigure}[t]{0.46\textwidth}
        \includegraphics[width=8cm,valign=t]{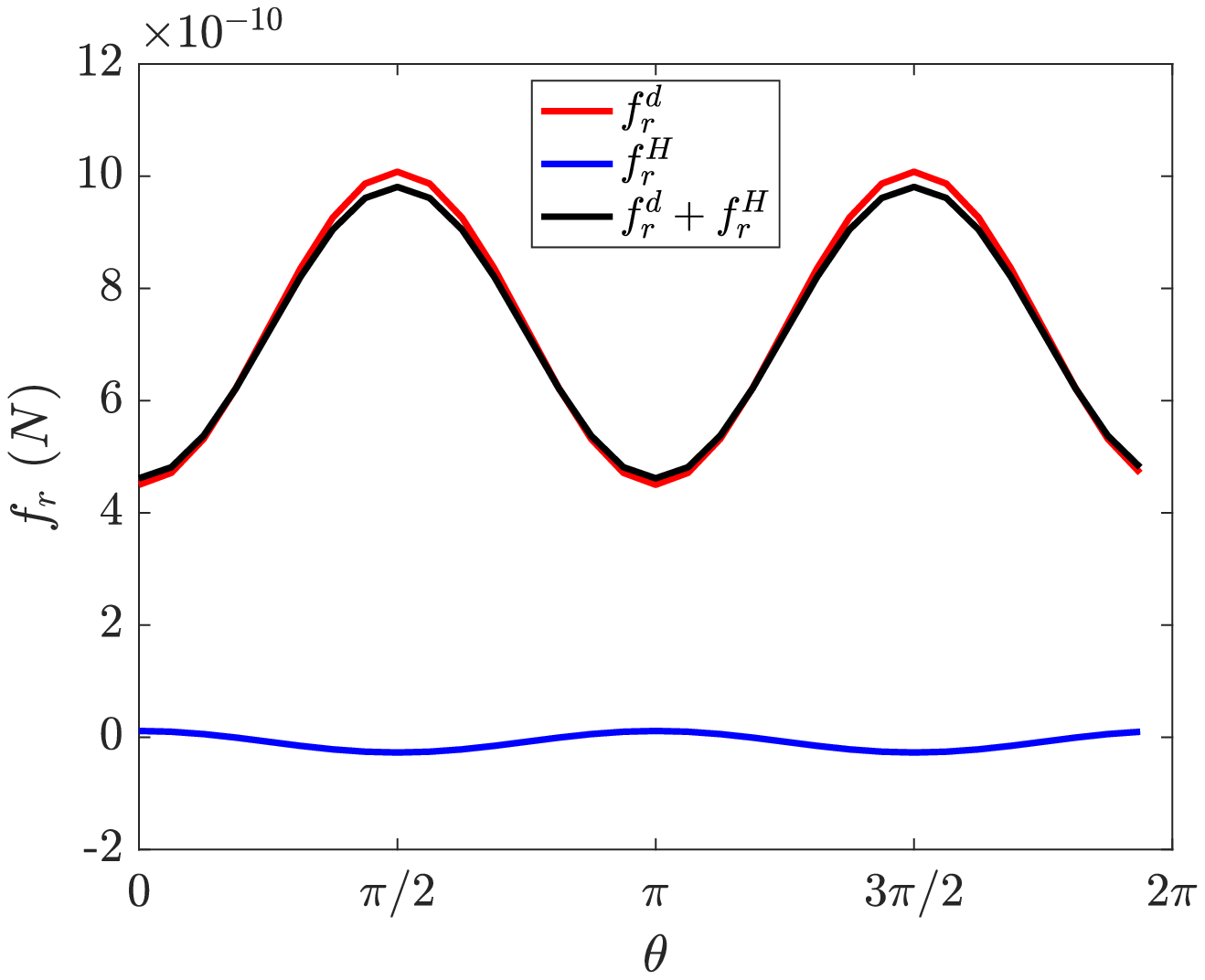}
        \end{subfigure}\hfill
        \begin{subfigure}[t]{0.03\textwidth}
        \textbf{(b)}
        \end{subfigure}
        \begin{subfigure}[t]{0.46\textwidth}
        \includegraphics[width=8cm,valign=t]{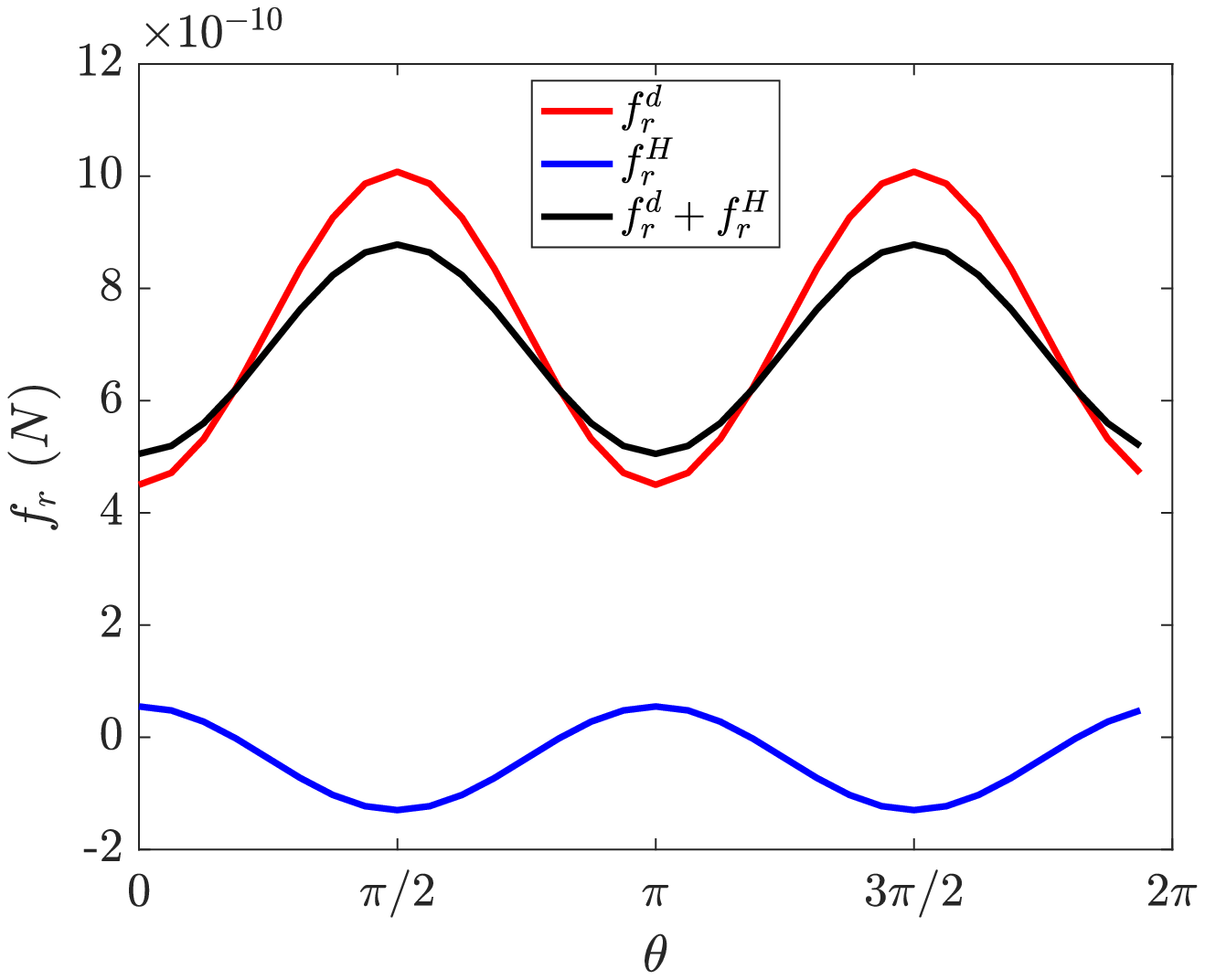}
        \end{subfigure}\hfill
        \begin{subfigure}[t]{0.03\textwidth}
        \textbf{(c)}
        \end{subfigure}
        \begin{subfigure}[t]{0.46\textwidth}
        \includegraphics[width=8cm,valign=t]{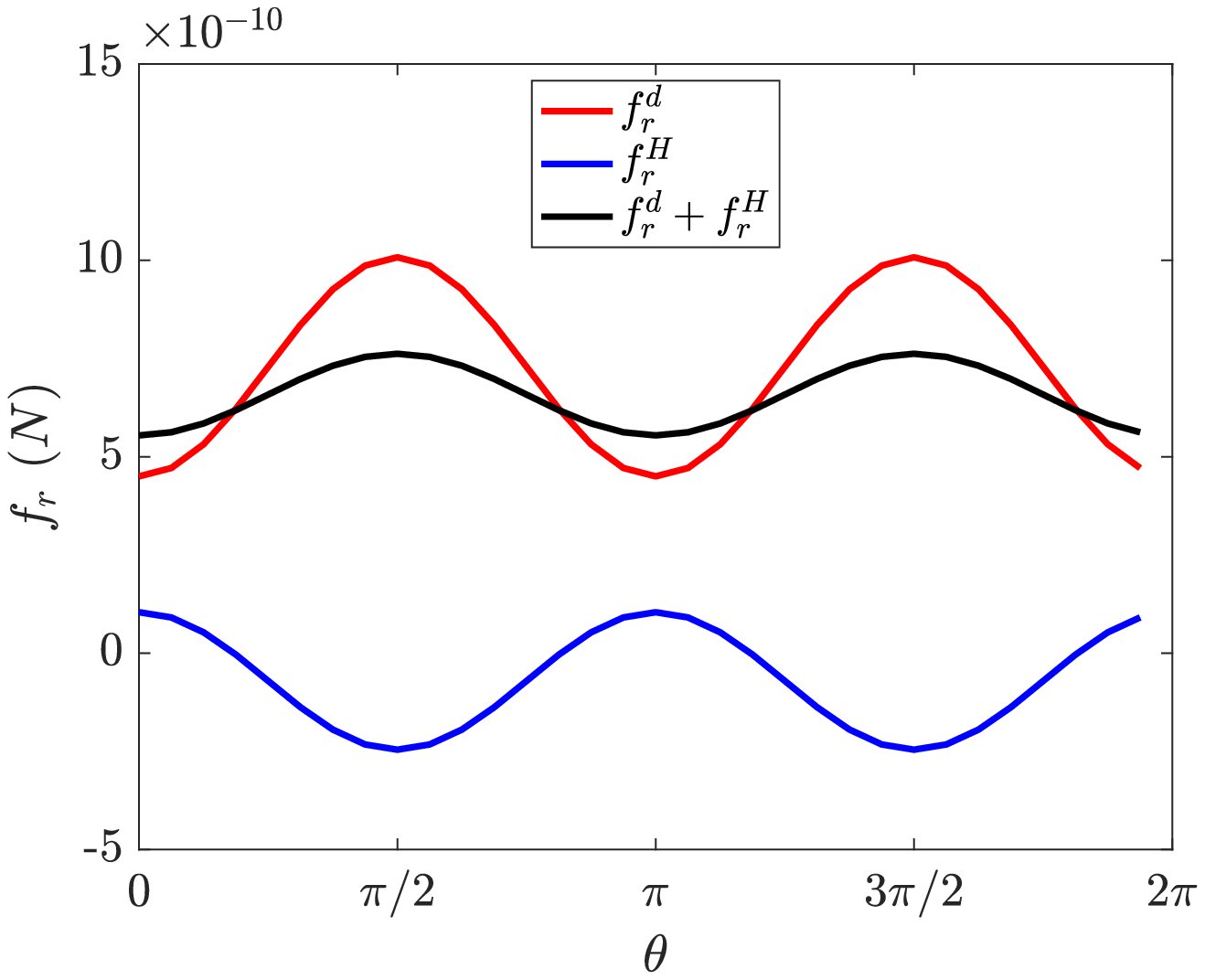}
        \end{subfigure}\hfill
        \begin{subfigure}[t]{0.03\textwidth}
        \textbf{(d)}
        \end{subfigure}
        \begin{subfigure}[t]{0.46\textwidth}
        \includegraphics[width=8cm,valign=t]{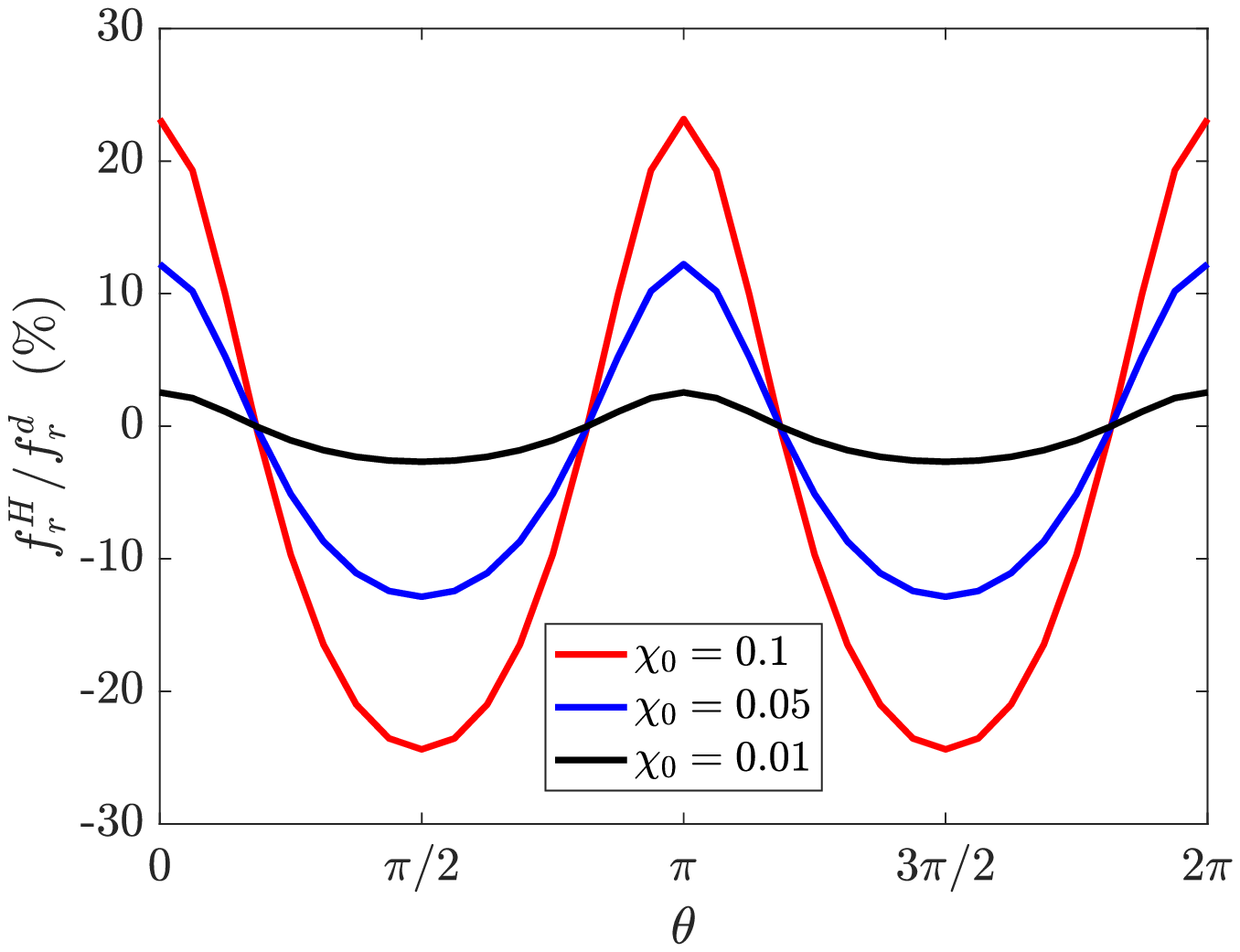}
        \end{subfigure}
        \caption{The radial components of the H-induced and dislocation-induced nodal forces as well as their summation for a circular loop of radius $R_0 = 1000b$, discretized into 32 segments and with (a) $\chi_0 = 0.01$; (b) $\chi_0 = 0.05$; and (c) $\chi_0 = 0.1$. A positive value implies a nodal force pointing towards the center of the loop. (d) The ratio between the radial components of the H-induced and dislocation-induced nodal forces as a function of the reference H mole fraction.}
         \label{fig:glideloopcomp}
\end{figure}

Here, a circular glide dislocation loop lying on the $xy-$plane, having a radius $R_0 = 1000b$ and a Burgers vector $b[100]$ is considered, as shown in Fig. \ref{fig:gloop_sch}. In the absence of an externally applied load, the loop would shrink due to the loop's line tension. The H-induced, $f_r^H$, and dislocation induced, $f_r^{d}$, radial nodal forces calculated for a steady state H distribution with respect to the dislocation stress field for three reference H mole fractions: $\chi_0=0.1, 0.05$ and $0.01$, are shown in Fig. \ref{fig:glideloopcomp} as a function of the dislocation node orientation angle, $\theta$, measured with respect to the $x-$axis. For all simulated cases, both $f_r^H$ and $f_r^d$ always have opposite signs on the endpoints of the loop's screw segments (i.e. corresponding to $\theta=\pi/2$ and $3\pi/2$), thus, H acts to reduce the total radial nodal forces on these dislocation nodes. On the other hand, both $f_r^H$ and $f_r^d$ always have the same sign at the endpoints of edge segments (i.e. corresponding to $\theta=0$ and $\pi$), which leads to an increase in the magnitudes of the total radial nodal forces at these dislocation nodes. It is clear that both $f_r^d$ and $f_r^H$ are oscillating functions with respect to $\theta$ and are both antiphase with respect to one another. This indicates that a H concentration in steady state equilibrium with the dislocation loop leads to a destructive interference as evidenced by the decrease in the total variation of the resultant radial nodal forces at all dislocation nodes, leading to a reduction in the effective speed at which the dislocation glide loop would shrink. This emphasizes a H shielding effect. Furthermore, it is observed from Fig. \ref{fig:glideloopcomp}(d) that with increasing reference H mole fraction this shielding effect is enhanced.

%---------------------------
%---------------------------

%---------------------------------------
\begin{figure}[!htb]
        \centering
        \begin{subfigure}[t]{0.08\textwidth}
        \end{subfigure}
        \begin{subfigure}[t]{0.3\textwidth}
        \hspace{8em}\text{$t=\SI{0.5}{\nano\second}$}
        \end{subfigure}\hfill
        \begin{subfigure}[t]{0.3\textwidth}
        \hspace{5.5em}\text{$t=\SI{1.0}{\nano\second}$}
        \end{subfigure}\hfill
        \begin{subfigure}[t]{0.3\textwidth}
        \hspace{5em}\text{$t=\SI{1.5}{\nano\second}$}
        \end{subfigure}\\
        \vspace{1em}
        \begin{subfigure}[t]{0.08\textwidth}
        \textbf{(a)}
        \end{subfigure}
        \begin{subfigure}[t]{0.3\textwidth}
        \includegraphics[width=4.2cm,valign=t]{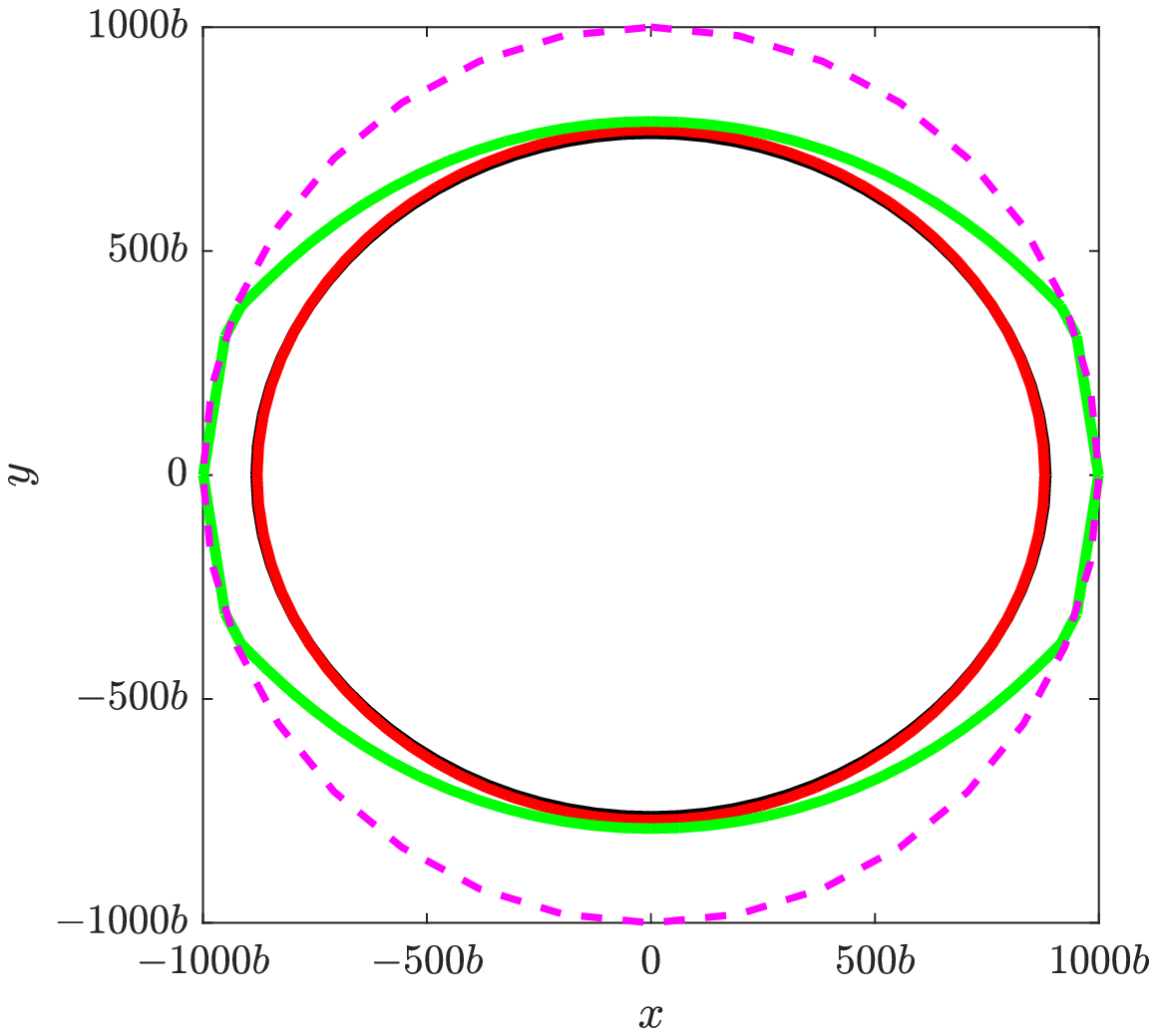}
        \end{subfigure}\hfill
        \begin{subfigure}[t]{0.3\textwidth}
        \includegraphics[width=4.2cm,valign=t]{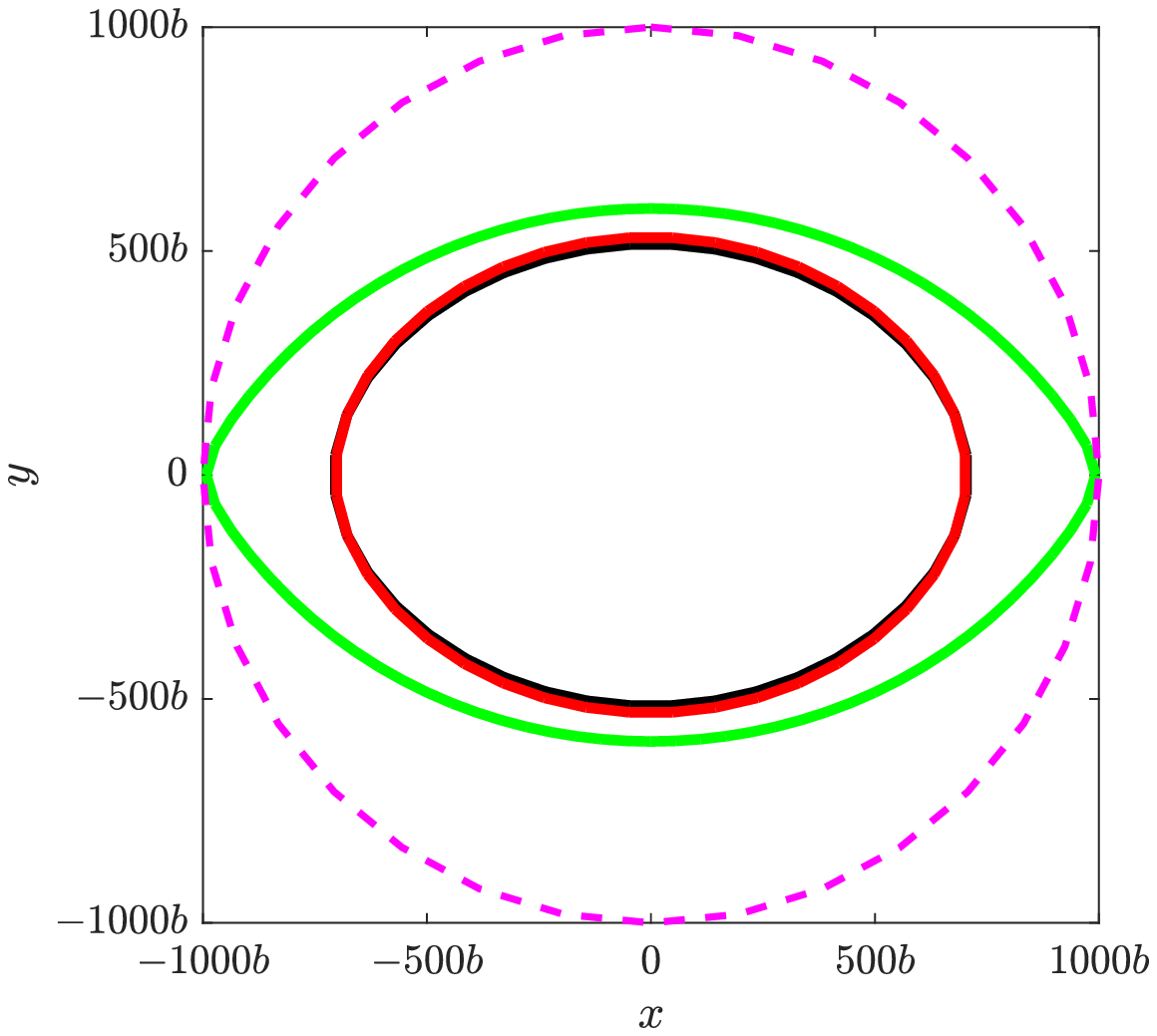}
        \end{subfigure}\hfill
        \begin{subfigure}[t]{0.3\textwidth}
        \includegraphics[width=4.2cm,valign=t]{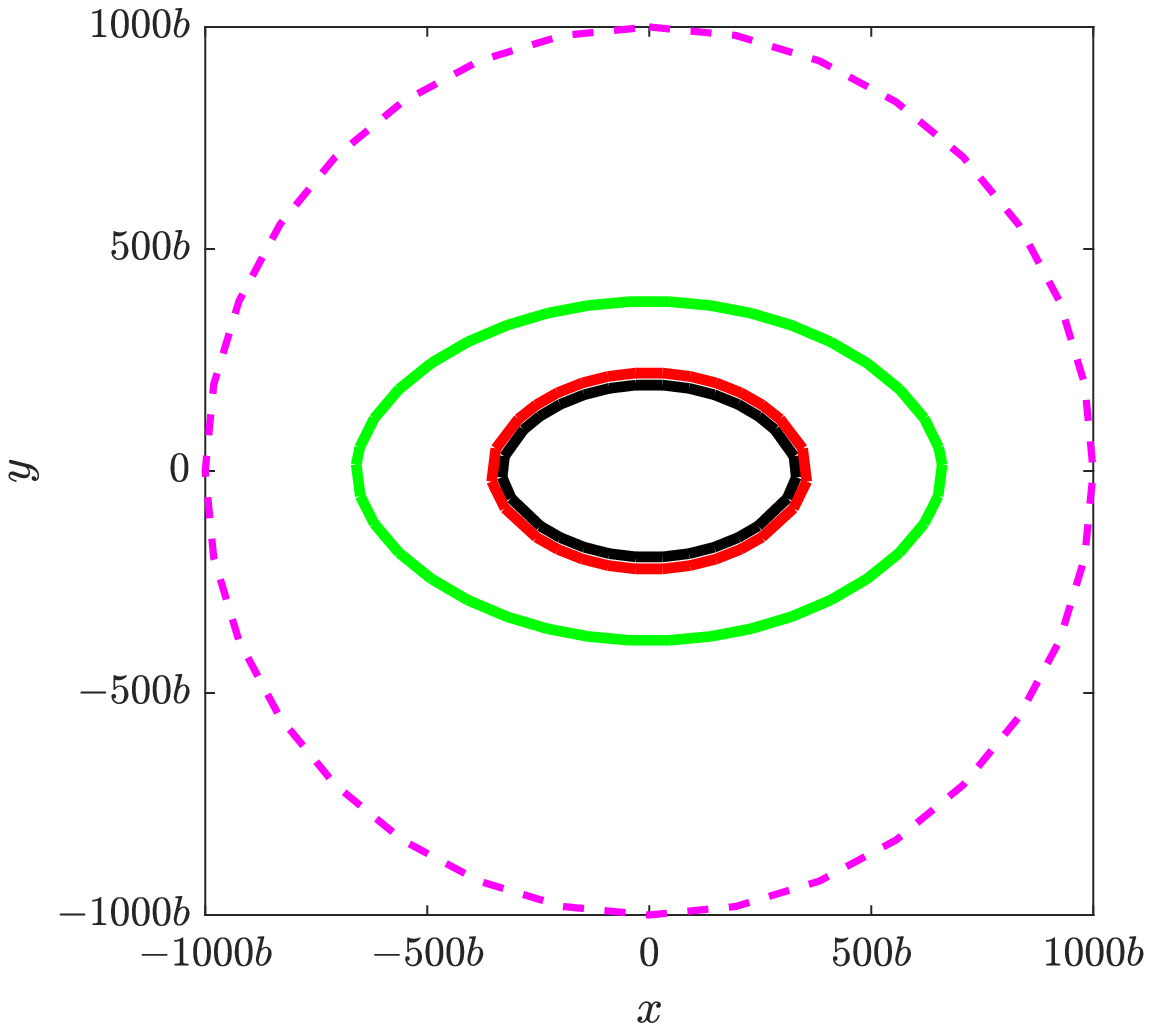}
        \end{subfigure}\hfill
        
        \begin{subfigure}[t]{0.08\textwidth}
        \textbf{(b)}
        \end{subfigure}
        \begin{subfigure}[t]{0.3\textwidth}
        \includegraphics[width=4.2cm,valign=t]{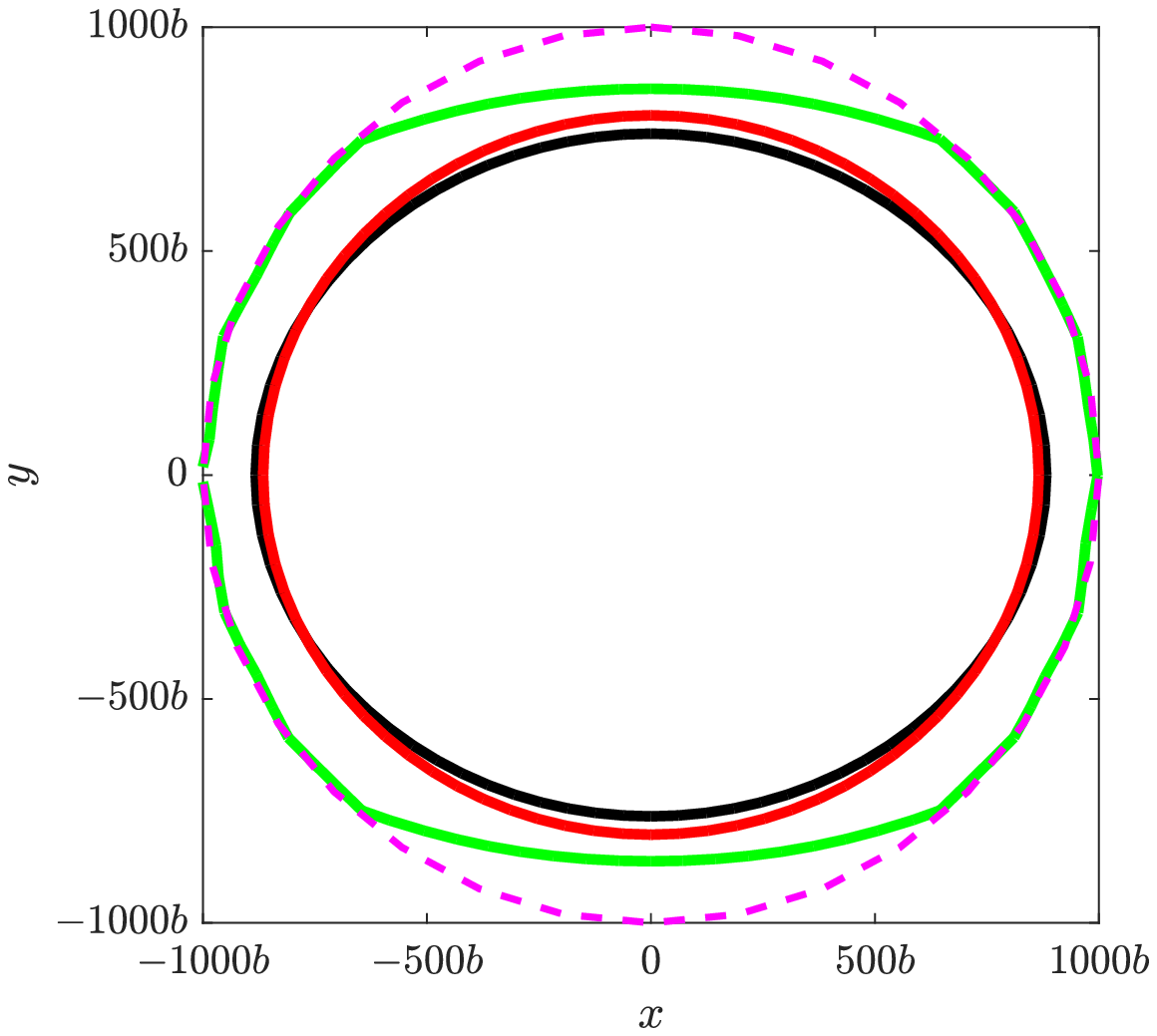}
        \end{subfigure}\hfill
        \begin{subfigure}[t]{0.3\textwidth}
        \includegraphics[width=4.2cm,valign=t]{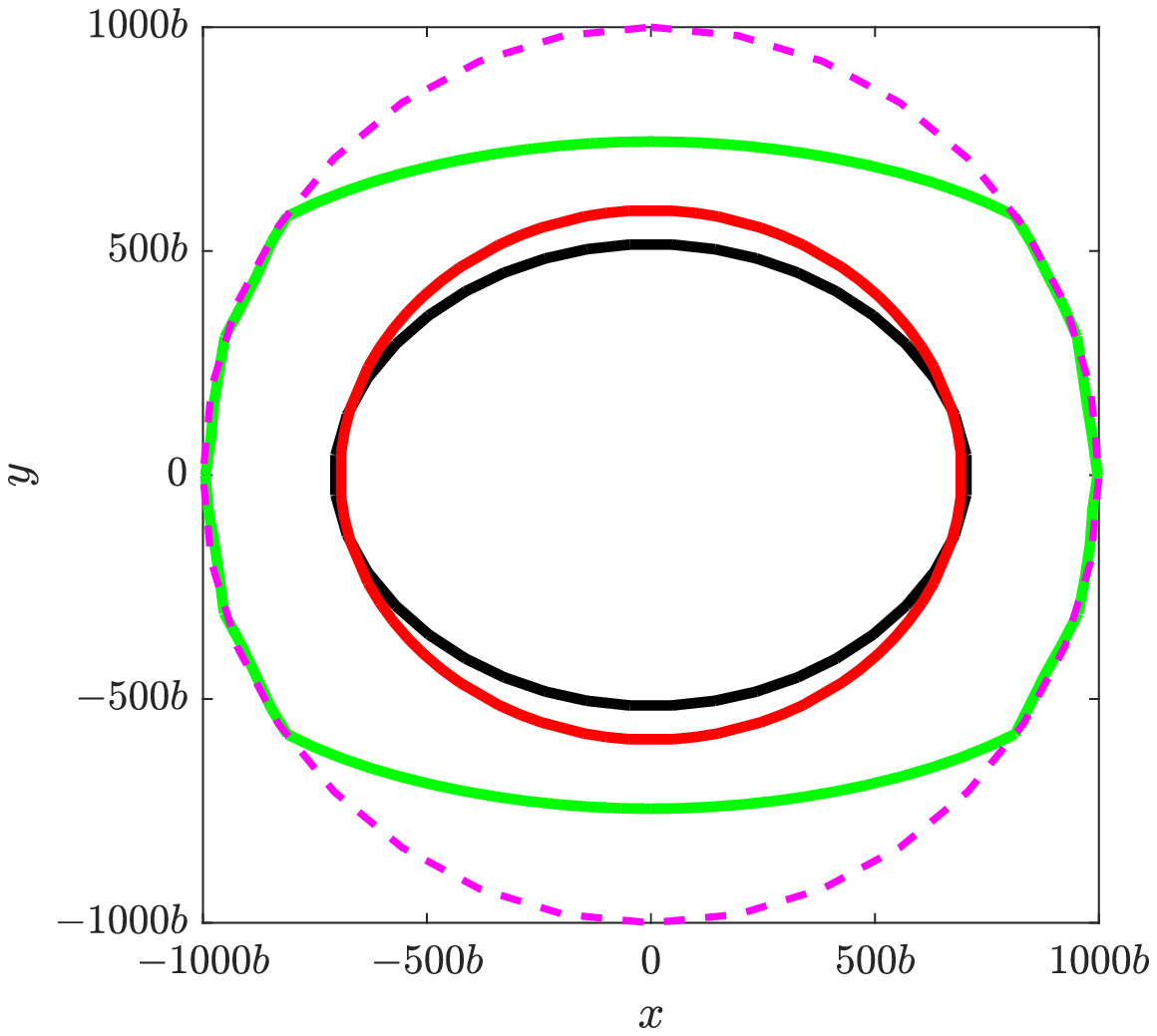}
        \end{subfigure}\hfill
        \begin{subfigure}[t]{0.3\textwidth}
        \includegraphics[width=4.2cm,valign=t]{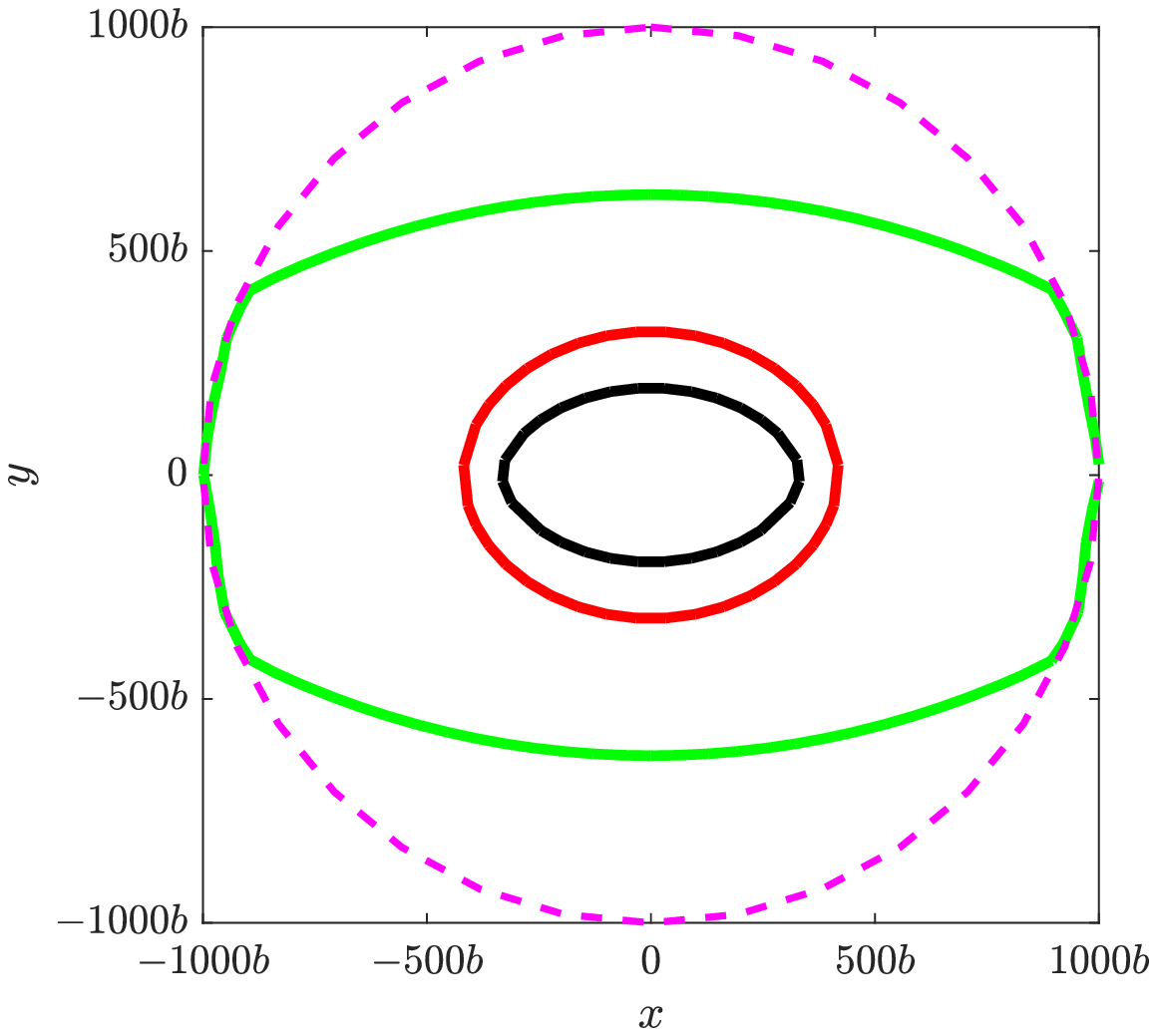}
        \end{subfigure}\hfill
        
        \begin{subfigure}[t]{0.08\textwidth}
        \textbf{(c)}
        \end{subfigure}
        \begin{subfigure}[t]{0.3\textwidth}
        \includegraphics[width=4.2cm,valign=t]{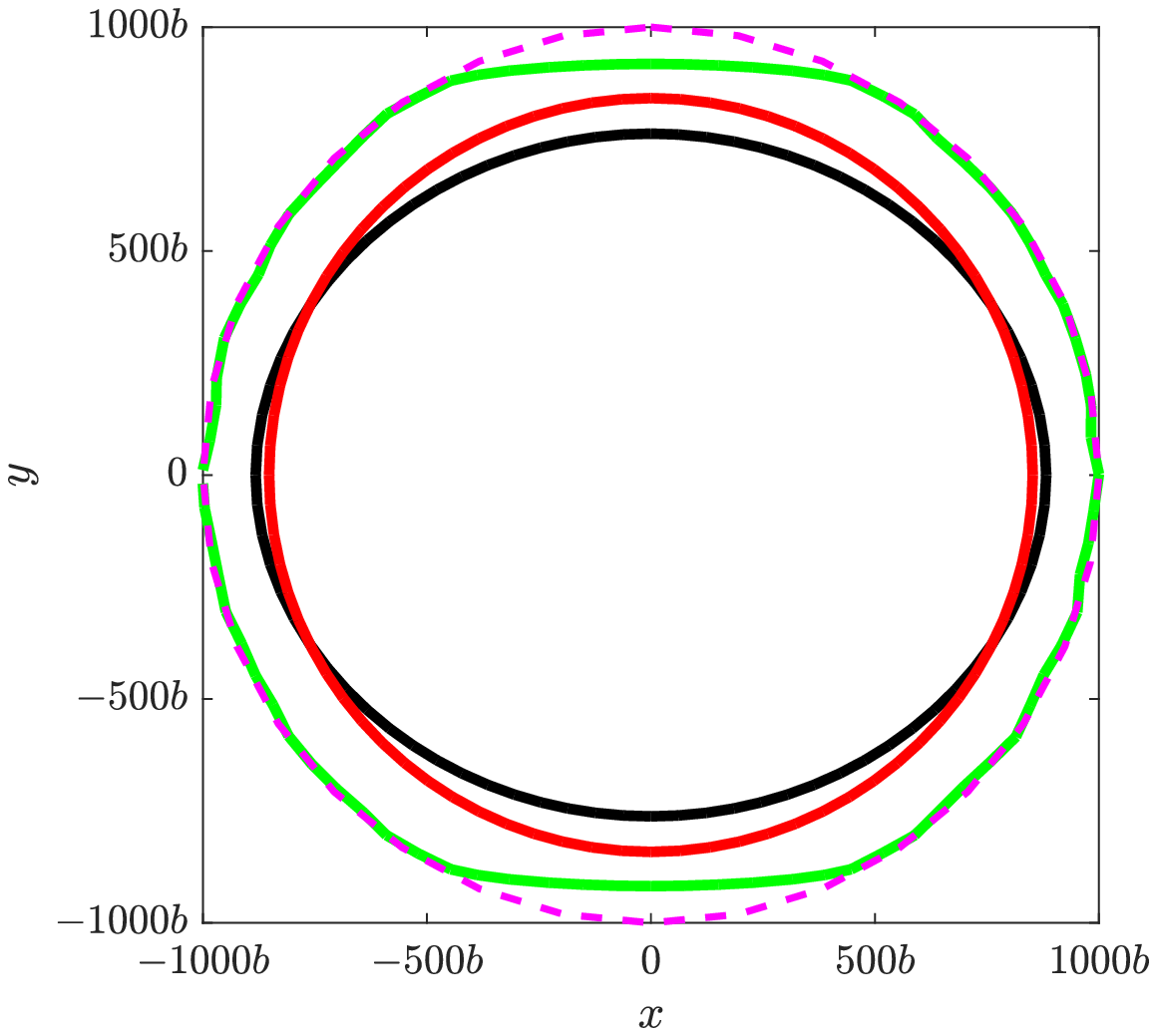}
        \end{subfigure}\hfill
        \begin{subfigure}[t]{0.3\textwidth}
        \includegraphics[width=4.2cm,valign=t]{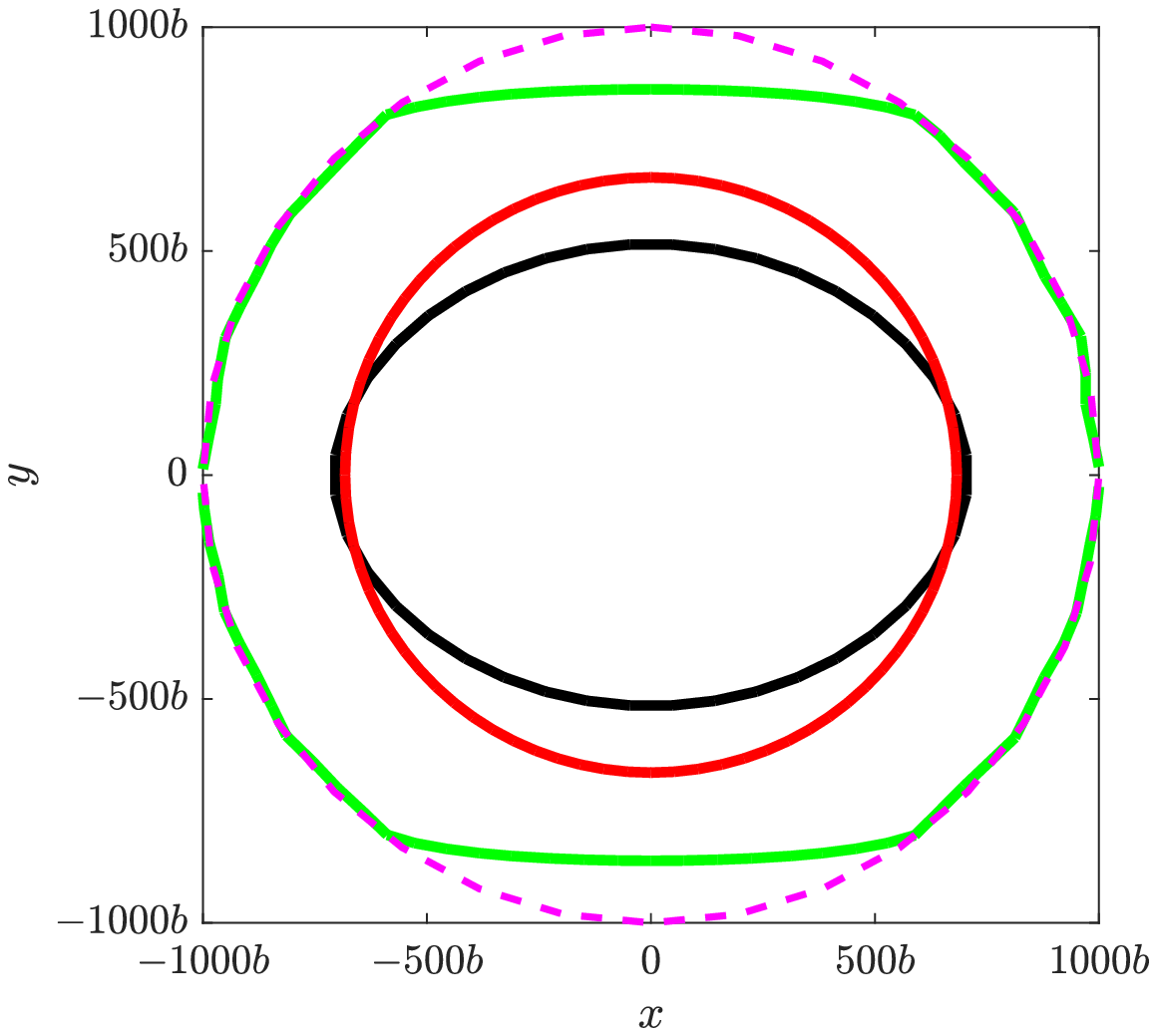}
        \end{subfigure}\hfill
        \begin{subfigure}[t]{0.3\textwidth}
        \includegraphics[width=4.2cm,valign=t]{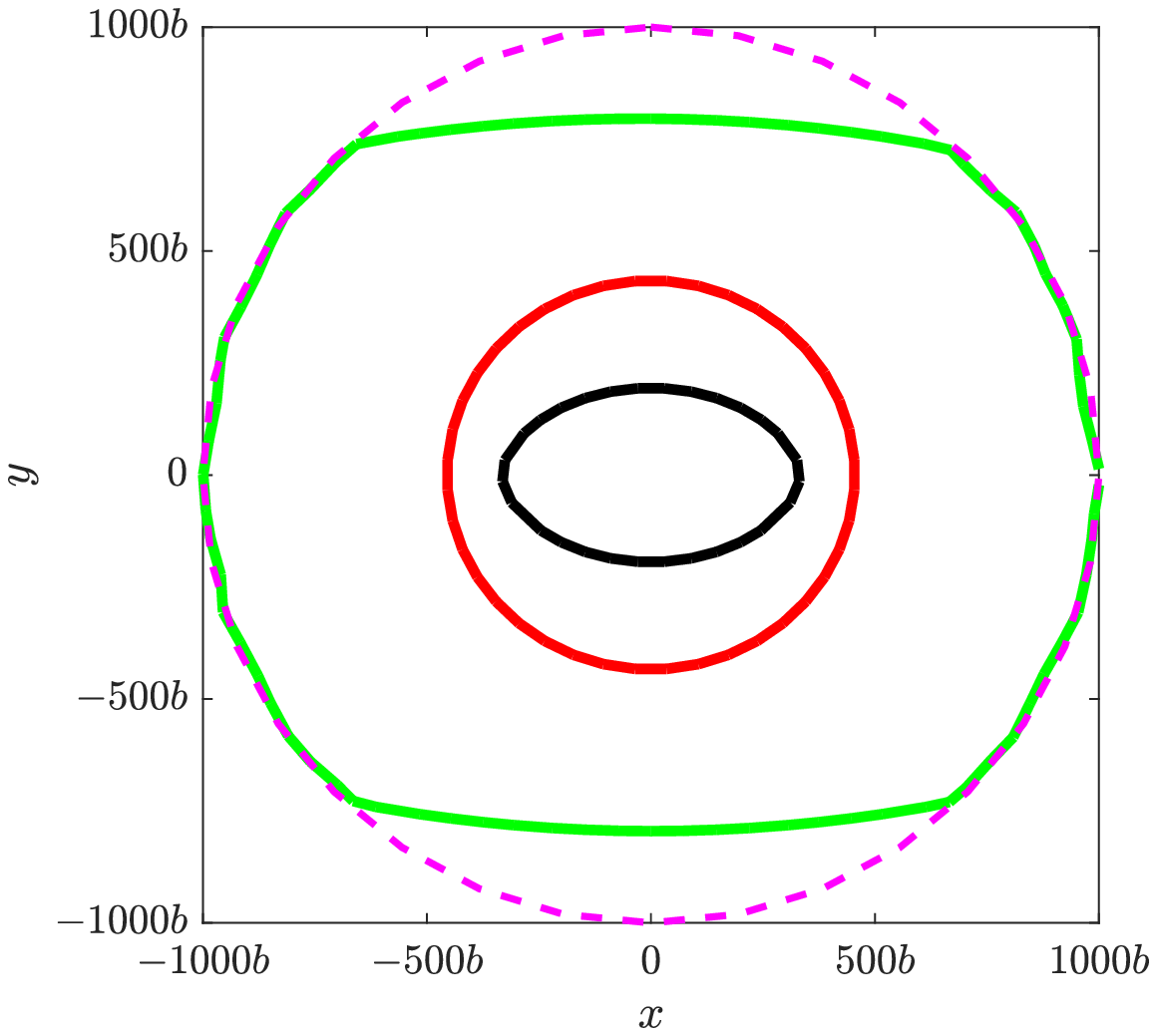}
        \end{subfigure}\hfill
        \caption{The dislocation glide loop shape at three different times for: (a) $\chi_0 = 0.01$; (b) $\chi_0 = 0.05$; and (c) $\chi_0 = 0.1$. The dashed magenta lines represent the initial position of the dislocation loop. The benchmark case in the absence of H is also shown by black solids lines. Simulations with a large $D_H$ (i.e. Case \RN{1}) and a small $D_H$ (i.e. Case \RN{2}) are shown by the red and green solid lines, respectively.}
         \label{fig:gloop_dd}
\end{figure}

\begin{figure}[!htb]
        \centering        
        \begin{subfigure}[t]{0.025\textwidth}
        \hspace{-1em}\textbf{(a)}
        \end{subfigure}\hfill   
        \begin{subfigure}[t]{0.47\textwidth}
        \includegraphics[width=8.5cm,valign=t]{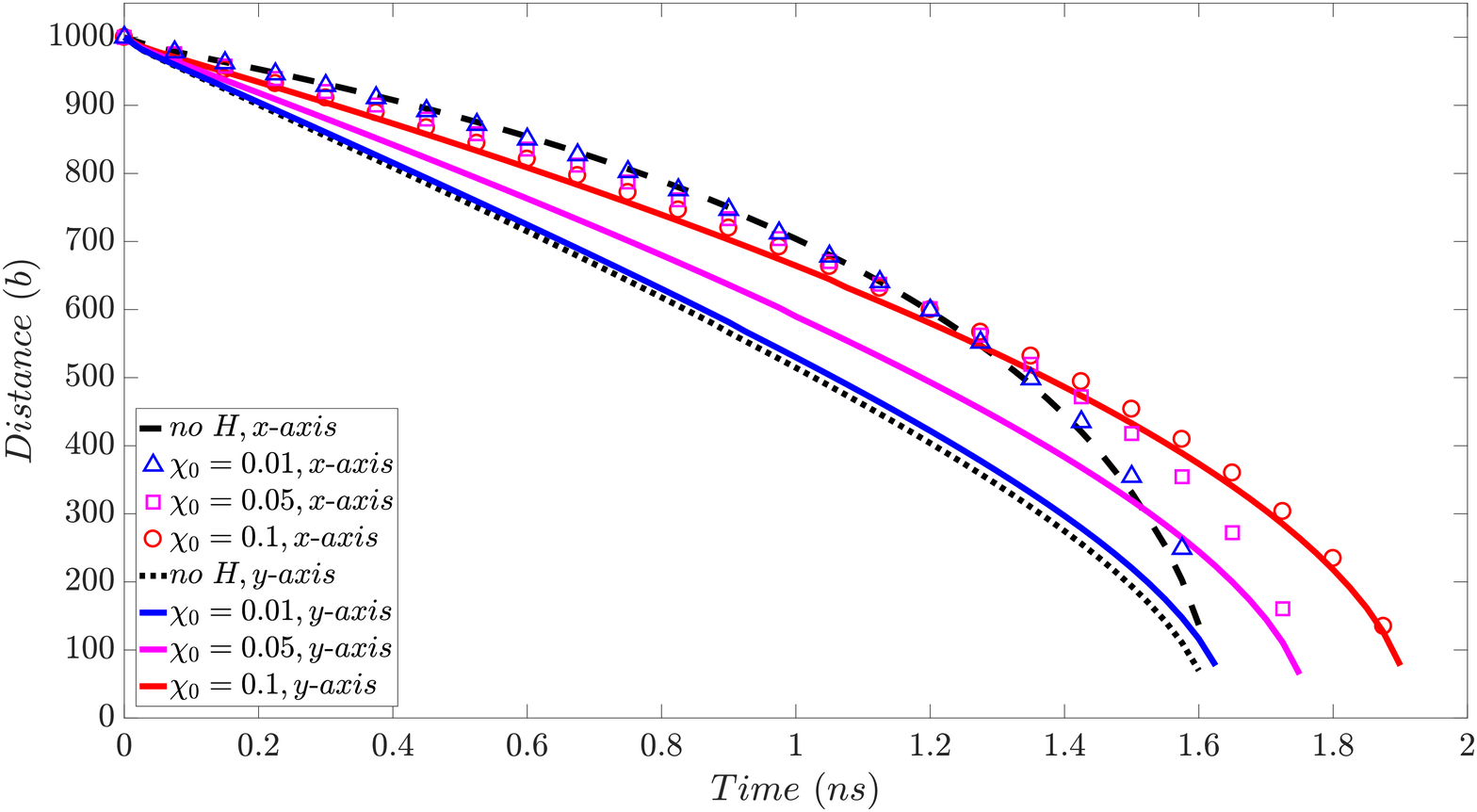}
        \end{subfigure}\hfill
        \begin{subfigure}[t]{0.028\textwidth}
        \textbf{(b)}
        \end{subfigure}\hfill   
        \begin{subfigure}[t]{0.47\textwidth}
        \includegraphics[width=8.5cm,valign=t]{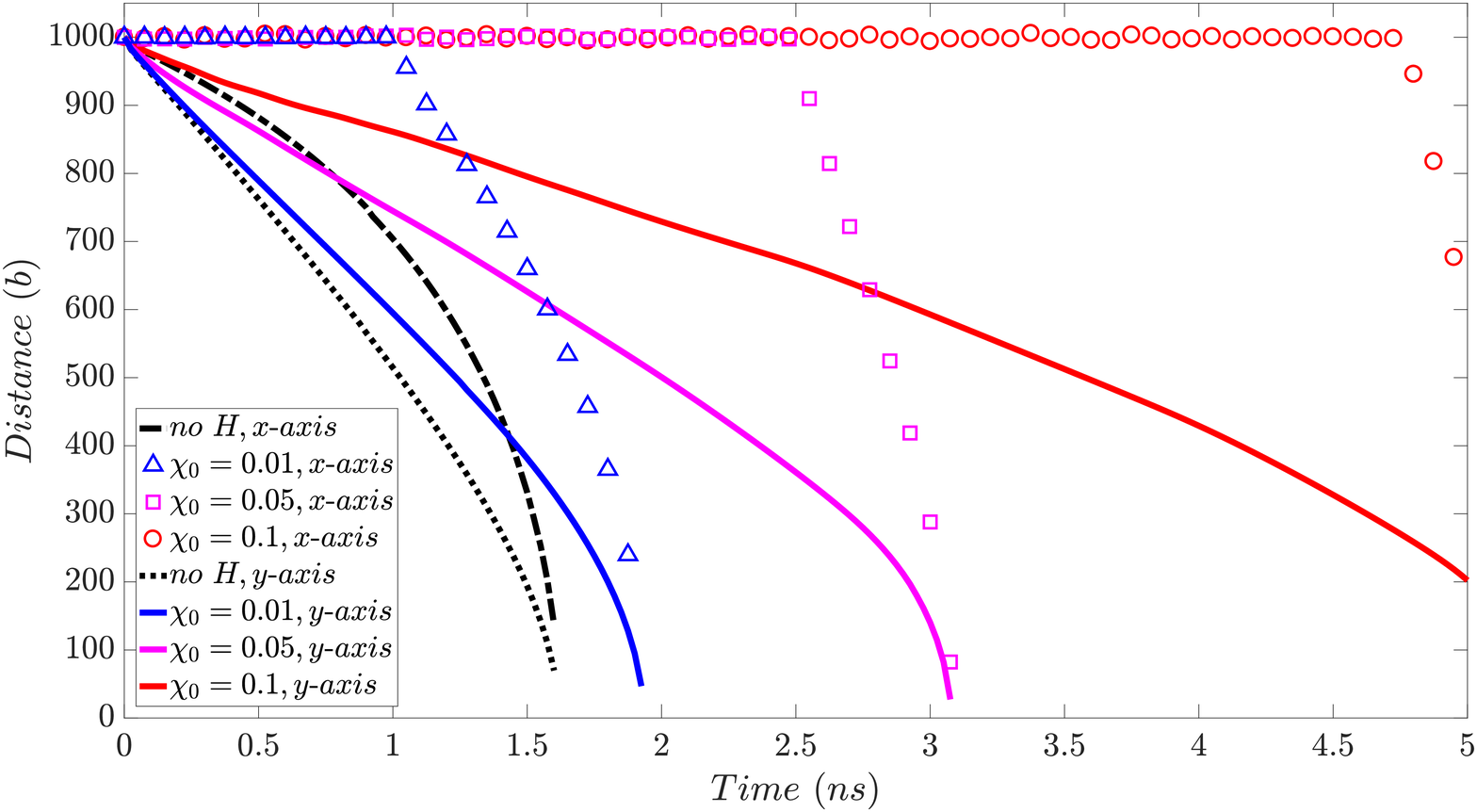}
        \end{subfigure}
        \caption{The position of the two extrema points along the $x-$axis and $y-$axis of the dislocation loop relative to the loop center as a function of time for: (a) Case \RN{1} with a large $D_H$; and (b) Case \RN{2} with a small $D_H$.}
         \label{fig:glideloop_dis_comp}
\end{figure}

%---------------------------------------
The glide loop shapes at three different simulated times are shown in Fig. \ref{fig:gloop_dd} for the three different reference H mole fractions. The benchmark case in the absence of H is also shown for comparison. Two conditions are simulated in these dynamic simulations, namely, a large $D_H$ (i.e. Case \RN{1}) and a small $D_H$ (i.e. Case \RN{2}). It is clear that the glide loop shrinks fastest in the absences of H, and the H shielding effect is more apparent with increasing reference H mole fraction. To quantify this effect, the position of the two extrema points along the $x-$axis and $y-$axis of the dislocation loop relative to the loop center as a function of time are shown in Fig. \ref{fig:glideloop_dis_comp} for all simulated cases. This figure indicates the distance traversed by the pure edge and pure screw components of the glide loop. For the simulations with large $D_H$, the H-shielding effect is observed to homogenize the shrinkage process leading to the loop maintaining its circular shape through out the simulation. On the other hand, for the case of a small $D_H$ the response is considerably different. In this case, for the three different reference H mole fractions simulated here, the screw segments move first towards the loop center, while the edge segments are pinned at their initial positions until the line tension is strong enough to overcome the H pinning effect. This H-induced pinning effect was commonly observed from MD simulations of H interactions with edge dislocations in metals having a small $D_H$ (e.g. Ref. \cite{song2014,tang2012}).

%---------------------------------------
%---------------------------------------

%**********************************************************
\subsection{Effect of hydrogen on the dislocation spacing in an array of parallel edge dislocation}

\begin{figure}[!htb]
        \centering
        {\includegraphics[width=15cm]{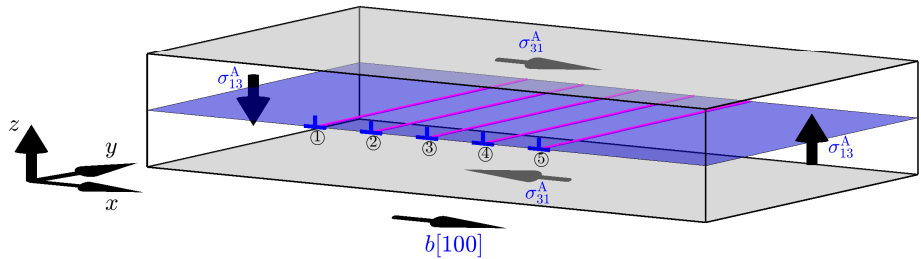}\label{dd-sch}}
         \caption{Schematic of the simulations of a dislocation array under an applied shears stress, $\sigma^{\rm A}_{13}$.}
         \label{fig:dd-sch}
\end{figure}

Five straight edge dislocations having a Burgers vector $\mathbf{b}=b[100]$ are initially equidistantly distributed at the center of the $(001)$ slip plane located at $z = 0$ in a rectangular simulation cell having dimensions $10,000b\times1000b\times1000b$, as shown in Fig. \ref{fig:dd-sch}. The initial distance between neighboring dislocations is set to be $250b$. To quantify the effect of H on the dislocation spacing, DDD simulations are performed for three different conditions. First, a benchmark simulation is performed in the absence of H. The simulations were then performed for four different reference H mole fractions, namely, $\chi_0 = 0.001, 0.005, 0.05,$ and $0.1$. Each simulation was repeated  with a large (Case \RN{1}) and small (Case \RN{2}) $D_H$ assumption. It should be noted that the simulation surfaces are all considered to be free surfaces. However, the effect of image fields are ignored for simplicity.
%------------------------------------------
%------------------------------------------

\begin{figure}[!htb]
        \centering
        \begin{subfigure}[t]{0.03\textwidth}
        \end{subfigure}
        \begin{subfigure}[t]{0.32\textwidth}
        \hspace{7em}\text{$t=\SI{0.25}{\nano\second}$}
        \end{subfigure}\hfill
        \begin{subfigure}[t]{0.32\textwidth}
        \hspace{6.5em}\text{$t=\SI{1.00}{\nano\second}$}
        \end{subfigure}\hfill
        \begin{subfigure}[t]{0.32\textwidth}
        \hspace{6em}\text{$t=\SI{1.50}{\nano\second}$}
        \end{subfigure}\\
        \vspace{2.5em}
        \begin{subfigure}[t]{0.03\textwidth}
        \textbf{(a)}
        \end{subfigure}
        \begin{subfigure}[t]{0.3\textwidth}
        \includegraphics[width=4.5cm,valign=t]{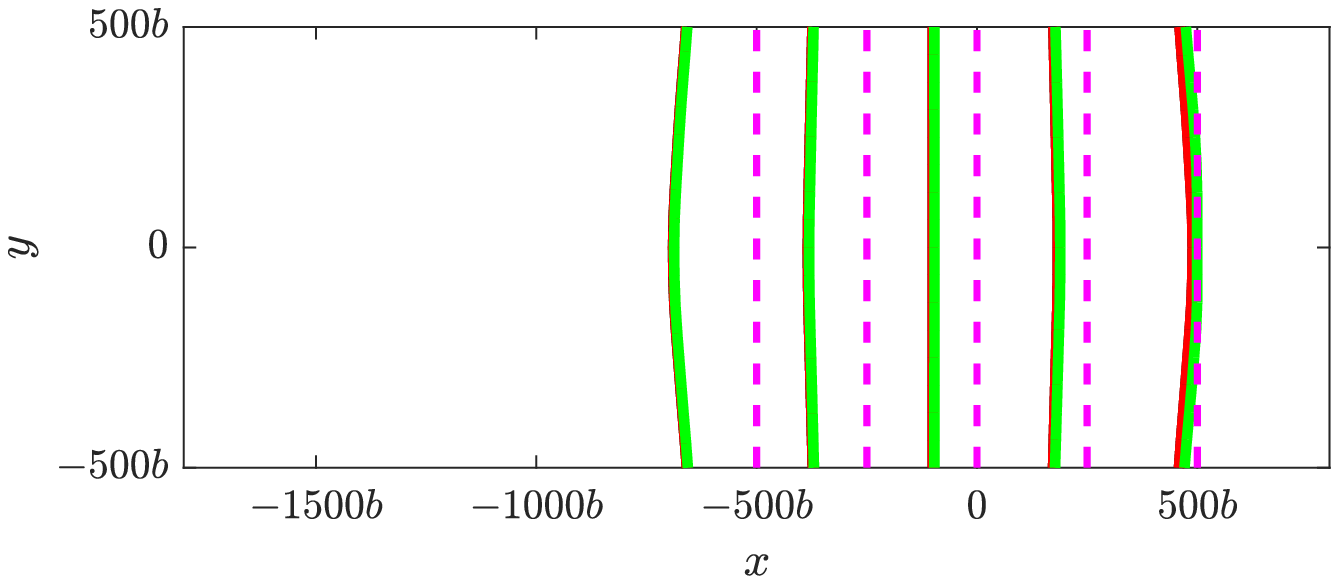}
        \end{subfigure}\hfill
        \begin{subfigure}[t]{0.3\textwidth}
        \includegraphics[width=4.5cm,valign=t]{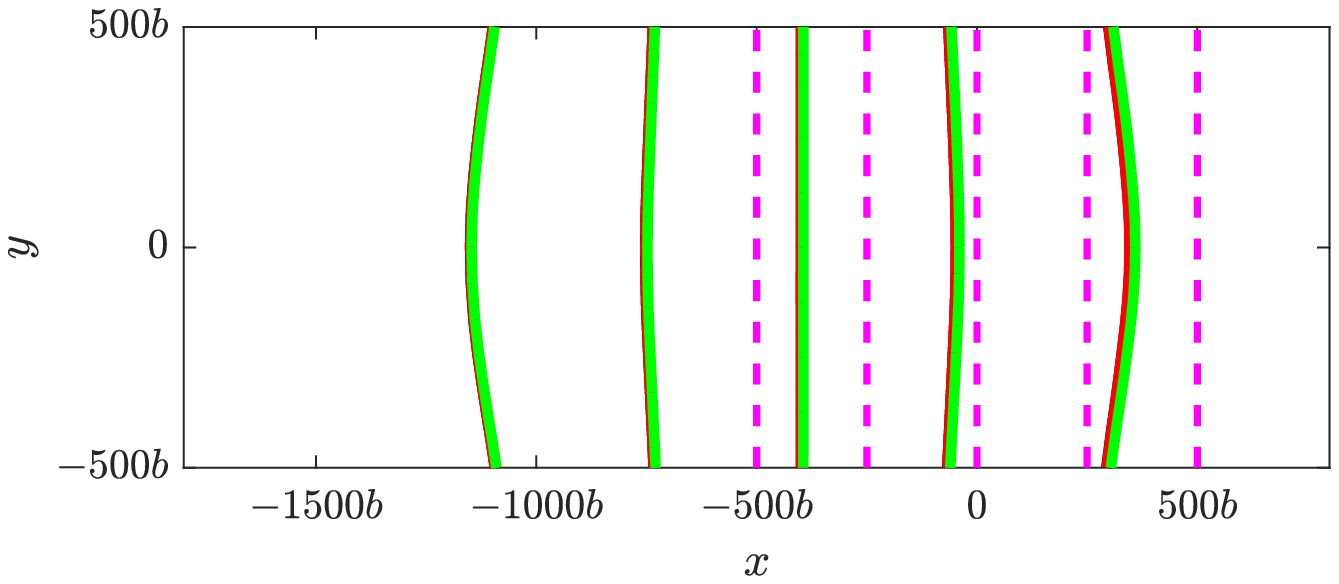}
        \end{subfigure}\hfill
        \begin{subfigure}[t]{0.3\textwidth}
        \includegraphics[width=4.5cm,valign=t]{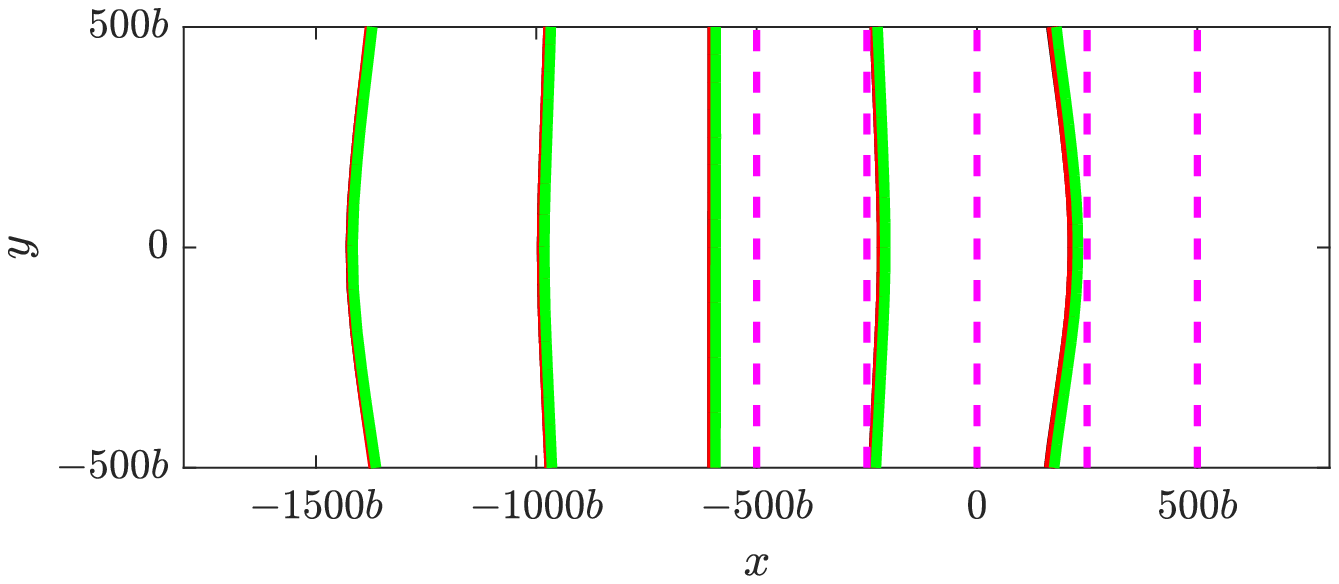}
        \end{subfigure}\\
        \vspace{2.5em}
        
        \begin{subfigure}[t]{0.03\textwidth}
        \textbf{(b)}
        \end{subfigure}
        \begin{subfigure}[t]{0.3\textwidth}
        \includegraphics[width=4.5cm,valign=t]{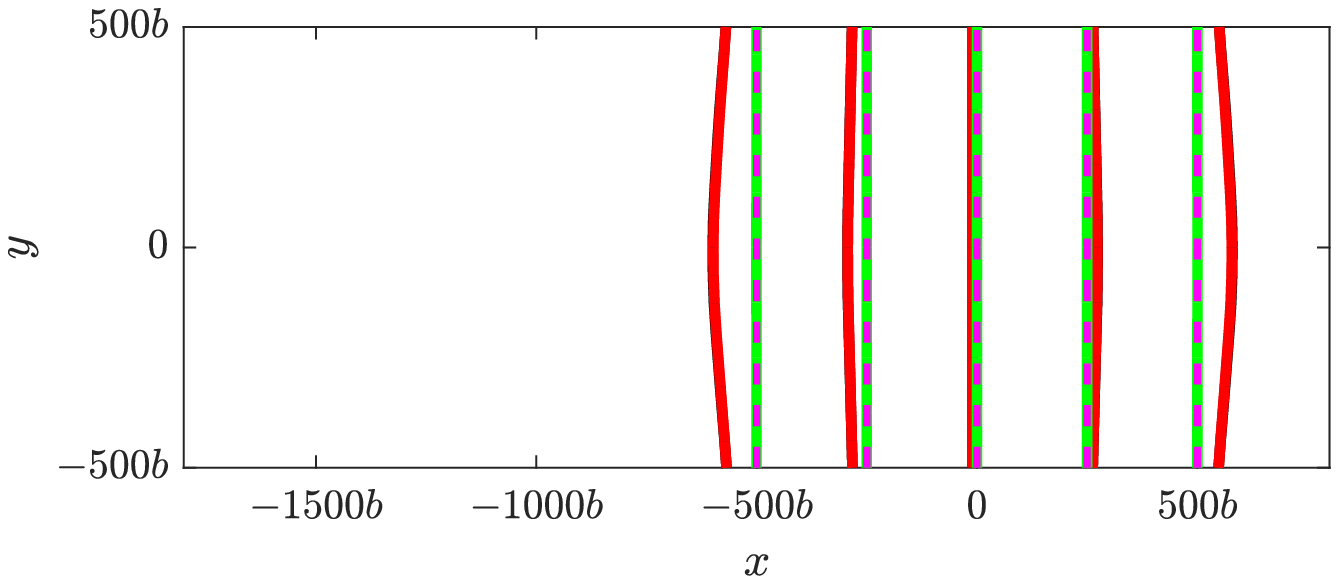}
        \end{subfigure}\hfill
        \begin{subfigure}[t]{0.3\textwidth}
        \includegraphics[width=4.5cm,valign=t]{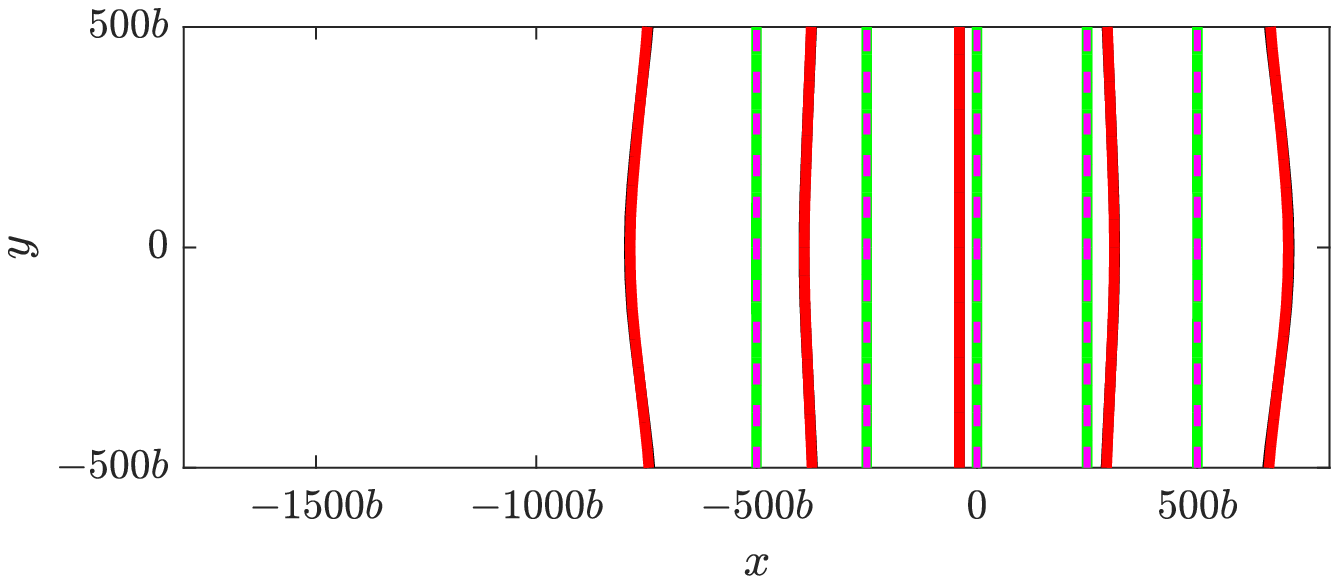}
        \end{subfigure}\hfill
        \begin{subfigure}[t]{0.3\textwidth}
        \includegraphics[width=4.5cm,valign=t]{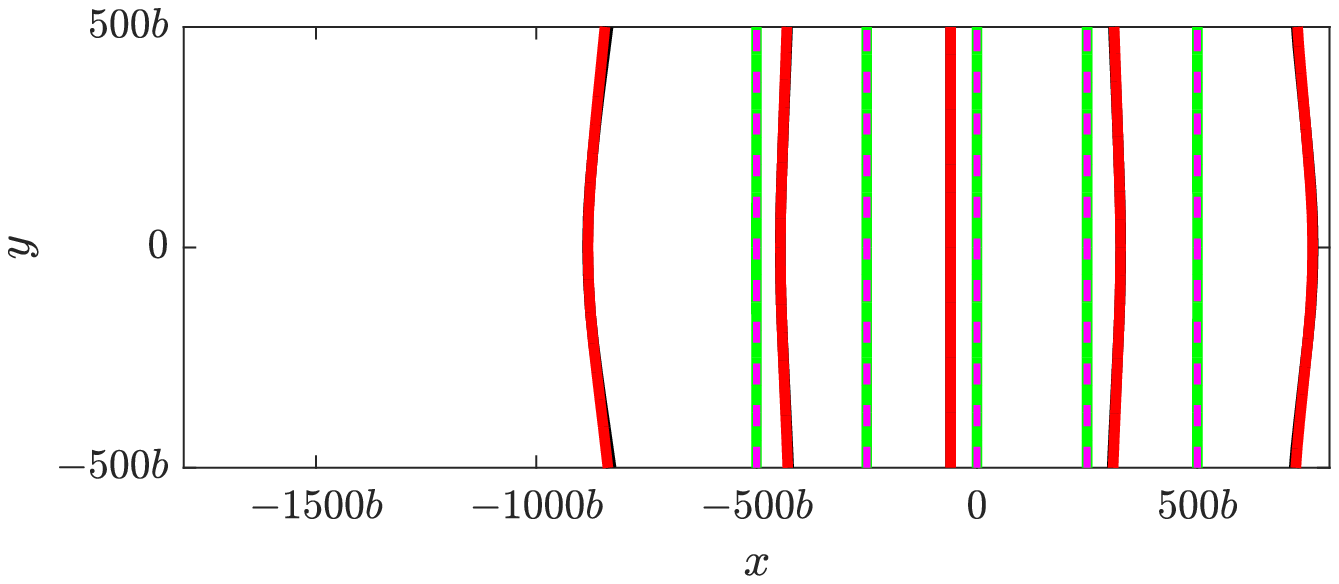}
        \end{subfigure}\\
       \vspace{2.5em}
        \begin{subfigure}[t]{0.03\textwidth}
        \textbf{(c)}
        \end{subfigure}
        \begin{subfigure}[t]{0.3\textwidth}
        \includegraphics[width=4.5cm,valign=t]{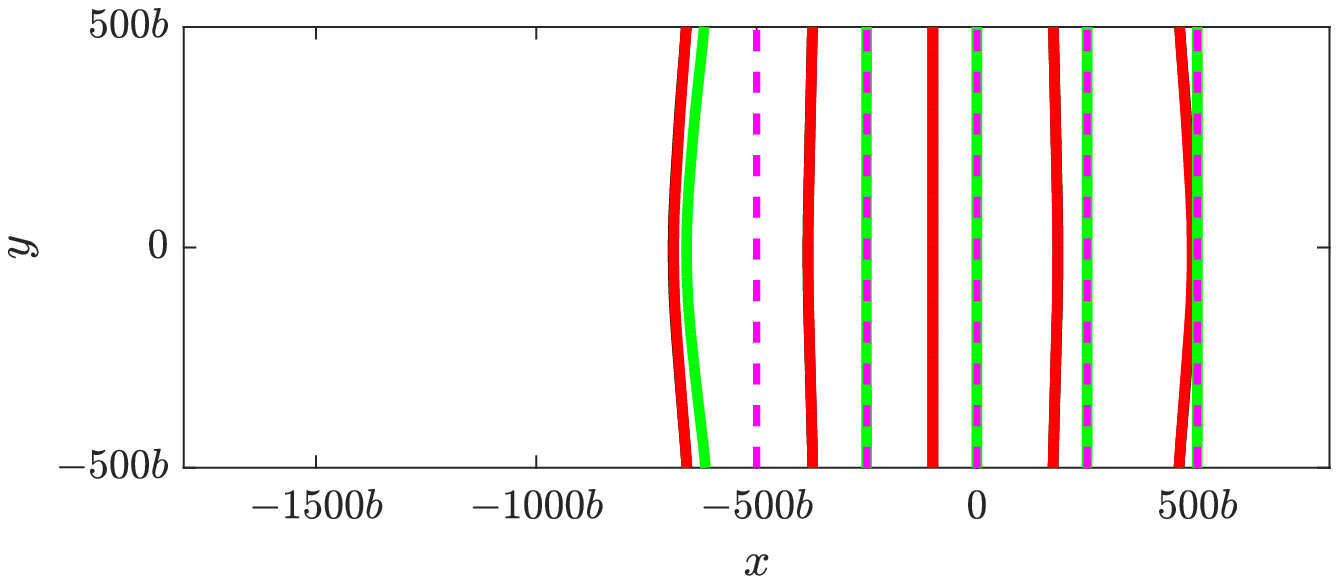}
        \end{subfigure}\hfill
        \begin{subfigure}[t]{0.3\textwidth}
        \includegraphics[width=4.5cm,valign=t]{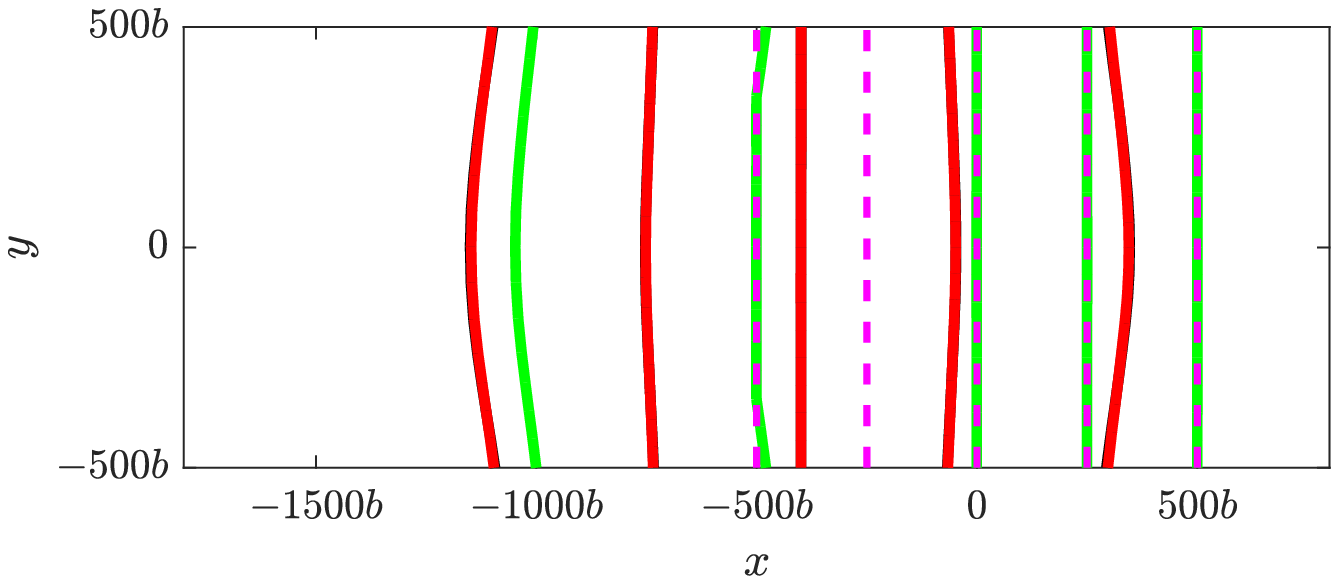}
        \end{subfigure}\hfill
        \begin{subfigure}[t]{0.3\textwidth}
        \includegraphics[width=4.5cm,valign=t]{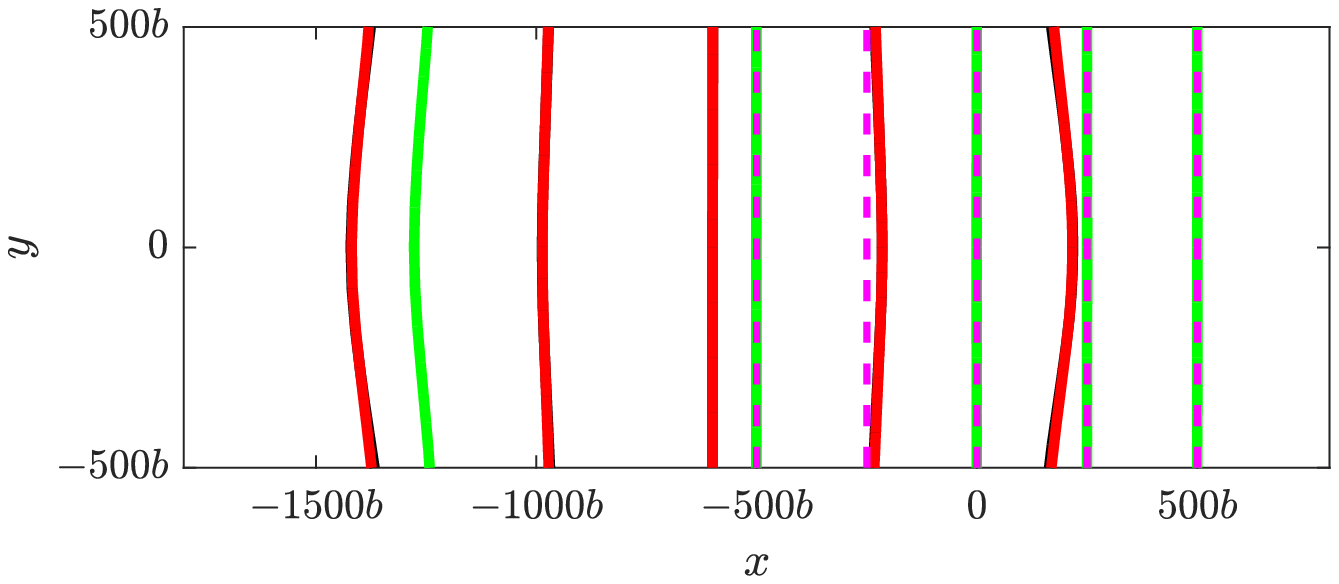}
        \end{subfigure}\\
       \vspace{2.5em}
       
        \begin{subfigure}[t]{0.03\textwidth}
        \textbf{(d)}
        \end{subfigure}
        \begin{subfigure}[t]{0.3\textwidth}
        \includegraphics[width=4.5cm,valign=t]{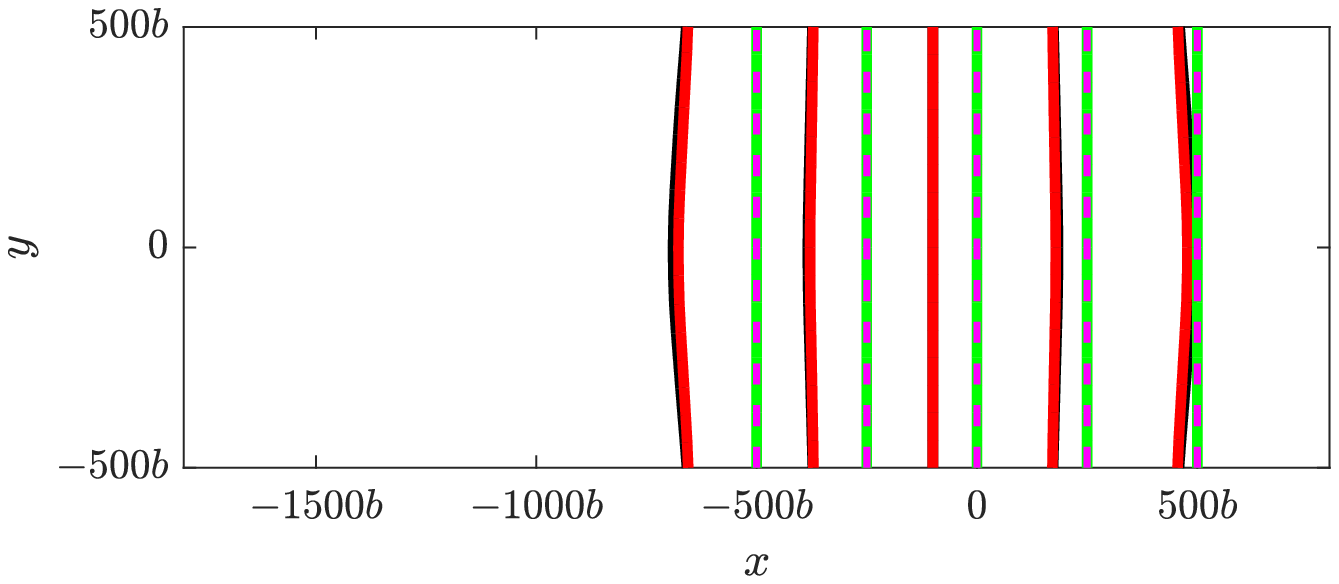}
        \end{subfigure}\hfill
        \begin{subfigure}[t]{0.3\textwidth}
        \includegraphics[width=4.5cm,valign=t]{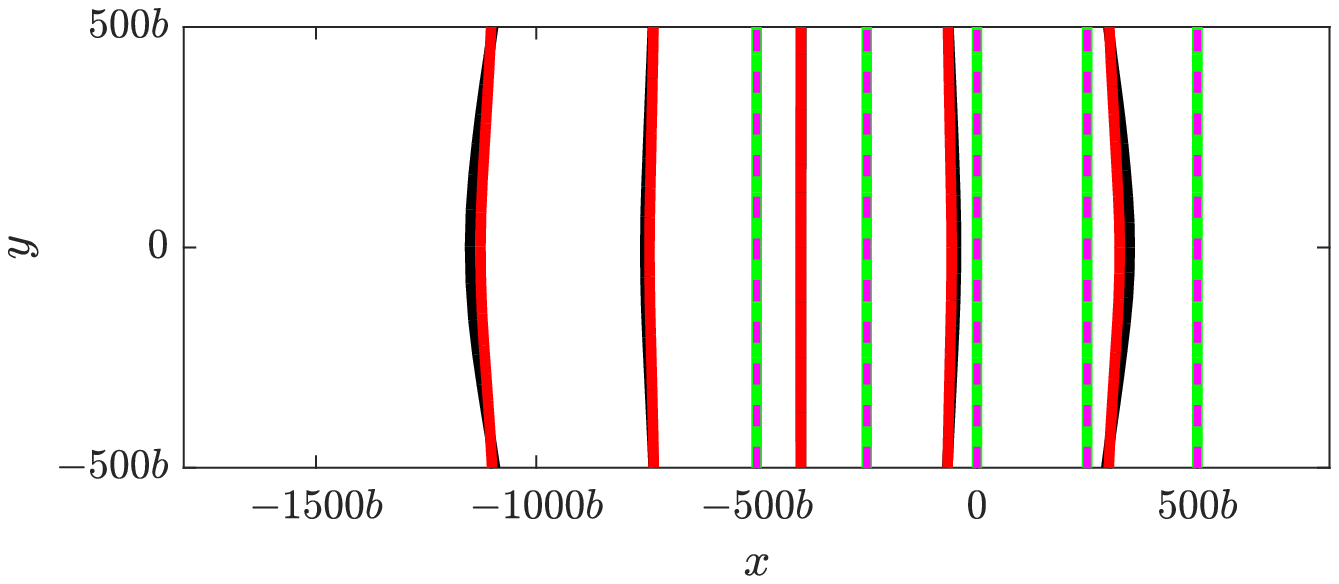}
        \end{subfigure}\hfill
        \begin{subfigure}[t]{0.3\textwidth}
        \includegraphics[width=4.5cm,valign=t]{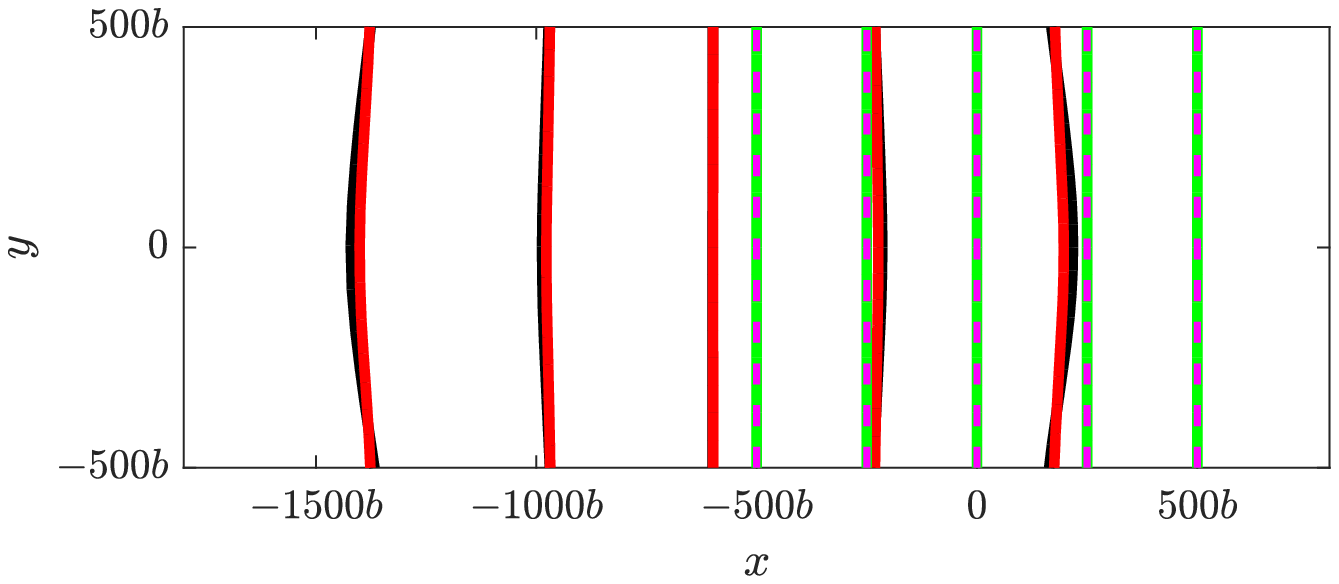}
        \end{subfigure}\\
       \vspace{2.5em}
        \begin{subfigure}[t]{0.03\textwidth}
        \textbf{(e)}
        \end{subfigure}
        \begin{subfigure}[t]{0.3\textwidth}
        \includegraphics[width=4.5cm,valign=t]{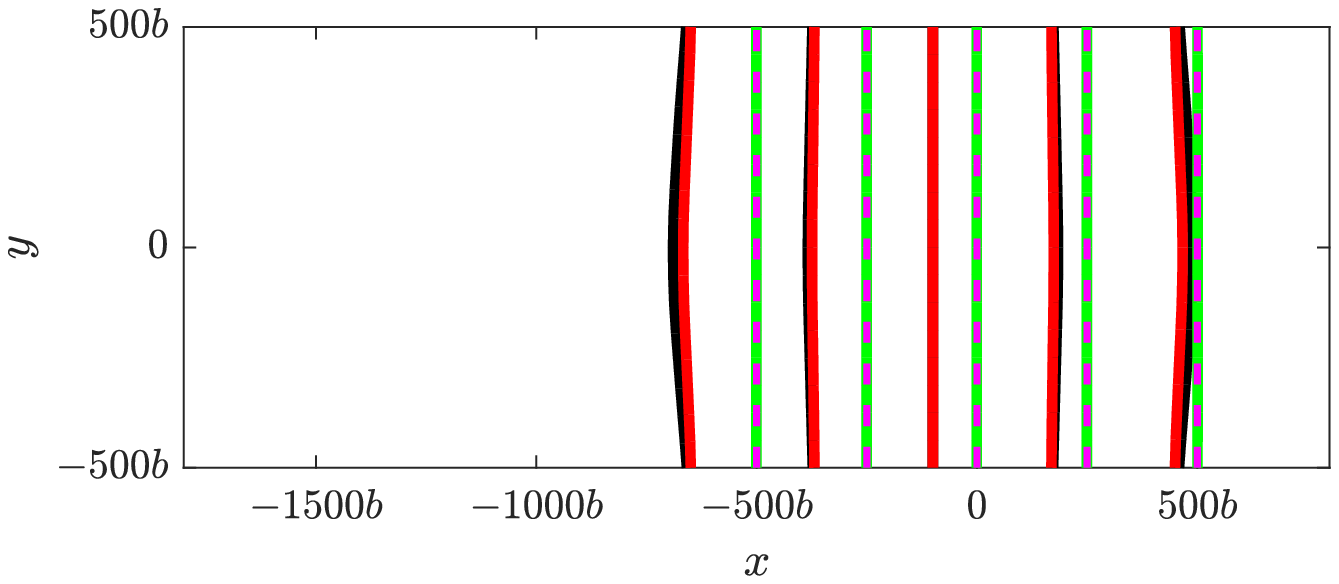}
        \end{subfigure}\hfill
        \begin{subfigure}[t]{0.3\textwidth}
        \includegraphics[width=4.5cm,valign=t]{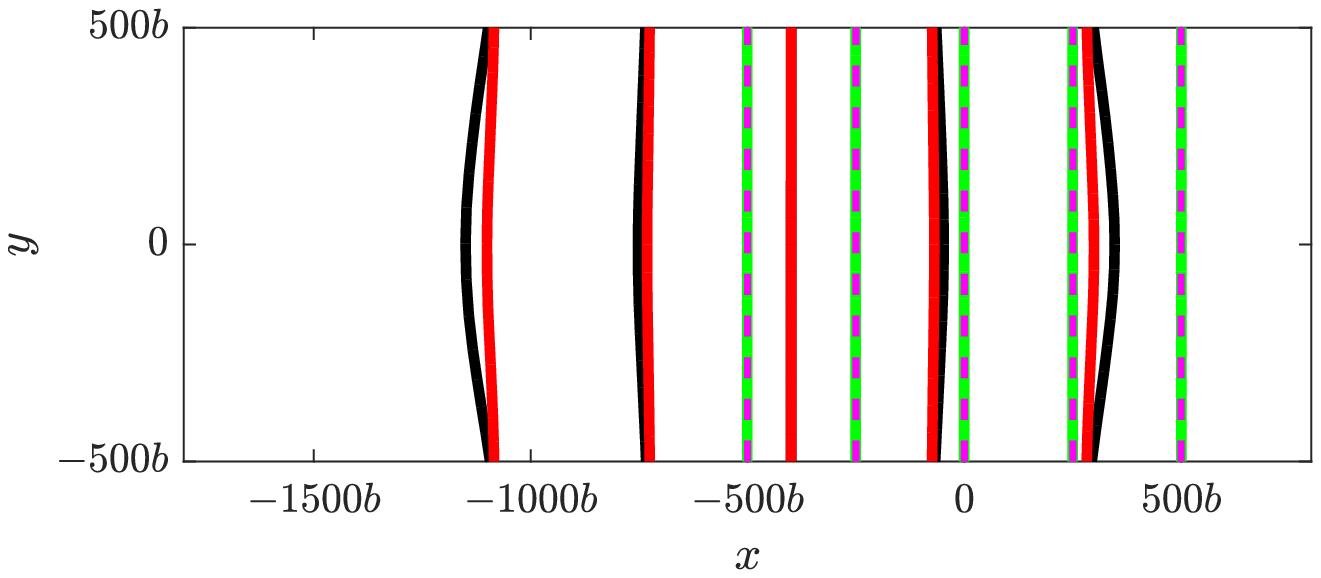}
        \end{subfigure}\hfill
        \begin{subfigure}[t]{0.3\textwidth}
        \includegraphics[width=4.5cm,valign=t]{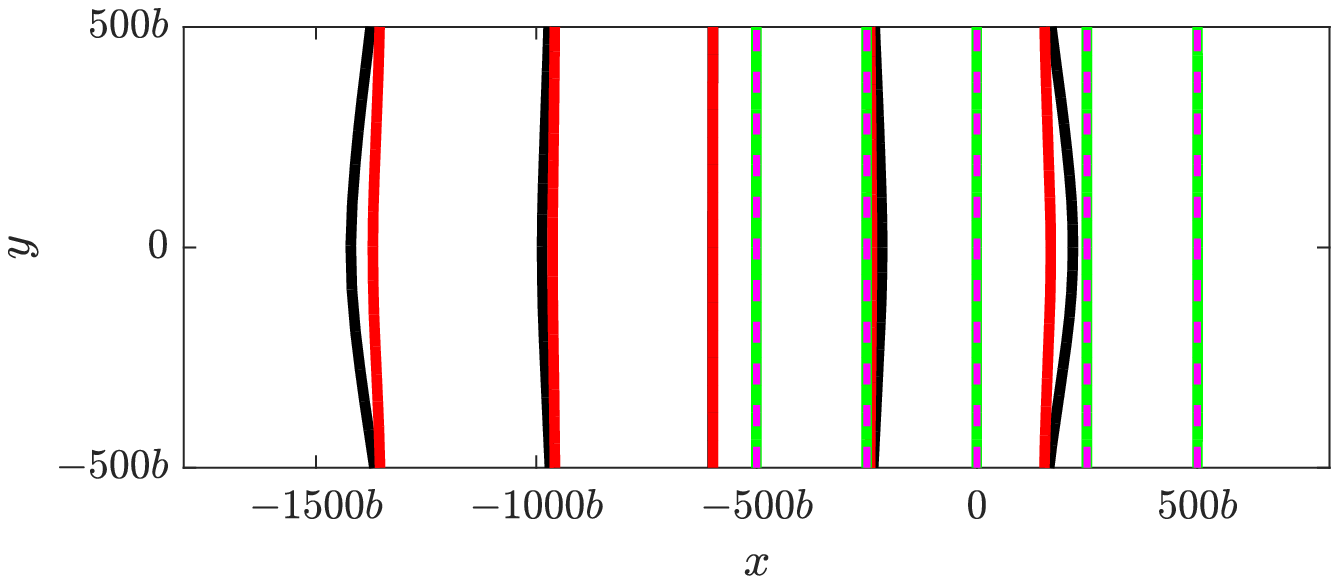}
        \end{subfigure}\hfill
        \caption{A dislocation array evolution from simulations with: (a) $\chi_0=0.001, \sigma_{13}^{\rm A}=\SI{100}{\MPa}$; (b) $\chi_0=0.005, \sigma_{13}^{\rm A}=\SI{10}{\MPa}$; (c) $\chi_0=0.005, \sigma_{13}^{\rm A}=\SI{100}{\MPa}$; (d) $\chi_0=0.05, \sigma_{13}^{\rm A}=\SI{100}{\MPa}$; and (e) $\chi_0=0.1, \sigma_{13}^{\rm A}=\SI{100}{\MPa}$. The magenta dotted lines are the initial positions of dislocations. The black dislocation lines are the benchmark case in the absence of H. The lines in red are with a large $D_H$ (Case \RN{1}) and those in green are with a small $D_H$ (Case \RN{2}).}
        \label{fig:dd-results}
\end{figure}

\begin{figure}[!htb]
        \centering
        \begin{subfigure}[t]{0.03\textwidth}
        \end{subfigure}
        \begin{subfigure}[t]{0.46\textwidth}
        \hspace{11em} \textbf{Case \RN{1}}
        \end{subfigure}\hfill
        \begin{subfigure}[t]{0.46\textwidth}
        \hspace{9.5em} \textbf{Case \RN{2}}
        \end{subfigure}\hfill
        
        \begin{subfigure}[t]{0.03\textwidth}
        \textbf{(a)}
        \end{subfigure}
        \begin{subfigure}[t]{0.46\textwidth}
        \hspace{9.5em}\text{$t=\SI{0.00}{\nano\second}$}
        \end{subfigure}\hfill
        \begin{subfigure}[t]{0.03\textwidth}
        \textbf{(b)}
        \end{subfigure}
        \begin{subfigure}[t]{0.46\textwidth}
        \hspace{9.5em}\text{$t=\SI{0.00}{\nano\second}$}
        \end{subfigure}\hfill
        \begin{subfigure}[t]{0.03\textwidth}
        \hspace{2em}
        \end{subfigure}
        \begin{subfigure}[t]{0.46\textwidth}
        \includegraphics[width=8cm,valign=t]{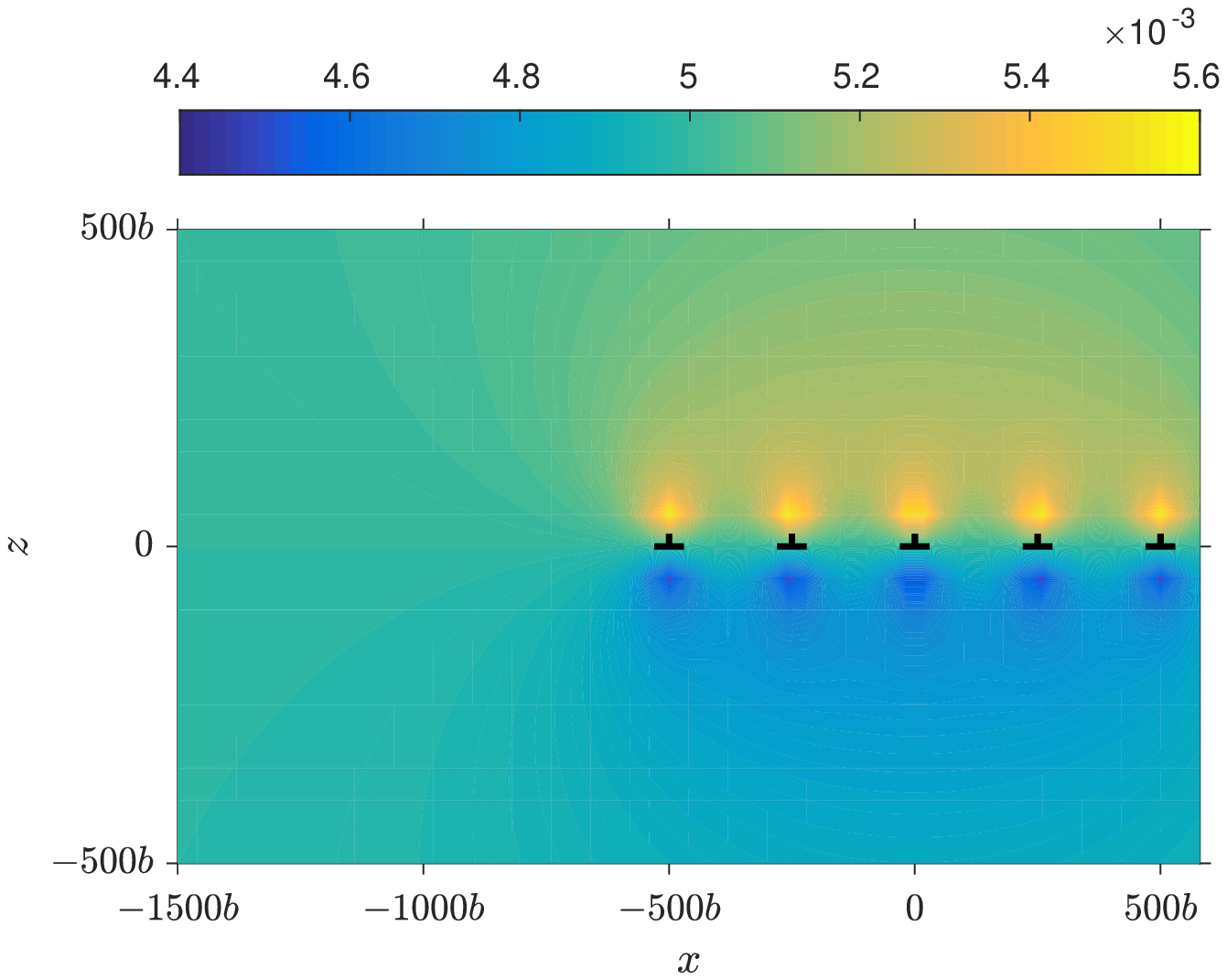}
        \end{subfigure}\hfill
        \begin{subfigure}[t]{0.03\textwidth}
        \hspace{2em}
        \end{subfigure}
        \begin{subfigure}[t]{0.46\textwidth}
        \includegraphics[width=8cm,valign=t]{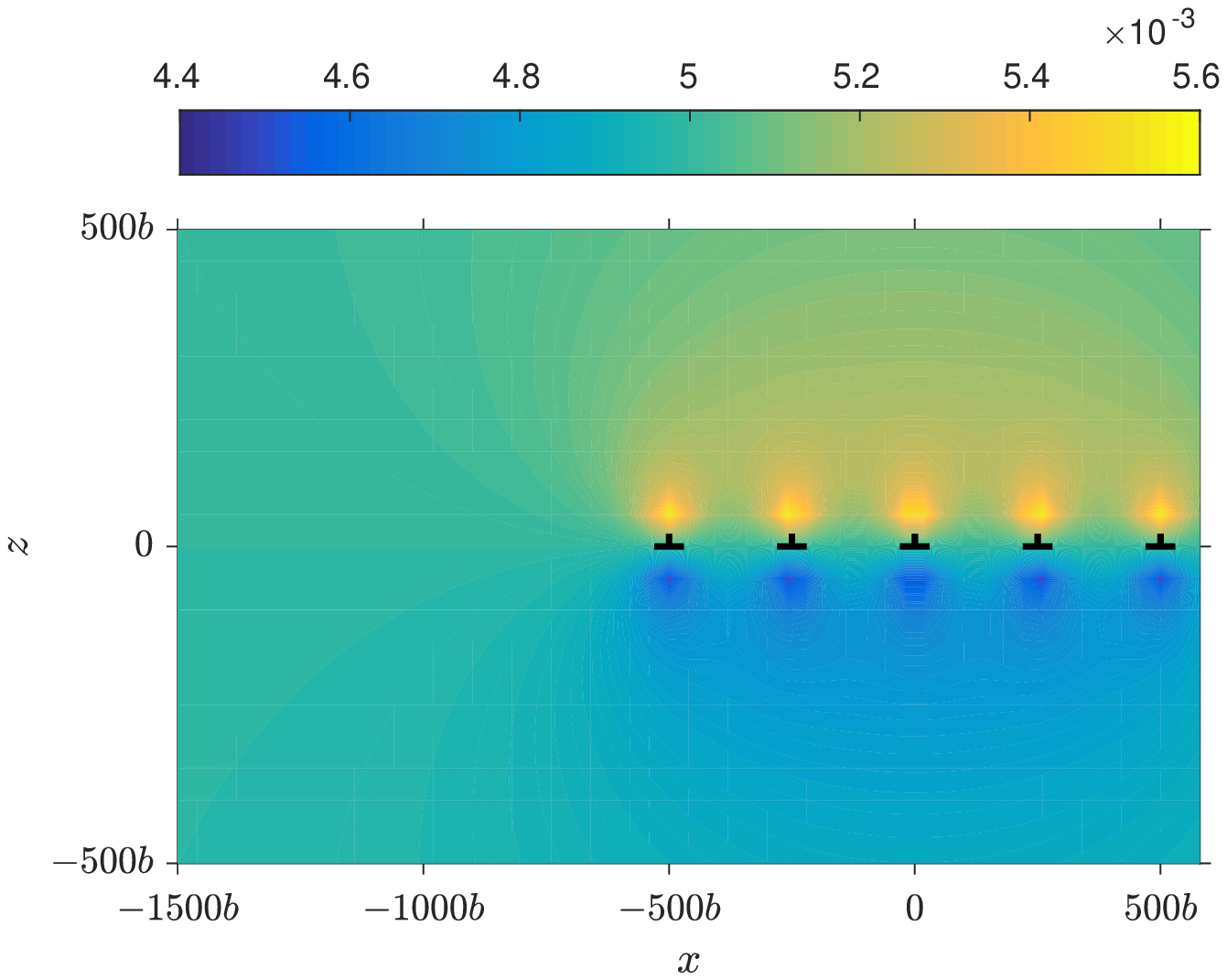}
        \end{subfigure}\hfill
        \begin{subfigure}[t]{0.03\textwidth}
        \hspace{2em}
        \end{subfigure}
        \begin{subfigure}[t]{0.46\textwidth}
        \hspace{9.5em}\text{$t=\SI{0.25}{\nano\second}$}
        \end{subfigure}\hfill
        \begin{subfigure}[t]{0.03\textwidth}
        \hspace{2em}
        \end{subfigure}
        \begin{subfigure}[t]{0.46\textwidth}
        \hspace{9.5em}\text{$t=\SI{0.25}{\nano\second}$}
        \end{subfigure}\hfill 
        \vspace{-1.8em}        
        \begin{subfigure}[t]{0.03\textwidth}
        \hspace{2em}
        \end{subfigure}
        \begin{subfigure}[t]{0.46\textwidth}
        \includegraphics[width=8cm,valign=t]{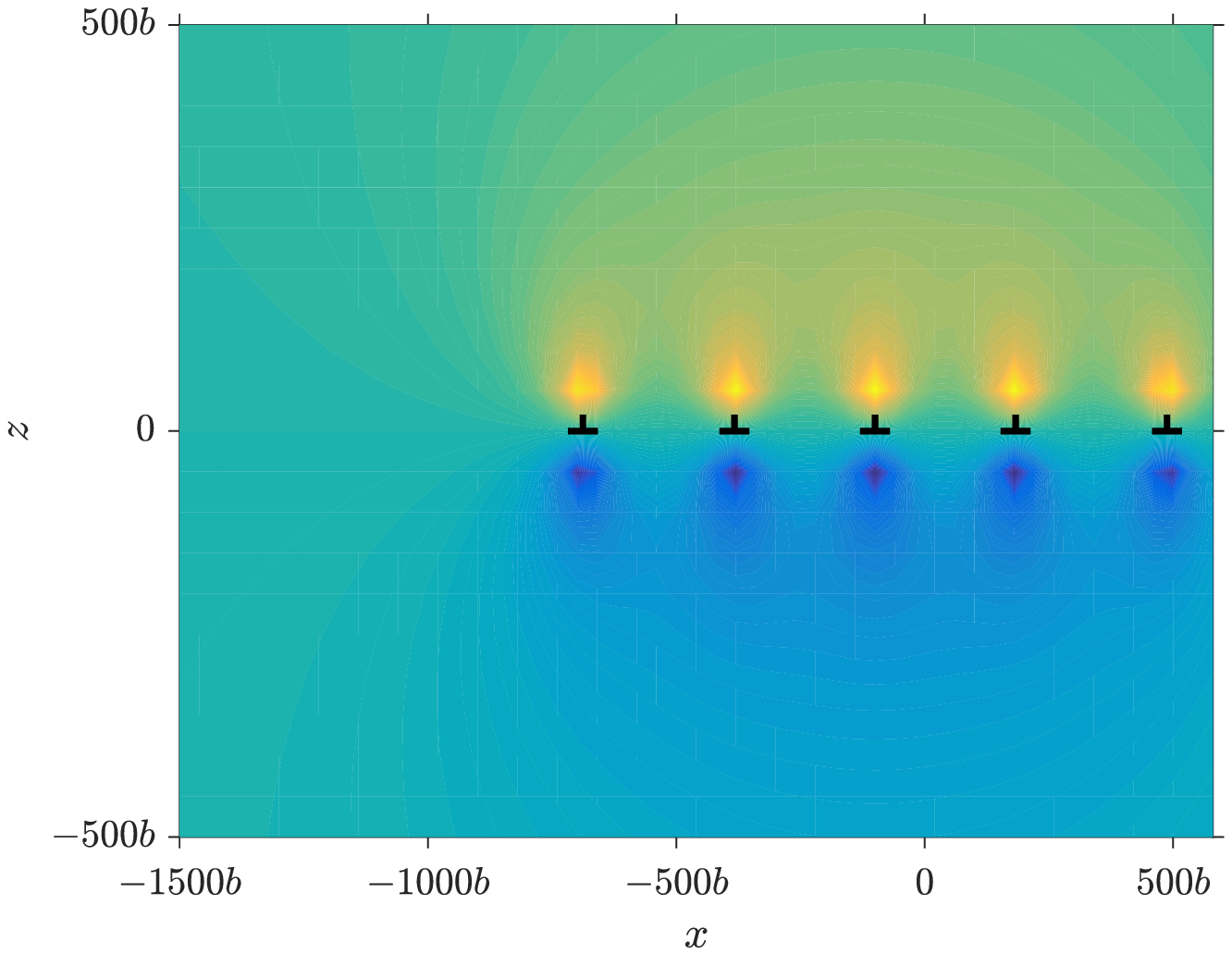}
        \end{subfigure}\hfill
        \begin{subfigure}[t]{0.03\textwidth}
        \hspace{2em}
        \end{subfigure}
        \begin{subfigure}[t]{0.46\textwidth}
        \includegraphics[width=8cm,valign=t]{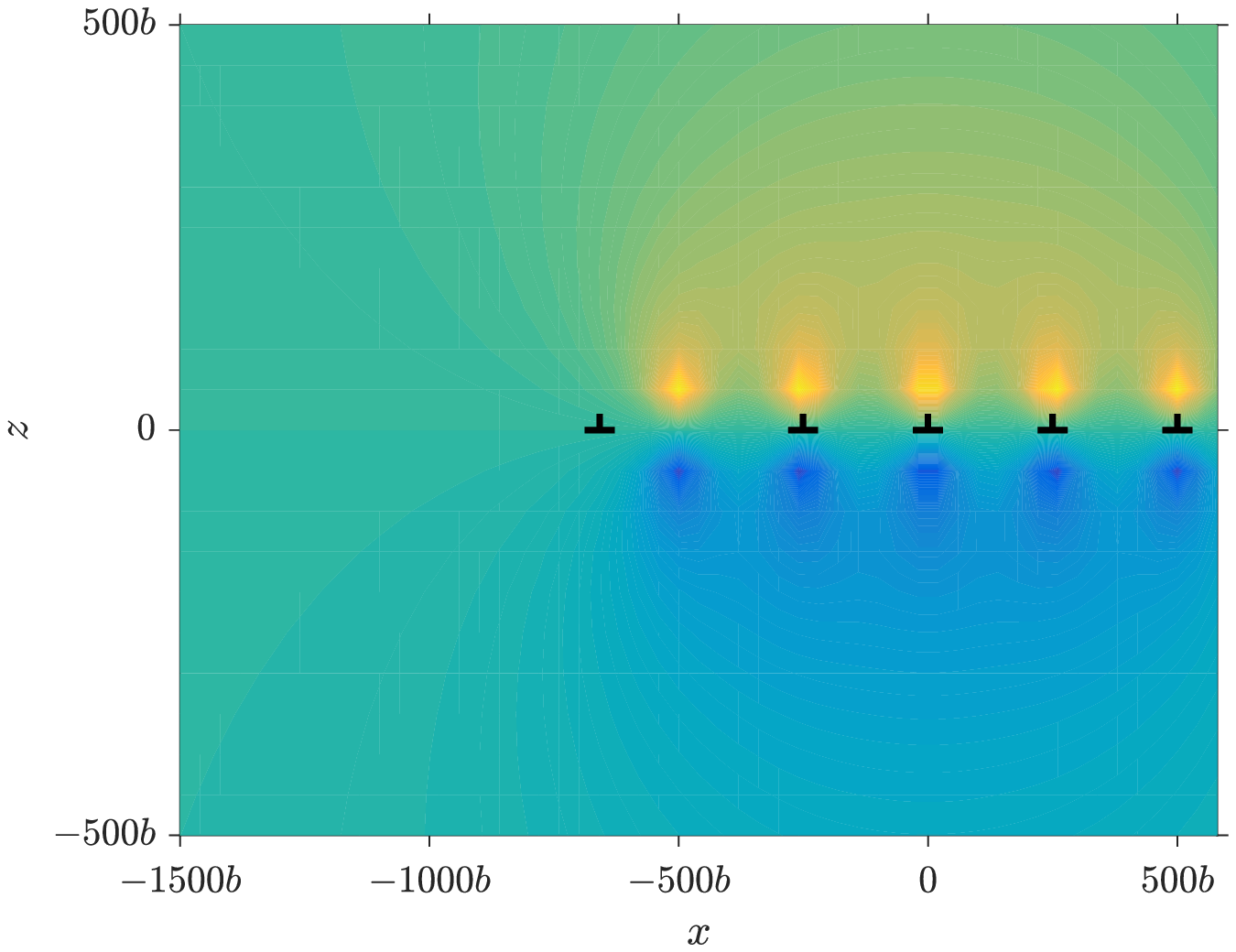}
        \end{subfigure}\hfill
        \begin{subfigure}[t]{0.03\textwidth}
        \hspace{2em}
        \end{subfigure}
        \begin{subfigure}[t]{0.46\textwidth}
        \hspace{9.5em}\text{$t=\SI{1.00}{\nano\second}$}
        \end{subfigure}\hfill
        \begin{subfigure}[t]{0.03\textwidth}
        \hspace{2em}
        \end{subfigure}
        \begin{subfigure}[t]{0.46\textwidth}
        \hspace{9.5em}\text{$t=\SI{1.00}{\nano\second}$}
        \end{subfigure}\hfill        
        \vspace{-1.8em}
        \begin{subfigure}[t]{0.03\textwidth}
        \hspace{2em}
        \end{subfigure}
        \begin{subfigure}[t]{0.46\textwidth}
        \includegraphics[width=8cm,valign=t]{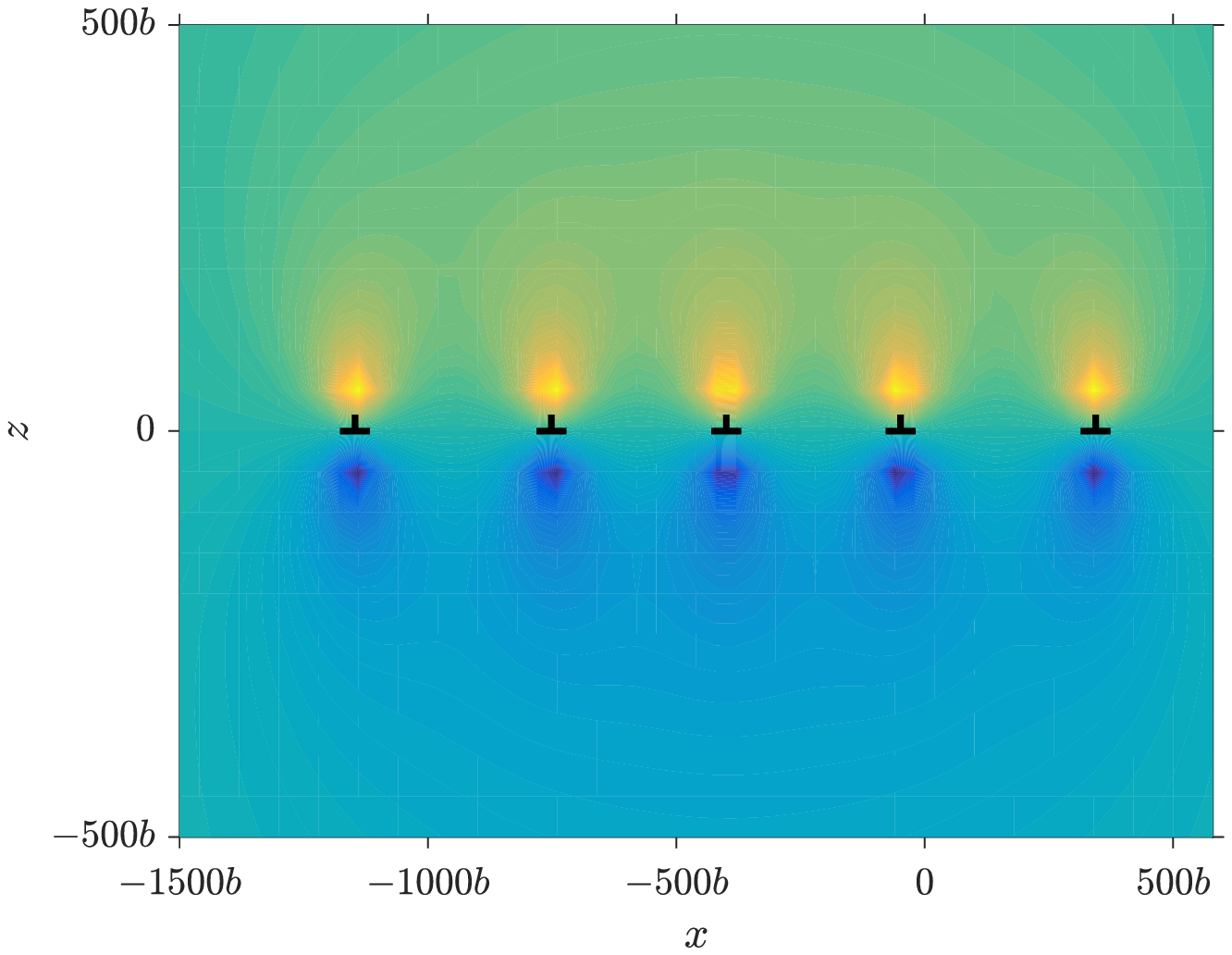}
        \end{subfigure}\hfill
        \begin{subfigure}[t]{0.03\textwidth}
        \hspace{2em}
        \end{subfigure}
        \begin{subfigure}[t]{0.46\textwidth}
        \includegraphics[width=8cm,valign=t]{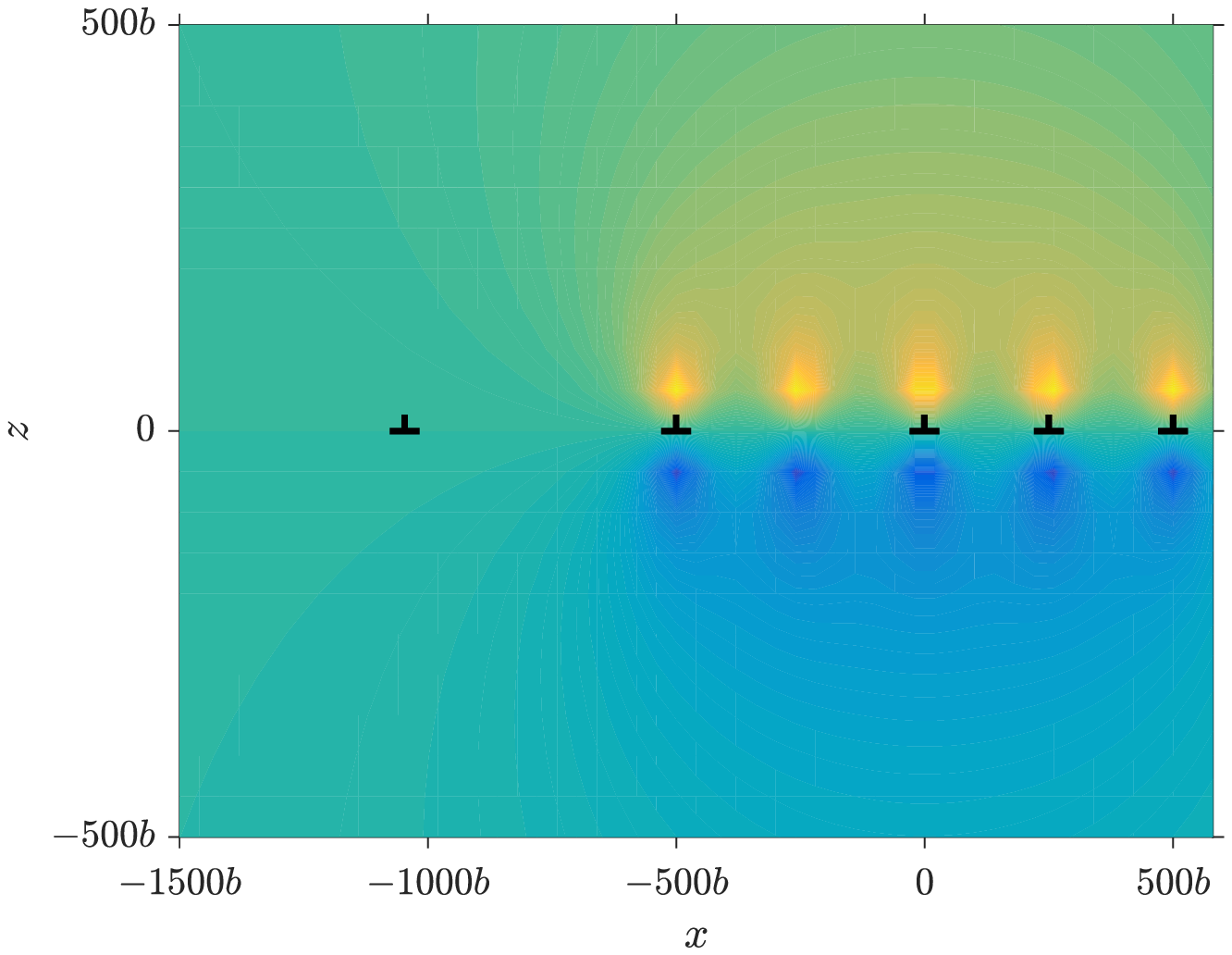}
        \end{subfigure}\hfill
        \caption{ Hydrogen mole fraction distribution on the $y=0$ plane with $\chi_0=0.005$ and $\sigma_{13}^{\rm A}=\SI{100}{\MPa}$: (a) with a large $D_H$ (Case \RN{1}), and (b) with a small $D_H$ (Case \RN{2}) at $t=\SI{0.00}{\nano\second}, \SI{0.25}{\nano\second}$ and $\SI{1.00}{\nano\second}$.}
         \label{fig:c-comp}
\end{figure}

The dislocation microstructure at different times as predicted from the current 3D DDD simulations are shown in Fig. \ref{fig:dd-results} for a select number of positive applied shear stresses $\sigma_{13}^{\rm A}$. It should be noted that a positive applied shear stress will result in a Peach-Kohler force on each dislocation node, which would have a resolved component on the glide plane pointing in the negative $x-$direction.  For a relevantly small $\chi_0=0.001$ and a relevantly large $\sigma_{13}^{\rm A}=\SI{100}{\MPa}$ (Fig. \ref{fig:dd-results}(a)), the H shielding effect is very weak resulting in almost overlapping dislocation lines, regardless of magnitude of the H-diffusion coefficient. For a moderate $\chi_0=0.005$ (Figs \ref{fig:dd-results}(b) and \ref{fig:dd-results}(c)) and a large $D_H$ the dislocation lines almost overlap with those in the benchmark case, regardless of the applied resolved shear stress, indicating a negligible H effect. The H mole fraction distribution profile on the $y=0$ plane for this case is shown in Fig. \ref{fig:c-comp}(a) at three different time steps: $t= 0,~0.25$ and $\SI{1.0}{\nano\second}$. Here, the H distribution is always in equilibrium with the dislocations stress field at each time step. On the other hand, for a moderate $\chi_0=0.005$ and a small $D_H$, it is observed that H pins the dislocations, and a small applied resolved shear stress of $\sigma_{13}^{\rm A}=\SI{10}{\MPa}$ is not sufficient to overcome this pinning effect resulting in all the dislocations being trapped at their initial position (Fig. \ref{fig:dd-results}(b)). However, with a higher resolved shear stress of $\sigma_{13}^{\rm A}=\SI{100}{\MPa}$, the first two leading dislocations in the array are able to break away from their initial position aided by the repulsive stress field of the dislocation array (Fig. \ref{fig:dd-results}(c)). While the first leading dislocation glides freely afterwards, the second leading dislocation is eventually trapped by the high H-concentration left behind by the first leading dislocation. At this point the total force on this dislocation is not large enough to overcome the H pinning effect. Furthermore, the three trailing dislocations are all trapped at their initial positions. The H mole fraction distribution profile on the plane $y=0$ for this case is shown in Fig. \ref{fig:c-comp}(b) at three time steps: $t= 0,~0.25$ and $\SI{1.0}{\nano\second}$. Here, the H distribution is in equilibrium with the dislocations stress field at the first time step, then due to the small $D_H$, the H atoms are not able to diffuse any considerable distance within the simulated time.  

For the two largest reference H mole fractions studied here, $\chi_0=0.05$ and $0.1$ (Figs \ref{fig:dd-results}(d) and \ref{fig:dd-results}(e), respectively), strong H-shielding effect is observed in the simulations with a large $D_H$, while a strong H-pinning effect is observed for simulations with a small $D_H$. The H shielding effect reported here is in excellent agreement with experimental observation indicating a decrease in the separation distance between dislocations in a pileup in 310s stainless steel (i.e. a material with a large $D_H$)~\cite{ferreira1998}. Furthermore, the H-induced pinning effect is also in agreement with MD simulation result ~\cite{tang2012}.
\newline

\begin{figure}[!htb]
        \centering
        \begin{subfigure}[t]{0.03\textwidth}
        \textbf{(a)}
        \end{subfigure}
        \begin{subfigure}[t]{0.46\textwidth}
        \includegraphics[width=8cm,valign=t]{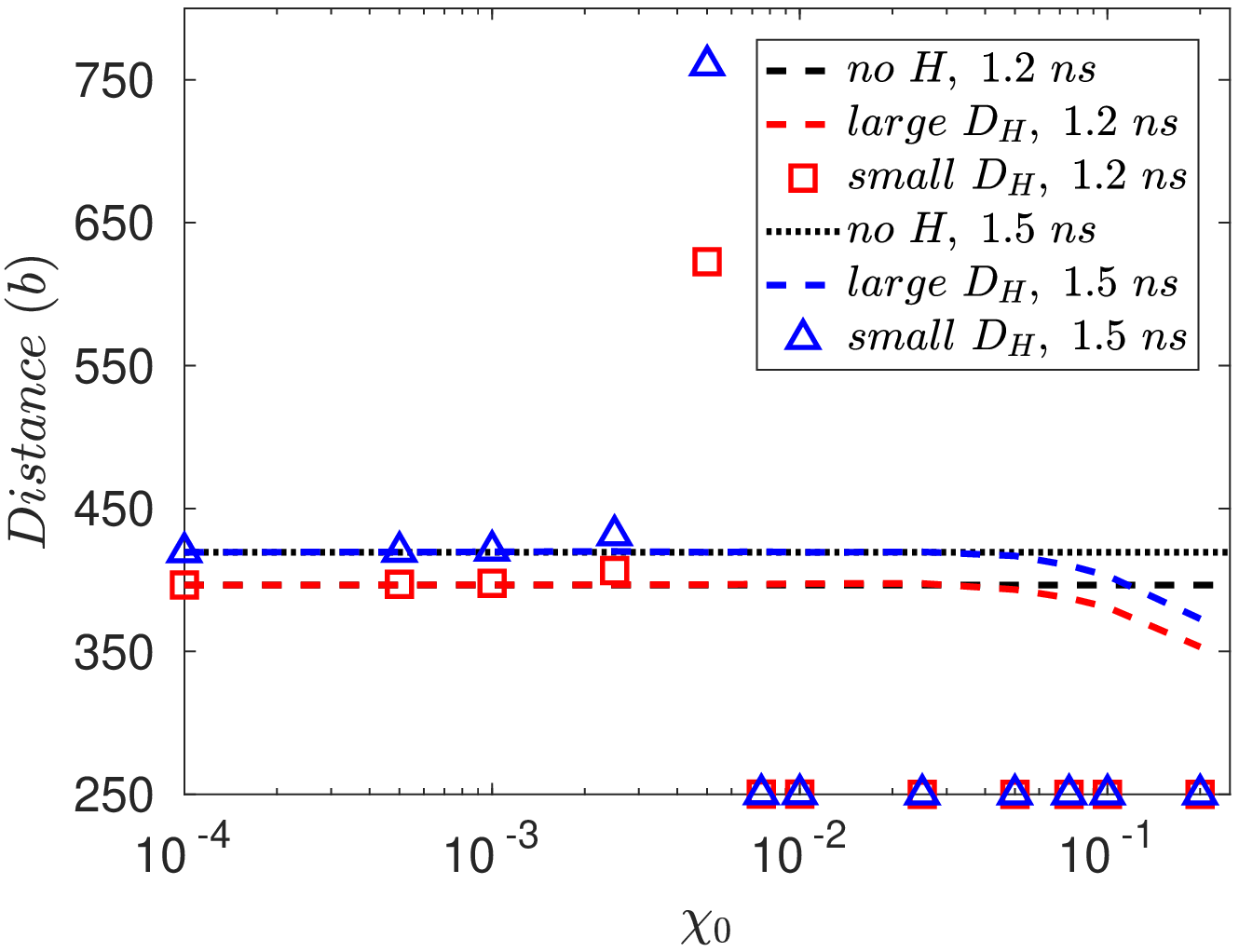}
        \end{subfigure}\hfill
        \begin{subfigure}[t]{0.03\textwidth}
        \textbf{(b)}
        \end{subfigure}
        \begin{subfigure}[t]{0.46\textwidth}
        \includegraphics[width=8cm,valign=t]{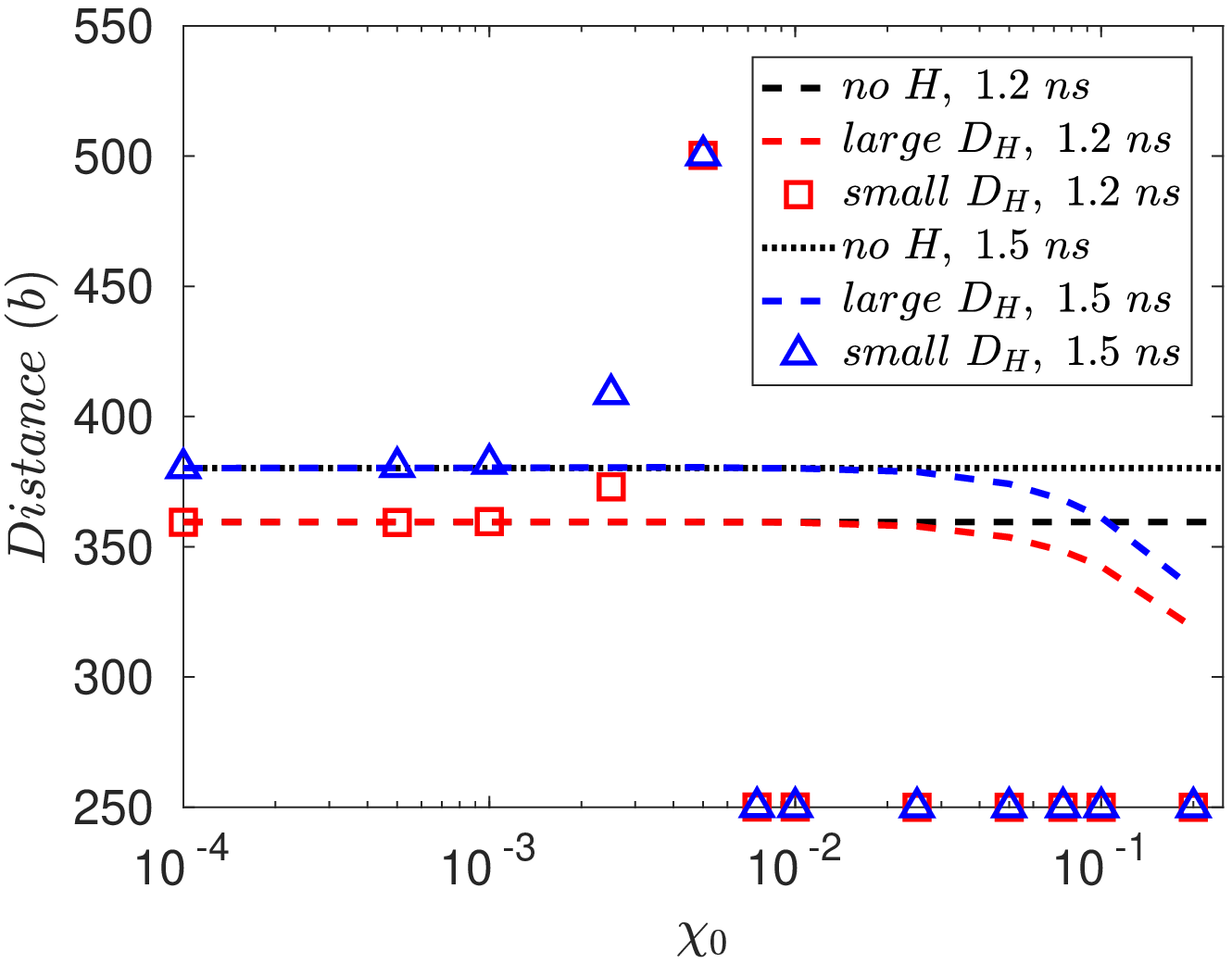}
        \end{subfigure}\hfill
        \caption{Average separation distance between: (a) dislocations \textcircled{1} and \textcircled{2}; (b) dislocations \textcircled{2} and \textcircled{3} as a function of $\chi_0$ and with $\sigma_{13}^{\rm A}=\SI{100}{\MPa}$.}
         \label{fig:dd-comp}
\end{figure}

To quantify the effect of H on the spacing between the dislocations in the array, the spacings between dislocations \textcircled{1} and \textcircled{2} as well as that between dislocations \textcircled{2} and \textcircled{3} are shown in Fig.~\ref{fig:dd-comp} as a function of reference H mole fraction for $\sigma_{13}^{\rm A}=\SI{100}{\MPa}$ at two different simulation times: $t=\SI{1.2}{\nano\second}$ and $\SI{1.5}{\nano\second}$. The spacings for the benchmark case are shown for reference by dashed and dotted lines for the two times, respectively. It should be noted that the spacing between dislocations \textcircled{1} and \textcircled{2} is larger than that between dislocations \textcircled{2} and \textcircled{3} due to the larger repulsive forces applied on dislocation \textcircled{1} than dislocation \textcircled{2} or \textcircled{3}. Furthermore, for simulations with a large $D_H$, the shielding effect, characterized by the decrease in the spacing between the dislocations, increases with increasing $\chi_0$ as a consequence of increasing magnitudes of H-induced stresses. On the other hand, for simulations with a small $D_H$, the applied shear stress is large enough for dislocations \textcircled{1} and \textcircled{2} to overcome the H-induced pinning effect at their initial positions. However, dislocation \textcircled{2} is subsequently trapped due to the high H concentration still remaining at the initial position of dislocation \textcircled{1}. This corresponds to a \textit{partial} pinning effect, which is responsible for the continuously increasing distance between dislocations \textcircled{1} and \textcircled{2} as $\chi_0$ increases in Fig.~\ref{fig:dd-comp}(a) and the constant distance between dislocations \textcircled{2} and \textcircled{3} ($500b$ after $\SI{1.2}{\nano\second}$) in Fig.~\ref{fig:dd-comp}(b). A \textit{complete} pinning effect is observed if we further increase the value of $\chi_0$ (greater than $0.0075$ in this simulation), where all of the dislocations are pinned by H and the distance is always identical to the initial spacing of $250b$.

\begin{figure}[!htb]
        \centering
        \includegraphics[width=8cm]{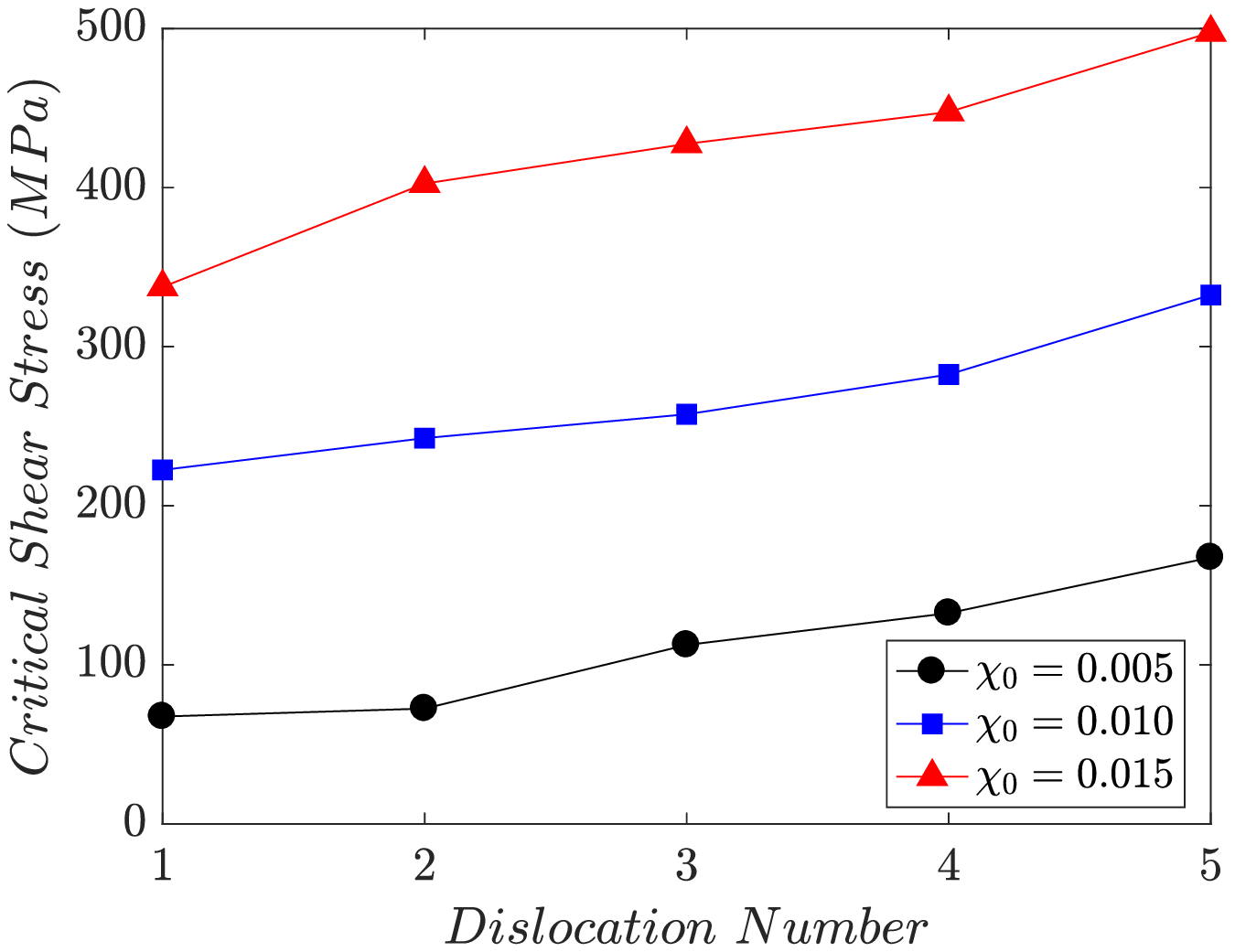}
        \caption{Critical shear stress for each dislocation to overcome its initial local pinning for different reference H mole fractions with a small $D_H$.}
        \label{fig:cstress}
\end{figure}

To better understand the role of H-induced pinning effects in metals having a small H diffusion coefficient, the critical shear stress for each dislocation to break away from its initial site for three reference H mole fractions are shown in Fig. \ref{fig:cstress}. It is clear that with increasing reference H mole fraction the critical shear stress for each dislocation to break away from its initial site increases. Furthermore, every time a dislocation is able to overcome its local pinning effect, the remaining pinned dislocations experience weaker repulsive forces. Therefore, a larger critical shear stress is required for the remaining dislocations to overcome their local H-induced pinning. 

It is also possible to quantify the pinning effect analytically for an infinitely long straight edge dislocation, where the H concentration is in equilibrium with the initial dislocation position (i.e. a material with a small $D_H$). Assume the dislocation line direction is $[010]$ and the Burgers vector is $b[100]$, the total shear stress on the dislocation is
\begin{align}
\sigma_{13}^{tot}=\sigma_{13}^{\rm A}+\sigma_{13}^{d}+\sigma_{13}^{H},
\label{eq:stress-critic}
\end{align}
For a single infinitely long dislocation, the self-stress term $\sigma_{13}^{d}$ vanishes. Hence, the critical shear stress for this dislocation corresponds to the case $\sigma_{13}^{\rm A}=\sigma_{13}^{H}$, or
\begin{align}
\sigma_{13}^{\rm A}=&\frac{G^2(1+\nu)^2\Delta V^2}{18\pi(1-\nu)^2k_{\rm B}T}c_{max}\chi_0(1-\chi_0).
\label{eq:stress-an}
\end{align}
By utilizing the material parameters for Ni (listed in Table \ref{table:parameter}), it is predicted from Eq. \eqref{eq:stress-an} that $\sigma_{13}^{\rm A}=\SI{80.61}{\MPa}$ for $\chi_0=0.005$ and $T=\SI{300}{\K}$, which is comparable with the values predicted in Fig.~\ref{fig:cstress} for the stress required for dislocation \textcircled{1} to break away.

If the applied shear stress is below this critical value, the attractive force produced by H atoms is large enough to trap the dislocation around its initial position. Once the applied shear stress is beyond this critical value, the dislocation will overcome the H-induced attractive force leading to dislocation glide. 

\section{Conclusions}\label{sec:conc}
In this work, a new formulation to account for hydrogen effects in the framework of three-dimensional discrete dislocation dynamics simulations was developed. The contribution of hydrogen in the DDD framework can be divided into two aspects. The first aspect is the H-dislocation interaction, which is accounted for here by the first order elastic interaction energy term associated with the volume change induced by the H-inclusion. The second is predicting the hydrogen distribution in the simulation cell at every time step, which is computed through a continuum representation. The most important assumption in the current approach is that the H-induced stress field is derived for a hydrogen concentration that is in equilibrium with the pre-existing dislocation induced stress field. This assumption works well for materials having a large or small hydrogen diffusion coefficients. However, to precisely model the hydrogen diffusion process for materials having an intermediate hydrogen diffusion coefficients, it is inevitable to solve the diffusion equation, which is somewhat a numerically expensive calculation due to the need for a high resolution near the dislocation core. 

Another simplification made in the current analysis is that only the first order elastic interaction energy, which is associated with the volume change induced by the H-atom inclusion, is accounted for. This assumption is consistent with the Eshelby inclusion model, in which the elastic constants do not changed with the presence of an inclusion.  Nevertheless, other interaction energy terms such as the second order elastic interaction energy associated with moduli change as well as higher order hydrogen-hydrogen interaction energy terms could have an effect on the dislocation dynamics. The formulas to include these interaction energies will be reported elsewhere. This limitation also exists in the discussion of the H pinning effect. To accurately describe this effect on dislocations, the dislocation core structures and the binding energy landscapes of H atoms around dislocations should be considered, which needs to be informed from atomic simulations. In this paper, only the pinning effect due to the first order elastic interaction energy is considered.

Furthermore, even though the formulations for H-induced stresses are derived in an infinite medium, the impact of the H-induced image fields arising from the finite medium on dislocation microstructures may be neglected with small errors for large simulation cells (e.g. micropillars having diameters larger than \SI{1.0}{\um} \cite{el2008}) since the H-induced stress field has similar expressions as the dislocation stress field. Nevertheless, it is still possible to account for free surface boundary conditions in a similar manner to the superposition method \cite{fivel1999} by coupling with FEM \cite{tang2003} or BEM \cite{el2008}.

The newly developed framework was utilized to quantify the effect of hydrogen on the dynamics of a glide loop and an array of parallel edge dislocations. For a glide loop in a material having a large hydrogen diffusion coefficient, the loop shrinkage process is homogenized due to hydrogen shielding effects and the glide loop maintains its circular shape. For a glide loop in a material with a small hydrogen diffusion coefficient, the screw segments move towards the loop center more easily than edge segments since edge segments experience a pinning effect due to the high hydrogen concentration surrounding these edge segments at its initial position. Furthermore, in an array of parallel edge dislocations, the dislocation separation distance is observed to decrease with increasing hydrogen concentration in materials having a large hydrogen diffusion coefficient. However, dislocations are completely or partially pinned by hydrogen in materials having a small hydrogen diffusion coefficient. 

Finally, this new framework can open the door for further large scale studies on the effect of hydrogen on the different aspects of dislocation-mediated plasticity in metals. With further modification, the method can also be generalized to model the effect of other inclusion-induced (e.g. C-induced) stress fields.  
 
\section*{Acknowledgement}
This work was supported by the National Science Foundation CAREER Award \#CMMI-1454072. The authors would like to thank Prof. Wei Cai of Stanford University for helpful discussions regarding the hydrogen concentration distributions.

\appendix
\setcounter{section}{-1}
\renewcommand{\thesection}{Appendix}
\setcounter{equation}{0}
\renewcommand{\theequation}{A.\arabic{equation}}
\setcounter{figure}{0}
\setcounter{table}{0}
\section{Derivation of the hydrogen-induced three-dimensional stress field}
\label{sec:app}
Here, derivation of the H-induced 3D stress field is presented for completeness. The total strain can be derived from the displacement field $\mathbf{u}$ according to
\begin{align}
\varepsilon_{ij}^{\rm tot}=\frac{1}{2}(\frac{\partial u_i}{\partial x_j}+\frac{\partial u_j}{\partial x_i}).
\label{eq:app1}
\end{align}

The elastic strain is the difference between the total strain $\varepsilon_{ij}$ and hydrogen-induced strain $e^H_{ij}$, where the hydrogen-induced strain has a local expression as
\begin{align}
e^H_{ij}=\frac{1}{3}(c-c_0)\Delta V\delta_{ij}.
\label{eq:app2}
\end{align}

Thus, the H-induced stress is
\begin{align}
\sigma_{ij}^H=2G(\varepsilon^{tot}_{ij}-e^H_{ij})+\frac{2G\nu}{1-2\nu}(\varepsilon^{tot}_{kk}-e^H_{kk}).
\label{eq:app3}
\end{align}

The equilibrium condition for this H-induced stress field in the absence of body forces is
\begin{align}
\frac{\partial \sigma^H_{ij}}{\partial x_j}=0.
\label{eq:app4}
\end{align}

Substituting Eqs~\eqref{eq:app1},~\eqref{eq:app2} and ~\eqref{eq:app3} into Eq.~\eqref{eq:app4}, the force balance equation becomes
\begin{align}
\nabla^2 u_i+\frac{1}{1-2\nu}\frac{\partial u_{k}}{\partial x_k\partial x_i}=\frac{1+\nu}{1-2\nu}\frac{2\Delta V}{3}\frac{\partial (c-c_0)}{\partial x_i}.
\label{eq:app5}
\end{align}

By introducing a scalar potential $B_0$ such that (this expression is valid in an infinite medium):
\begin{align}
\mathbf{u}=-\frac{1}{4\pi(1-\nu)}\nabla B_0,
\end{align}
and substituting into Eq.~\eqref{eq:app5}, the following Poisson's equation is obtained
\begin{align}
\nabla^2B_{0,i}=-\frac{4(1+\nu)}{3}\Delta V\frac{\partial (c-c_0)}{\partial x_i}.
\end{align}

By integrating Eq.~\eqref{eq:app5} once with respect to $x_i$ then:
\begin{align}
\nabla^2B_0=-\frac{4(1+\nu)}{3}(c-c_0)\Delta V.
\label{eq:app6}
\end{align}

The H-induced stress field can then be expressed in terms of this scalar potentials by solving Eqs~\eqref{eq:app1} through ~\eqref{eq:app6} along with Eq.~\eqref{eq:app8} such that:
\begin{align}
\sigma_{ij}^H=-\frac{G}{2(1-\nu)}B_{0,ij}-\frac{2G(1+\nu)}{3(1-\nu)}(c-c_0)\Delta V\delta_{ij}.
\label{eq:app7}
\end{align}

Now, consider the fundamental solution $F(\mathbf{x})$ for the Laplace equation:
\begin{align}
\nabla^2 F=\delta(\mathbf{x}),
\end{align}
where
\begin{align}
F(\mathbf{x})=\frac{1}{4\pi \sqrt{x_1^2+x_2^2+x_3^2}}.
\end{align}

The solution to Eq.~\eqref{eq:app6} can be explicitly expressed by convolving the right-hand side term $-\frac{4(1+\nu)}{3}(c-c_0)\Delta V$ with the fundamental solution $F(\mathbf{x})$ as
\begin{align}
B_0=-\frac{(1+\nu)}{3\pi}\int_{\mathbb{R}^3}\frac{c(\mathbf{y})-c_0}{\sqrt{(x_1-y_1)^2+(x_2-y_2)^2+(x_3-y_3)^2}}\ud y_1\ud y_2\ud y_3.
\end{align}

Substituting into Eq.~\eqref{eq:app7} the H-induced stress tensor in 3D space can be expressed as
\begin{align}
\sigma_{ij}^H(\mathbf{x})=&-\frac{G(1+\nu)\Delta V}{6\pi(1-\nu)}\int_{\mathbb{R}^3}\left(c(\mathbf{y})-c_0)\right)\cdot \Big[\frac{\delta_{ij}}{[(x_1-y_1)^2+(x_2-y_2)^2+(x_3-y_3)^2]^{3/2}}\nonumber\\
&\hspace{8em}-\frac{3\left(x_i-y_i\right)\left(x_j-y_j\right)}{[(x_1-y_1)^2+(x_2-y_2)^2+(x_3-y_3)^2]^{5/2}}\Big]\ud y_1\ud y_2\ud y_3\nonumber\\
&-\frac{2G(1+\nu)}{3(1-\nu)}(c(\mathbf{x})-c_0)\Delta V\delta_{ij}.
\label{eq:app8}
\end{align}

It should be noted that by converting this 3D stress field into a 2D formulation, the formulation would be in agreement with the 2D analytical solution derived by Cai et al.~\cite{cai2014}, but differs slightly from the 2D stress field derived by Sorfronis and Birnbaum~\cite{sofronis1995b}. The ratio of the two results is $2(1+\nu)/3$, which comes from the dimension difference when calculating the hydrostatic stress. A reference concentration $c_0$ is subtracted in the current 3D formulation as well as the 2D formulation developed by Cai et al., but it was not explicitly accounted for in the 2D formulation developed by in Sorfronis and Birnbaum.
%where a cylindrical coordinate is used, where $c$ is uniform along $z-$axis. By taking the outer cut-off radius $R$ and the inner cut-off radius $r_{\rm cut},$ the shear stress has contribution to dislocation glide is
%\begin{align}
%\tau^{\rm H}(\mathbf{x})\approx-\frac{G_0(1+\nu_0)}{3\pi(1-\nu_0)}\frac{V_{\rm H}}{N_{\rm A}}\int_0^{2\pi}\int_{r_{\rm cut}}^{R}\left(c(r\cos\theta,r\sin\theta)-c_0\right)\frac{\sin(2\theta)}{r}\ud r\ud \theta
%\label{eq:tau-cont_inf2d}
%\end{align}

Note that in practice, the computation cost of the integral in Eq.~\eqref{eq:app8} for a general hydrogen concentration is expensive due to the high resolution requirement at dislocation cores and to track moving dislocation lines. Proper numerical schemes are essential for solving Eq.~\eqref{eq:app8}, which is beyond the scope of the current paper and will be addressed elsewhere.

\bibliography{Gu_he_bbl,Gu_dislocations_bbl}

\begin{thebibliography}{10}

\bibitem{arsenlis2007}
A.~Arsenlis, W.~Cai, M.~Tang, M.~Rhee, T.~Oppelstrup, G.~Hommes, T.~G. Pierce,
  and V.~V. Bulatov.
\newblock Enabling strain hardening simulations with dislocation dynamics.
\newblock {\em Model. Simul. Mater. Sci. Eng.}, 15(6):553, 2007.

\bibitem{bammann2005}
D.~J. Bammann and P.~Sofronis.
\newblock A coupled dislocation-hydrogen based model of inelastic deformation.
\newblock In {\em ICF11, Italy 2005}.

\bibitem{barnoush2008}
A.~Barnoush and H.~Vehoff.
\newblock In situ electrochemical nanoindentation: A technique for local
  examination of hydrogen embrittlement.
\newblock {\em Corros. Sci.}, 50(1):259--267, 2008.

\bibitem{barnoush2010}
A.~Barnoush and H.~Vehoff.
\newblock Recent developments in the study of hydrogen embrittlement: Hydrogen
  effect on dislocation nucleation.
\newblock {\em Acta Mater.}, 58(16):5274--5285, 2010.

\bibitem{beachem1972}
C.~D. Beachem.
\newblock A new model for hydrogen-assisted cracking (hydrogen
  “embrittlement”).
\newblock {\em Metall. Mater. Trans. B}, 3(2):441--455, 1972.

\bibitem{birnbaum1994}
H.~K. Birnbaum and P.~Sofronis.
\newblock Hydrogen-enhanced localized plasticity$-$a mechanism for
  hydrogen-related fracture.
\newblock {\em Mater. Sci. Eng. A Struct.}, 176(1-2):191--202, 1994.

\bibitem{bulatov2006}
V.~V. Bulatov and W.~Cai.
\newblock {\em Computer simulations of dislocations}, volume~3.
\newblock Oxford University Press on Demand, 2006.

\bibitem{bulatov_2006}
V.~V. Bulatov, L.~L. Hsiung, M.~Tang, A.~Arsenlis, M.~C. Bartelt, W.~Cai, J.~N.
  Florando, M.~Hiratani, M.~Rhee, G.~Hommes, T.~Pierce, and T.~D. Rubia.
\newblock Dislocation multi-junctions and strain hardening.
\newblock {\em Nature}, 440(7088):1174--1178, 2006.

\bibitem{cai2006}
W.~Cai, A.~Arsenlis, C.~R. Weinberger, and V.~V. Bulatov.
\newblock A non-singular continuum theory of dislocations.
\newblock {\em J. Mech. Phys. Solids}, 54(3):561--587, 2006.

\bibitem{cai2014}
W.~Cai, R.~B. Sills, D.~M. Barnett, and W.~D. Nix.
\newblock Modeling a distribution of point defects as misfitting inclusions in
  stressed solids.
\newblock {\em J. Mech. Phys. Solids}, 66:154--171, 2014.

\bibitem{crone2014}
J.~C. Crone, P.~W. Chung, K.~W. Leiter, J.~Knap, S.~Aubry, G.~Hommes, and
  A.~Arsenlis.
\newblock A multiply parallel implementation of finite element-based discrete
  dislocation dynamics for arbitrary geometries.
\newblock {\em Model. Simul. Mater. Sci. Eng.}, 22(3):035014, 2014.

\bibitem{el2008}
J.~A. El-Awady, S.~B. Biner, and N.~M. Ghoniem.
\newblock A self-consistent boundary element, parametric dislocation dynamics
  formulation of plastic flow in finite volumes.
\newblock {\em J. Mech. Phys. Solids}, 56(5):2019--2035, 2008.

\bibitem{el2016}
J.~A. El-Awady, H.~Fan, and A.~M. Hussein.
\newblock Advances in discrete dislocation dynamics modeling of size-affected
  plasticity.
\newblock In {\em Multiscale Materials Modeling for Nanomechanics}, pages
  337--371. Springer, 2016.

\bibitem{el2009}
J.~A. El-Awady, M.~Wen, and N.~M. Ghoniem.
\newblock The role of the weakest-link mechanism in controlling the plasticity
  of micropillars.
\newblock {\em J. Mech. Phys. Solids}, 57(1):32--50, 2009.

\bibitem{eshelby1955}
J.~D. Eshelby.
\newblock The elastic interaction of point defects.
\newblock {\em Acta Metall.}, 3(5):487--490, 1955.

\bibitem{eshelby1957}
J.~D. Eshelby.
\newblock The determination of the elastic field of an ellipsoidal inclusion,
  and related problems.
\newblock {\em Proc. R. Soc. Lond. A Math. Phys. Sci.}, 241(1226):376--396,
  1957.

\bibitem{eshelby1959}
J.~D. Eshelby.
\newblock The elastic field outside an ellipsoidal inclusion.
\newblock {\em Proc. R. Soc. Lond. A Math. Phys. Sci.}, 252(1271):561--569,
  1959.

\bibitem{eshelby1961}
J.~D. Eshelby.
\newblock Elastic inclusions and inhomogeneities.
\newblock {\em Prog. Solid Mech.}, 2(1):89--140, 1961.

\bibitem{fan2015}
H.~Fan, S.~Aubry, A.~Arsenlis, and J.~A. El-Awady.
\newblock The role of twinning deformation on the hardening response of
  polycrystalline magnesium from discrete dislocation dynamics simulations.
\newblock {\em Acta Mater.}, 92:126--139, 2015.

\bibitem{ferreira1998}
P.~J. Ferreira, I.~M. Robertson, and H.~K. Birnbaum.
\newblock Hydrogen effects on the interaction between dislocations.
\newblock {\em Acta Mater.}, 46(5):1749--1757, 1998.

\bibitem{fivel1999}
M.~C. Fivel, T.~J. Gosling, and G.~R. Canova.
\newblock Developing rigorous boundary conditions to simulations of discrete
  dislocation dynamics.
\newblock {\em Model. Simul. Mater. Sci. Eng.}, 7(5):753, 1999.

\bibitem{gahr1977}
S.~Gahr, M.~L. Grossbeck, and H.~K. Birnbaum.
\newblock Hydrogen embrittlement of \text{Nb I—}macroscopic behavior at low
  temperatures.
\newblock {\em Acta Metall.}, 25(2):125--134, 1977.

\bibitem{ghoniem2000}
N.~M. Ghoniem, S.-H. Tong, and L.~Z. Sun.
\newblock Parametric dislocation dynamics: A thermodynamics-based approach to
  investigations of mesocopic plastic deformation.
\newblock {\em Phys. Rev. B}, 61:913--927, 2000.

\bibitem{gu2015}
Y.~Gu, Y.~Xiang, S.~S. Quek, and D.~J. Srolovitz.
\newblock Three-dimensional formulation of dislocation climb.
\newblock {\em J. Mech. Phys. Solids}, 83:319--337, 2015.

\bibitem{haftbaradaran2011}
H.~Haftbaradaran, J.~Song, W.~A. Curtin, and H.~Gao.
\newblock Continuum and atomistic models of strongly coupled diffusion, stress,
  and solute concentration.
\newblock {\em J. of Power Sources}, 196(1):361--370, 2011.

\bibitem{hirth1982theory}
J.~P. Hirth and J.~Lothe.
\newblock {\em Theory of Dislocations}.
\newblock John Wiley \& Sons, 1982.

\bibitem{huang2017}
S.~Huang, D.~Chen, J.~Song, D.~L. McDowell, and T.~Zhu.
\newblock Hydrogen embrittlement of grain boundaries in nickel: an atomistic
  study.
\newblock {\em npj Comp. Mater.}, 3(1):28, 2017.

\bibitem{HUSSEIN2015}
A.~M. Hussein, S.~I. Rao, M.~D. Uchic, D.~M. Dimiduk, and J.~A. El-Awady.
\newblock Microstructurally based cross-slip mechanisms and their effects on
  dislocation microstructure evolution in fcc crystals.
\newblock {\em Acta Mater.}, 85:180--190, 2015.

\bibitem{itakura2013}
M.~Itakura, H.~Kaburaki, M.~Yamaguchi, and T.~Okita.
\newblock The effect of hydrogen atoms on the screw dislocation mobility in bcc
  iron: A first-principles study.
\newblock {\em Acta Mater.}, 61(18):6857--6867, 2013.

\bibitem{johnson1874}
W.~H. Johnson.
\newblock On some remarkable changes produced in iron and steel by the action
  of hydrogen and acids.
\newblock {\em Proc. R. Soc. Lond.}, 23(156-163):168--179, 1874.

\bibitem{keller2010}
C.~Keller, E.~Hug, R.~Retoux, and X.~Feaugas.
\newblock Tem study of dislocation patterns in near-surface and core regions of
  deformed nickel polycrystals with few grains across the cross section.
\newblock {\em Mech. Mater.}, 42(1):44--54, 2010.

\bibitem{Keralavarma2012}
S.~M. Keralavarma, T.~Cagin, A.~Arsenlis, and A.~A. Benzerga.
\newblock Power-law creep from discrete dislocation dynamics.
\newblock {\em Phys. Rev. Lett.}, 109:265504, 2012.

\bibitem{kubin1993}
L.~P. Kubin.
\newblock Dislocation patterning during multiple slip of fcc crystals. a
  simulation approach.
\newblock {\em Phys. Status Solidi A}, 135(2):433--443, 1993.

\bibitem{lu2001}
G.~Lu, Q.~Zhang, N.~Kioussis, and E.~Kaxiras.
\newblock Hydrogen-enhanced local plasticity in aluminum: an ab initio study.
\newblock {\em Phys. Rev. Lett.}, 87(9):095501, 2001.

\bibitem{martin2011}
M.~L. Martin, I.~M. Robertson, and P.~Sofronis.
\newblock Interpreting hydrogen-induced fracture surfaces in terms of
  deformation processes: a new approach.
\newblock {\em Acta Mater.}, 59(9):3680--3687, 2011.

\bibitem{martin2013}
M.~L. Martin, P.~Sofronis, I.~M. Robertson, T.~Awane, and Y.~Murakami.
\newblock A microstructural based understanding of hydrogen-enhanced fatigue of
  stainless steels.
\newblock {\em Int. J. Fatigue}, 57:28--36, 2013.

\bibitem{martin2012}
M.~L. Martin, B.~P. Somerday, R.~O. Ritchie, P.~Sofronis, and I.~M. Robertson.
\newblock Hydrogen-induced intergranular failure in nickel revisited.
\newblock {\em Acta Mater.}, 60(6):2739--2745, 2012.

\bibitem{nedelcu2002}
S.~Nedelcu and P.~Kizler.
\newblock Molecular dynamics simulation of hydrogen--edge dislocation
  interaction in bcc iron.
\newblock {\em Phys. Status Solidi A}, 193(1):26--34, 2002.

\bibitem{neuber1934}
H.~Neuber.
\newblock Ein neuer ansatz zur l{\"o}sung r{\"a}umlicher probleme der
  elastizit{\"a}tstheorie. der hohlkegel unter einzellast als beispiel.
\newblock {\em J. Appl. Math. Mech./Z. Angew. Math. Mech.}, 14(4):203--212,
  1934.

\bibitem{novak2009}
P.~M. Novak.
\newblock A dislocation-based constitutive model for hydrogen-deformation
  interactions and a study of hydrogen-induced intergranular fracture, 2009.

\bibitem{oriani1972}
R.~Oriani.
\newblock A mechanistic theory of hydrogen embrittlement of steels.
\newblock {\em Ber. Bunsenges. Phys. Chem.}, 76(8):848--857, 1972.

\bibitem{oriani1979}
R.~A. Oriani and P.~H. Josephic.
\newblock Hydrogen-enhanced load relaxation in a deformed medium-carbon steel.
\newblock {\em Acta Metall.}, 27(6):997--1005, 1979.

\bibitem{pap1932}
P.~F. Papkovich.
\newblock The representation of the general integral of the fundamental
  equations of elasticity theory in terms of harmonic functions.
\newblock {\em Izv. Akad. Nauk SSSR, Phys. Math. Ser.}, 10(1425):90, 1932.

\bibitem{pundt2006}
A.~Pundt and R.~Kirchheim.
\newblock Hydrogen in metals: microstructural aspects.
\newblock {\em Annu. Rev. Mater. Res.}, 36:555--608, 2006.

\bibitem{Quek2013}
S.~S. Quek, R.~Ahluwalia, and D.~J. Srolovitz.
\newblock Deconstructing the high-temperature deformation of phase separating
  alloy.
\newblock {\em Model. Simul. Mater. Sci. Eng.}, 21:07501, 2013.

\bibitem{robertson2001}
I.~M. Robertson.
\newblock The effect of hydrogen on dislocation dynamics.
\newblock {\em Eng. Fract. Mech.}, 64(5):649--673, 1999.

\bibitem{sills2016}
R.~B. Sills and W.~Cai.
\newblock Solute drag on perfect and extended dislocations.
\newblock {\em Phil. Mag.}, 96(10):895--921, 2016.

\bibitem{sofronis1995b}
P.~Sofronis and H.~K. Birnbaum.
\newblock Mechanics of the hydrogen-dislocation-impurity
  interactions--\text{I}. increasing shear modulus.
\newblock {\em J. Mech. Phys. Solids}, 43(1):49--90, 1995.

\bibitem{sofronis2001}
P.~Sofronis, Y.~Liang, and N.~Aravas.
\newblock Hydrogen induced shear localization of the plastic flow in metals and
  alloys.
\newblock {\em Eur. J. Mech. A Solids}, 20(6):857--872, 2001.

\bibitem{song2011}
J.~Song and W.~A. Curtin.
\newblock A nanoscale mechanism of hydrogen embrittlement in metals.
\newblock {\em Acta Mater.}, 59(4):1557--1569, 2011.

\bibitem{song2013}
J.~Song and W.~A. Curtin.
\newblock Atomic mechanism and prediction of hydrogen embrittlement in iron.
\newblock {\em Nature Mater.}, 12(2):145--151, 2013.

\bibitem{song2014}
J.~Song and W.~A. Curtin.
\newblock Mechanisms of hydrogen-enhanced localized plasticity: an atomistic
  study using $\alpha$-\text{Fe} as a model system.
\newblock {\em Acta Mater.}, 68:61--69, 2014.

\bibitem{takano1974}
S.~Takano and T.~Suzuki.
\newblock An electron-optical study of $\beta$-hydride and hydrogen
  embrittlement of vanadium.
\newblock {\em Acta Matell.}, 22(3):265--274, 1974.

\bibitem{tang2003}
M.~Tang, G.~Xu, W.~Cai, and V.~V. Bulatov.
\newblock Dislocation image stresses at free surfaces by the finite element
  method.
\newblock {\em Mat. Res. Soc. Symp. Proc.}, 795, 2003.

\bibitem{tang2012}
Y.~Tang and J.~A. El-Awady.
\newblock Atomistic simulations of the interactions of hydrogen with
  dislocations in fcc metals.
\newblock {\em Phys. Rev. B}, 86(17):174102, 2012.

\bibitem{thomas1983}
G.~Thomas and W.~Drotning.
\newblock Hydrogen induced lattice expansion in nickel.
\newblock {\em Metallurgical Transactions A}, 14(8):1545--1548, 1983.

\bibitem{volkl1975}
J.~V{\"o}lkl and G.~Alefeld.
\newblock 5 - hydrogen diffusion in metals.
\newblock In A.~S. Nowick and J.~J. Burton, editors, {\em Diffusion in Solids},
  pages 231--302. Academic Press, 1975.

\bibitem{von2011}
J.~von Pezold, L.~Lymperakis, and J.~Neugebeauer.
\newblock Hydrogen-enhanced local plasticity at dilute bulk \text{H}
  concentrations: The role of \text{H-H} interactions and the formation of
  local hydrides.
\newblock {\em Acta Mater.}, 59(8):2969--2980, 2011.

\bibitem{wang2014}
S.~Wang, M.~L. Martin, P.~Sofronis, S.~Ohnuki, N.~Hashimoto, and I.~M.
  Robertson.
\newblock Hydrogen-induced intergranular failure of iron.
\newblock {\em Acta Mater.}, 69:275--282, 2014.

\bibitem{wen2011}
M.~Wen, A.~Barnoush, and K.~Yokogawa.
\newblock Calculation of all cubic single-crystal elastic constants from single
  atomistic simulation: Hydrogen effect and elastic constants of nickel.
\newblock {\em Comput. Phys. Commun.}, 182(8):1621--1625, 2011.

\bibitem{wen2009}
M.~Wen, L.~Zhang, B.~An, S.~Fukuyama, and K.~Yokogawa.
\newblock Hydrogen-enhanced dislocation activity and vacancy formation during
  nanoindentation of nickel.
\newblock {\em Phys. Rev. B}, 80(9):094113, 2009.

\bibitem{wolfer1985}
W.~G. Wolfer and M.~I. Baskes.
\newblock Interstitial solute trapping by edge dislocations.
\newblock {\em Acta Metall.}, 33(11):2005--2011, 1985.

\bibitem{zapffe1941}
C.~A. Zapffe and C.~E. Sims.
\newblock Hydrogen embrittlement, internal stress and defects in steel.
\newblock {\em Trans. AIME}, 145(1941):225--271, 1941.

\bibitem{zbib1998}
H.~M. Zbib, M.~Rhee, and J.~P. Hirth.
\newblock On plastic deformation and the dynamics of \text{3D} dislocations.
\newblock {\em Int. J. Mech. Sci.}, 40(2):113--127, 1998.

\bibitem{zhou1998}
G.~Zhou, F.~Zhou, X.~Zhao, W.~Zhang, N.~Chen, F.~Wan, and W.~Chu.
\newblock Molecular dynamics simulation of hydrogen enhancing dislocation
  emission.
\newblock {\em Sci. China E}, 41(2):176--181, 1998.

\end{thebibliography}
\end{document}